# Finite Strain Topology Optimization with Nonlinear Stability Constraints

Guodong Zhang[1], Kapil Khandelwal[2] and Tong Guo[3]

[1]Associate Professor, School of Civil Engineering, Southeast University, 319 Civil Engineering BLDG, Jiulonghu campus, Nanjing, China, 211189, gzhang@seu.edu.cn (Corresponding Author)

[2]Associate Professor, Dept. of Civil & Env. Engg. & Earth Sci., 156 Fitzpatrick Hall, University of Notre Dame, Notre Dame, IN 46556, United States, Email: kapil.khandelwal@nd.edu ORCID: 0000-0002-5748-6019, (Corresponding Author)

[3]Professor, School of Civil Engineering, Southeast University, 319 Civil Engineering BLDG, Jiulonghu campus, Nanjing, China, 211189.

*Preprint Submitted*

## Abstract

This paper proposes a computational framework for the design optimization of stable structures under large deformations by incorporating nonlinear buckling constraints. A novel strategy for suppressing spurious buckling modes related to low-density elements is proposed. The strategy depends on constructing a pseudo-mass matrix that assigns small pseudo masses for DOFs surrounded by only low-density elements and degenerates to an identity matrix for the solid region. A novel optimization procedure is developed that can handle both simple and multiple eigenvalues wherein consistent sensitivities of simple eigenvalues and directional derivatives of multiple eigenvalues are derived and utilized in a gradient-based optimization algorithm – the method of moving asymptotes. An adaptive linear energy interpolation method is also incorporated in nonlinear analyses to handle the low-density elements distortion under large deformations. The numerical results demonstrate that, for systems with either low or high symmetries, the nonlinear stability constraints can ensure structural stability at the target load under large deformations. Post-analysis on the B-spline fitted designs shows that the safety margin, i.e., the gap between the target load and the 1st critical load, of the optimized structures can be well controlled by selecting different stability constraint values. Interesting structural behaviors such as mode switching and multiple bifurcations are also demonstrated.

---

## 1 Introduction

Starting from the seminal paper by Bendsøe and Kikuchi [1] introducing homogenization-based topology optimization, the development of computational methods for structural topology optimization has undergone tremendous progress [2]. Besides the linear elastic structures on which the vast majority of the research is still focused [3], the extension to inelasticity [4, 5], finite deformations [6-8], and multi-physics [9] has also been addressed. To ensure that the optimized structures can operate under the defined loading conditions without failure, the design optimization must, accordingly, incorporate appropriate failure mechanisms. Moreover, the meaning of failure can be quite diverse depending on the context. For example, the failure can be related to the loss of structural stability [10], the emergence of plasticity in the material [11], and the degradation of the material mechanical properties [12, 13], among others. In this study, attention is focused on structures with hyperelastic materials that can sustain large strains within the elastic domain. Accordingly, the failure mechanism is defined as a "buckling-type" instability and for conservative systems, the energy criterion can be used for investigating stability [14], as compared to non-conservative systems which may require a more general Lyapunov criterion [15].

For elastic solids, the most commonly used optimization formulation is minimum compliance subject to material volume constraint also referred to as the stiffness design formulation [2, 6]. For stiffness design, the optimization process may generate slender members, and the stability considerations are then important in producing meaningful practical designs, especially under large deformations [16]. In past, the majority of the topology optimization studies with buckling constraints were devoted to truss structures, mostly limited to linear truss structures, with constraints ranging from simple local buckling constraints using the Euler buckling criterion on each member [17] to the global buckling constraints on the entire structure [18]. The extension to geometrically nonlinear buckling-constrained truss topology optimization was carried out by Li and Khandelwal [19]. In contrast to the truss structures, there are relatively fewer studies on the topology optimization of continuum structures with buckling constraints. For continuum, the first work was carried out by Neves et al. [20], where a linear buckling constraint was incorporated. The linear buckling analysis can be written as a generalized eigenvalue problem

$$(\boldsymbol{K}_0 - \lambda \boldsymbol{K}_\sigma)\boldsymbol{\phi} = \boldsymbol{0} \tag{1}$$

where $\boldsymbol{K}_0$ is the initial stiffness matrix, i.e., at zero displacements, $\boldsymbol{K}_\sigma$ is the stress stiffness matrix that is evaluated from the displacement field $\boldsymbol{u}$ solved from the linear equations $\boldsymbol{K}_0\boldsymbol{u} = \boldsymbol{P}$ where $\boldsymbol{P}$ is the applied force. The smallest eigenvalue is denoted as $\lambda_1$ represents the buckling load factor that gives the approximate buckling load $\lambda_1\boldsymbol{P}$, while the corresponding eigenvector $\boldsymbol{\phi}_1$ represents the buckling mode. It is noted that the linear buckling analysis in Eq. (1) implicitly assumes that the stress stiffness depends linearly on the loads and that the displacements at the critical point are small. Furthermore, for incorporating linear buckling constraints in a density-based topology optimization framework two main challenges have been identified: (a) the presence of spurious buckling modes in the low-density regions [10, 20]; (b) non-differentiability of multiple/repeated eigenvalues $\lambda$ [21].

In Neves et al. [20], spurious buckling modes were suppressed by ignoring the contributions from the low-density elements to the stress stiffness matrix, and the non-differentiability of multiple eigenvalues is circumvented using the generalized gradient concept. Later, in Bendsøe and Sigmund [2], different penalization schemes for the initial stiffness and stress stiffness matrices were proposed to handle the spurious buckling mode issue. Bruyneel [22] demonstrates the necessity of including enough representative buckling modes to handle the switching of buckling modes to avoid poor convergence in linear buckling optimization. The derivatives of repeated eigenvalues have been extensively studied by Seyranian et al. [21] for structural optimization, see also Refs. [23, 24]. The differentiability of the symmetric polynomials of the repeated eigenvalues has been employed in [25] for tackling the non-differentiability related to repeated eigenvalues. Other important contributions to the linear buckling topology optimization can be found in [26-28]. Recent years have seen an increasing interest in structural buckling topology optimization. For instance, Gao and Ma [29] avoided spurious buckling modes by the eigenvalue shift and mode identification via computing modal strain energy ratio; Dunning et al. [30] used an effective iterative block conjugate gradient method to solve large eigenvalue problems; Thomsen et al. [31] combined linear buckling with Bloch-wave analysis for the design of 2D periodic materials with improved buckling strength; Ferrari et al. [32] studied the use of aggregation functions for approximating the lowest eigenvalue in the buckling constraints, which was adopted by Russ and Waisman [33] for designing elastoplastic structures with damage and linear buckling constraints; Ferrari et al. [34] also developed a multilevel approach for reducing computational cost in large scale optimization problems with linearized buckling constraints; Gao et al. [35] combined linear buckling with stress constraints for the structural stiffness design with improved strength and stability.

The linear buckling analysis is only valid when the deformations are small in the pre-buckling stage and under large deformations, the linear buckling analysis can lead to erroneous results. However, the extension to nonlinear buckling analysis in the topology optimization is not yet fully investigated due to many challenges, e.g., accurate estimation of critical points or other stability indicators, nontrivial sensitivity analysis (even for simple eigenvalues) since the tangent stiffness matrix is a function of the displacement, low-density mesh distortions under large deformations, among others. For instance, Kemmler et al. [36] exploited the so-called extended system of equations to directly calculate the critical point which was constrained from below to minimize compliance. The difficulty is to find the solution to the critical point as the current solution point must be close enough which is not always guaranteed. As an alternative approach, Lindgaard and Dahl [37] proposed to approximate the critical point at a precritical load step by

$$(\boldsymbol{K}_0 + \boldsymbol{K}_u^n)\boldsymbol{\phi} = -\lambda \boldsymbol{K}_\sigma^n \boldsymbol{\phi} \tag{2}$$

where $\boldsymbol{K}_u^n$ and $\boldsymbol{K}_\sigma^n$ are the displacement stiffness and stress stiffness at the precritical $n^{\text{th}}$ step, respectively. As shown in [19], the accuracy of the approximate critical point depends on the proximity of the precritical step to the critical point. In [36, 37], concerns related to mesh distortions and multiple eigenvalues were not addressed. In a recent study by Dalklint et al. [10], the approach in [37] is extended by adopting a linear energy interpolation scheme [38] to address mesh distortion together with an aggregation function approach for tackling sensitivities of multiple eigenvalues.

On the other hand, as both are solving eigenvalue problems, structural eigenfrequency optimization shares a lot in common with structural buckling optimization. It is often formulated as a generalized eigenvalue problem with both mass matrix and stiffness matrices, as compared to the static buckling analysis where only the stiffness matrix is involved. Likewise, most works are confined to linear elasticity. For example, Pedersen [39] maximized the fundamental eigenfrequency in which a new interpolation for stiffness and mass is used in combination with a strategy to neglect nodes surrounded by only low-density elements to avoid spurious vibration modes; Du and Olhoff [40] adopted higher penalization on the mass of low-density elements to suppress spurious eigenmodes, and proposed a two-loop optimization procedure where the inner loop consider directional derivatives to handle the non-differentiability of multiple eigenvalues. This procedure was extended to the geometrically nonlinear case by Dalklint et al. [41]. Besides that, the geometrically nonlinear eigenfrequency optimization was studied by Yoon et al. [42] where the fundamental eigenfrequency of a structure at deformed configuration was maximized in an element-connectivity topological parameterization framework and it was shown that eigenfrequencies of nonlinear systems can be significantly affected by large deformations.

This study focuses on the topology optimization of structures with minimized end compliance while satisfying material volume and nonlinear stability constraints. The main contributions of this work are: (a) A novel strategy for removing spurious buckling modes based on the construction of a pseudo-mass matrix is proposed; (b) A new formulation of nonlinear stability analysis in topology optimization is considered by directly computing the eigenvalues of the tangent stiffness matrix where no other approximations are made; (c) The optimization problem is formulated to incorporate a fixed number of clusters of eigenvalues rather than a fixed number of eigenvalues so that it can handle arbitrary multiplicities of eigenvalues during the optimization process; (d) the adaptive linear energy interpolation proposed in Zhang et al. [16] which has shown robust performance in handling mesh distortions in the previous studies [13, 43, 44] is incorporated in the proposed framework; and (e) Finally, the post-analysis on the B-spline fitted optimized topologies is carried out to evaluated the stability performance of the optimized structures.

The rest of the paper is organized as follows. In Section 2, the density-based framework is briefly reviewed. The nonlinear finite element analysis with a fictitious domain approach used in topology optimization is presented in Section 3. A novel pseudo-mass matrix is developed to handle the spurious buckling modes in the fictitious region in Section 4 together with an illustrative example. Section 5 gives the derivation of the sensitivity analysis for both simple and multiple eigenvalues. In Section 6, a novel optimization formulation that is capable of handling both simple and multiple eigenvalue scenarios is presented. In Section 7, four numerical examples are carried out to demonstrate the effectiveness of the nonlinear buckling constraints in controlling the structural stability under large deformations. Finally, concluding remarks are given in Section 8.

## 2 Density-based Framework

In the density-based topology optimization [2], with finite element discretization (see Section 3), a design is parameterized by an element-wise constant density field $\rho(\boldsymbol{X})$ that indicates the presence ($\rho = 1$) or absence ($\rho = 0$) of the material in an element, where $\boldsymbol{X} \in \Omega_0$ denotes an arbitrary material point position in the undeformed reference configuration $\Omega_0$. To accommodate gradient-based optimization algorithms, the discrete density variables are relaxed to continuous values, i.e., $\rho \in [0,1]$, where $0 < \rho < 1$ represents the mixture of void and solid phases.

### *2.1 Material interpolation*

To link the density variable ($\rho$) to the structural performance measures that are to be optimized or constrained, the material properties are parameterized by $\rho$. For instance, for compressible elastic materials, Young's modulus ($E$) can be interpolated such that $E(\rho = 0) = E_{min}$, $E(\rho = 1) = E_0$ and $E_{min} < E(0 < \rho < 1) < E_0$, where $E_{min}$ is a small number to avoid singularity while $E_0$ represents Young's modulus of the solid material used in the topology optimization. In this study, hyperelastic media is considered in the topology optimization and the so-called Simplified Isotropic Material with Penalization (SIMP) approach [45, 46] is adopted, wherein Young's modulus $E(\rho)$ for an element with density $\rho$ is interpolated as

$$E(\rho) = [\epsilon + (1 - \epsilon)\rho^p]E \tag{3}$$

where $E$ represents the Young's modulus of the media inside solid element, $\epsilon = 10^{-8}$ is the lower bound that assigns void element a negligible stiffness $E_{min} = \epsilon E$ to avoid singularity, $p \geq 1$ is the penalization power that penalizes intermediate densities. Detailed descriptions of the topology optimization process are given in Section 6.

## 3 FEA considering finite deformations

Let $\Omega_0 \in \mathbb{R}^3$ be the reference configuration of a deformable continuum body with $\boldsymbol{X} \in \Omega_0$ denoting the position vector of an arbitrary material point in $\Omega_0$. It is assumed that a motion that carries the continuum body from its reference configuration to its current configuration $\Omega_t \in \mathbb{R}^3$ can be described by a smooth one-to-one mapping $\boldsymbol{\varphi}: \boldsymbol{X} \to \boldsymbol{x}$ with $\boldsymbol{u}(\boldsymbol{X}) = \boldsymbol{\varphi}(\boldsymbol{X}) - \boldsymbol{X}$, where $\boldsymbol{u}$ represents the displacement field. The associated local deformation gradient is defined by $\boldsymbol{F} \coloneqq \boldsymbol{\nabla}_{\boldsymbol{X}}\boldsymbol{\varphi}$ with $J \coloneqq \det \boldsymbol{F} > 0$, where $\boldsymbol{\nabla}_{\boldsymbol{X}}$ denotes the gradient w.r.t. the reference coordinates $\boldsymbol{X}$. The strong form of the quasi-static boundary value problem is to find the displacement field ($\boldsymbol{u}$) such that

$$\begin{cases} \boldsymbol{\nabla}_{\boldsymbol{X}}.\boldsymbol{P} = \boldsymbol{0} & \text{in } \Omega_0 \\ \boldsymbol{u} = \overline{\boldsymbol{u}} & \text{on } \partial\Omega_{0u} \\ \boldsymbol{P}.\widehat{\boldsymbol{N}} = \overline{\boldsymbol{T}} & \text{on } \partial\Omega_{0\sigma} \end{cases} \tag{4}$$

where the body force is ignored, the boundary $\partial\Omega_0$ is decomposed into the disjoint sets $\partial\Omega_{0u}$ and $\partial\Omega_{0\sigma}$ such that $\partial\Omega_0 = \partial\Omega_{0u} \cup \partial\Omega_{0\sigma}$ and $\partial\Omega_{0u} \cap \partial\Omega_{0\sigma} = \emptyset$, $\boldsymbol{P}$ is the 1st Piola-Kirchhoff (PK) stress tensor and is related to the deformation gradient $\boldsymbol{F}$ through a specified constitutive model, $\overline{\boldsymbol{u}}$ and $\overline{\boldsymbol{T}}$ are the prescribed displacement and 1st PK traction vectors, respectively, on the boundaries $\partial\Omega_{0u}$ and $\partial\Omega_{0\sigma}$ with unit outward normal $\widehat{\boldsymbol{N}}$.

The corresponding weak form is given by: find $\boldsymbol{u} \in \mathcal{U}$ such that

$$\int_{\Omega_0} \delta\boldsymbol{F}: \boldsymbol{P}\, dV - \int_{\partial\Omega_{0\sigma}} \delta\boldsymbol{u}: \overline{\boldsymbol{T}}\, dS = 0 \quad \forall\, \delta\boldsymbol{u} \in \mathcal{V} \tag{5}$$

where $\delta\boldsymbol{F} = \boldsymbol{\nabla}_{\boldsymbol{X}}\delta\boldsymbol{u}$ and the appropriate solution and variation spaces are

$$\mathcal{U} = \{\boldsymbol{u}(\boldsymbol{X}) \mid \boldsymbol{u}(\boldsymbol{X}) \in H^1(\Omega_0), \boldsymbol{u}(\boldsymbol{X}) = \overline{\boldsymbol{u}} \text{ for } \boldsymbol{X} \in \partial\Omega_{0u}\}$$
$$\mathcal{V} = \{\boldsymbol{v}(\boldsymbol{X}) \mid \boldsymbol{v}(\boldsymbol{X}) \in H^1(\Omega_0), \boldsymbol{v}(\boldsymbol{X}) = \boldsymbol{0} \text{ for } \boldsymbol{X} \in \partial\Omega_{0u}\} \tag{6}$$

in which $H^1(\Omega_0) = \{\boldsymbol{v} \mid v_i \in L^2(\Omega_0), \partial v_i/\partial X_j \in L^2(\Omega_0),\ i,j = 1,2,3\}$ and $L^2(\Omega_0)$ represents the space of square-integrable functions.

Using the Galerkin method, the same finite element (FE) mesh is used to construct finite-dimensional approximations of the spaces $\mathcal{U}$ and $\mathcal{V}$ by $\mathcal{U}^h$ and $\mathcal{V}^h$. This leads to the following system of nonlinear equilibrium equations

$$\boldsymbol{R}(\boldsymbol{u}) = \boldsymbol{F}_{int}(\boldsymbol{u}) - \boldsymbol{F}_{ext} = \boldsymbol{0}$$
$$\text{with } \boldsymbol{F}_{int}(\boldsymbol{u}) = \mathop{\mathcal{A}}_{e=1}^{n_{ele}} \boldsymbol{F}_{int}^e \text{ and } \boldsymbol{F}_{int}^e = \int_{\Omega_0^e} \boldsymbol{B}^T \boldsymbol{P}\, dV \text{ and } \boldsymbol{F}_{ext} = \mathop{\mathcal{A}}_{e\in S_\sigma} \int_{\partial\Omega_{0\sigma}^e} \boldsymbol{N}^T \overline{\boldsymbol{T}}\, dS \tag{7}$$

where the global residual $\boldsymbol{R}(\boldsymbol{u})$ is solved for the unknown displacement $\boldsymbol{u} \in \mathcal{U}^h$, and $\boldsymbol{N}$ is the shape function matrix, $\boldsymbol{B}$ is the gradient operation matrix. $\Omega_0^e$ represents the $e^{\text{th}}$ element integration domain and $n_{ele}$ is the total number of elements. $\partial\Omega_{0\sigma}^e$ represents a part of the boundary $\partial\Omega_{0\sigma}$ that is shared with $e^{\text{th}}$ element and $\partial\Omega_{0\sigma} = \bigcup_{e\in S_\sigma} \partial\Omega_{0\sigma}^e$. The external load $\boldsymbol{F}_{ext}$ is assumed to be independent of displacement. For all the examples considered in this study, the standard bilinear Q4 element for the 2D plane strain problem is used. With conforming FE mesh, Eq. (*7*) is solved using Newton-Raphson (NR) method with the tangent stiffness matrix $\boldsymbol{K}_T = \partial\boldsymbol{F}_{int}/\partial\boldsymbol{u}$ computed to achieve quadratic convergence. With non-conforming mesh (fictitious domain), a modification to $\boldsymbol{F}_{int}$ and its associated $\boldsymbol{K}_T$ is made to handle low-density elements distortion, see Section 3.2.

### *3.1 Constitutive model*

In this study, a hyperelastic media is considered in the topology optimization. In particular, a regularized neo-Hookean model is adopted with the free energy function given by

$$\psi(\boldsymbol{C}) = \frac{\kappa}{2}(J-1)^2 + \frac{\mu}{2}(\bar{I}_1 - 3) \tag{8}$$

where $\bar{I}_1 = \text{tr}(\bar{\boldsymbol{C}})$ with $\bar{\boldsymbol{C}} = J^{-2/3}\boldsymbol{C}$ and $\boldsymbol{C} = \boldsymbol{F}^T.\boldsymbol{F}$ is the right Cauchy-Green strain tensor; $\kappa$ and $\mu$ are the bulk and shear moduli which are related to Young's modulus $E$ and Poisson's ratio $\nu$ by $\kappa = E/\big(3(1-2\nu)\big)$ and $\mu = E/\big(2(1+\nu)\big)$. The 1st PK stress tensor is given by

$$\begin{gathered}\boldsymbol{P} = \frac{\partial\psi}{\partial\boldsymbol{F}} = \kappa(J-1)\frac{\partial J}{\partial\boldsymbol{F}} + \frac{\mu}{2}\frac{\partial\bar{I}_1}{\partial\boldsymbol{F}} \\ \text{with} \quad \frac{\partial J}{\partial\boldsymbol{F}} = J\boldsymbol{F}^{-T} \quad \text{and} \quad \frac{\partial\bar{I}_1}{\partial\boldsymbol{F}} = -\frac{2}{3}\bar{I}_1\boldsymbol{F}^{-T} + 2J^{-2/3}\boldsymbol{F}\end{gathered} \tag{9}$$

and the tangent modulus that is used in the calculation of the tangent stiffness matrix is

$$\begin{aligned}\mathbb{A} = \frac{\partial^2\psi}{\partial\boldsymbol{F}\partial\boldsymbol{F}} &= \kappa(J-1)J\frac{\partial\boldsymbol{F}^{-T}}{\partial\boldsymbol{F}} + \kappa(2J-1)J\boldsymbol{F}^{-T}\otimes\boldsymbol{F}^{-T} \\ &+ \frac{\mu}{2}\left(-\frac{2}{3}\bar{I}_1\frac{\partial\boldsymbol{F}^{-T}}{\partial\boldsymbol{F}} - \frac{2}{3}\boldsymbol{F}^{-T}\otimes\frac{\partial\bar{I}_1}{\partial\boldsymbol{F}} + 2J^{-\frac{2}{3}}\mathbb{I}_4 - \frac{4}{3}J^{-\frac{2}{3}}\boldsymbol{F}\otimes\boldsymbol{F}^{-T}\right)\end{aligned} \tag{10}$$

with $(\mathbb{I}_4)_{ijkl} = \delta_{ik}\delta_{jl}$ and

$$\left(\frac{\partial\boldsymbol{F}^{-T}}{\partial\boldsymbol{F}}\right)_{ijkl} = -(\boldsymbol{F}^{-1})_{li}(\boldsymbol{F}^{-1})_{jk} \tag{11}$$

*Remark:* Based on the topological parameterization using element density variable $\rho$ (see Section 2), the constitutive model is a function of the density variable $\rho$. Specifically, here the bulk ($\kappa$) and shear ($\mu$) moduli are functions of $\rho$ via the interpolation of the Young's modulus by $\rho$ in Eq. (3).

### *3.2 Fictitious domain with (adaptive) linear energy interpolation*

In density-based topology optimization, the fictitious domain approach is mostly used as it facilitates finite element analysis (FEA) of structures with evolving topologies. However, as pointed out in Refs. [6, 38], the fictitious domain approach can lead to mesh distortions of void (or low-density) elements under finite deformations which leads to numerical convergence issues during the solution process. Several strategies have been proposed to tackle this issue, e.g., neglect of the degrees of freedom (DOFs) related to void elements in the NR convergence [6], element removal and reintroduction strategy [47], novel topological parameterization by inter-element zero-length elastic links [48], element distortion indicator informed design update [49], linear energy interpolation scheme [16, 38, 50], etc. Among them, the *adaptive* linear energy interpolation method proposed by the authors [16] is adopted in this work. For completeness, this approach is briefly described in this section.

The main idea of the linear energy interpolation is to interpolate the element deformation energy by linear energy based on the density variable $\rho$ such that when $\rho \rightarrow 0$ small deformation (linear) theory is used while when $\rho \rightarrow 1$ finite deformation theory is used. To this end, the deformation gradient $\boldsymbol{F}$ is interpolated as

$$\boldsymbol{F} = \boldsymbol{I} + \eta(\rho)\boldsymbol{\nabla}_{\boldsymbol{X}}\boldsymbol{u} \quad \text{with} \quad \eta(\rho) = \frac{\exp(\beta\rho)}{\exp(c\beta) + \exp(\beta\rho)} \tag{12}$$

where $c$ and $\beta$ are interpolation parameters chosen as $\beta = 120$ and $c = c_0 = 0.08$. The element internal force in Eq. $(7)_2$ is accordingly modified to

$$\boldsymbol{F}_{int}^{e}(\boldsymbol{u}) = \int_{\Omega_0^e} \eta \boldsymbol{B}^T \boldsymbol{P} \, dV + \int_{\Omega_0^e} (1 - \eta^2) \boldsymbol{B}_L^T [\mathbb{C} : \boldsymbol{\varepsilon}] \, dV \tag{13}$$

where $\boldsymbol{B}_L$ denotes the strain-displacement matrix for the small strain measure, i.e., $[\boldsymbol{\varepsilon}] = \boldsymbol{B}_L \boldsymbol{u}$ with $\boldsymbol{\varepsilon} \coloneqq \boldsymbol{\nabla}_{\boldsymbol{X}}^{s}\boldsymbol{u}$ and $\boldsymbol{\nabla}_{\boldsymbol{X}}^{s}$ the symmetric gradient operator and $\mathbb{C}$ is the linear isotropic elastic moduli determined by the interpolated Young's modulus using Eq. (3) but with a higher penalization power $p_L$, i.e., $p_L > p$, in order to discourage using low-density values to exploit small deformation kinematics. Here, a bracket outside the tensor denotes the matrix-vector form of the tensor.

Note that in the considered adaptive strategy, the $c$ value is adaptively updated if FEA fails to converge after reaching a predefined minimum step size, i.e., $c \rightarrow c + \Delta c$ with $\Delta c$ = 0.05. As the $c$ is increased more linear kinematics is employed, and with this adaptive strategy, the FEA eventually converges when $c$ is sufficiently high. The parameter $c$ is reset to its default value $c_0$ after the convergence of FEA at the current optimization step. It is emphasized that this adaptive scheme on the interpolation parameter $c$ is crucial for improving the robustness of the linear energy interpolation method, as it is difficult to pre-select a single or a series of $c$ values that can work for all the intermediate topologies during the optimization process. The necessity of this adaptive scheme can be seen from the results in Section 7.1.

With $\boldsymbol{F}_{int}^{e}$ being replaced by Eq. (13), the global nonlinear equilibrium equation (7) is solved using the NR method with an adaptive step-size strategy. To achieve quadratic convergence, the tangent stiffness matrix is computed as

$$\boldsymbol{K}_T = \mathop{\mathcal{A}}_{e=1}^{n_{ele}} \boldsymbol{k}_T^e \quad \text{with } \boldsymbol{k}_T^e = \int_{\Omega_0^e} \eta^2 \boldsymbol{B}^T [\mathbb{A}] \boldsymbol{B} \, dV + \int_{\Omega_0^e} (1-\eta^2) \boldsymbol{B}_L^T [\mathbb{C}] \boldsymbol{B}_L \, dV \tag{14}$$

where $\mathbb{A}$ is the tangent moduli from the material subroutine, see Eq. (10) for its expression corresponding to the neo-Hookean model.

## 4 Nonlinear structural stability analysis with fictitious domain

For a conservative system, an equilibrium state is said to be stable if the potential energy of the system in that state is a proper minimum. After FE discretization with a conforming mesh, the structural stability can be assessed by examining the positive definiteness of the tangent stiffness matrix $\boldsymbol{K}_T$ (after the application of boundary conditions). That is, a structure loses stability at a critical point (limit or bifurcation) where the tangent stiffness matrix loses positive definiteness, i.e., one or more eigenvalues become zero. At a critical point that corresponds to a limit point, the magnitude of the external load that the structure can withstand cannot further increase. At the bifurcation point, the uniqueness of the solution is lost, i.e., there exist other solution branches beside the primary branch, and the stability of these branches can be similarly examined via the check of the positive definiteness of the tangent stiffness matrix.

Assuming that the eigenpairs of $\boldsymbol{K}_T$ are denoted by $(\lambda_q, \boldsymbol{\phi}_q)$, the eigenvalue problem is given by

$$(\boldsymbol{K}_T - \lambda_q \boldsymbol{I})\boldsymbol{\phi}_q = \boldsymbol{0}\,, \quad q = 1, 2, \dots, s \tag{15}$$

where $\lambda_q$ are listed in increasing order, i.e., $\lambda_1 \leq \lambda_2 \leq \cdots \leq \lambda_s$ and $s$ is the size of the tangent stiffness matrix $\boldsymbol{K}_T$. It is assumed that the boundary conditions are applied such that the rigid-body motion is suppressed and the resulting $\boldsymbol{K}_T$ is stable in the initial undeformed state. Therefore, the structure is stable at a given state with tangent stiffness $\boldsymbol{K}_T$ if $\lambda_1 > 0$. Suppose that the external load is scaled by its magnitude by a scalar $\gamma$ (often called load factor), i.e., $\boldsymbol{F}_{ext} = \gamma\widehat{\boldsymbol{P}}$ with $\widehat{\boldsymbol{P}}$ taking the expression of $\boldsymbol{F}_{ext}$ in Eq. (7), and the structure loses stability at the load factor $\gamma_{cr}$, the nature (limit or bifurcation) of the critical point ($\gamma_{cr}$) is determined by the condition

$$\boldsymbol{\phi}_c^T \widehat{\boldsymbol{P}} = \begin{cases} = 0 & \text{bifurcation point} \\ \neq 0 & \text{limit point} \end{cases} \tag{16}$$

where $\boldsymbol{\phi}_c$ is the eigenvector associated with zero eigenvalue ($\lambda_c = 0$) at a load factor $\gamma_{cr}$. A further classification of the type of instability requires higher-order derivatives [51], which is not pursued in this study.

As void and/or low-density elements are present in the fictitious domain that has relatively much smaller stiffness compared to the solid elements, spurious buckling modes in these elements can exist in the stability analysis. To suppress spurious buckling modes, a new strategy is proposed that uses a *diagonal* pseudo-mass matrix ($\boldsymbol{S}_M$) which has one for the DOFs attached to solid element(s) and a small number for the DOFs attached to void elements. The idea is similar to the interpolation strategy of a mass matrix in the dynamic case for removing spurious vibration modes [40]. Accordingly, the eigenanalysis of the tangent stiffness matrix is modified to

$$\left(\boldsymbol{K}_T - \lambda_q \boldsymbol{S}_M\right)\boldsymbol{\phi}_q = \boldsymbol{0} \ , \qquad q = 1, 2, \ldots \tag{17}$$

where for conforming mesh with all elements being solid, Eq. (17) degenerates to Eq. (15) as $\boldsymbol{S}_M = \boldsymbol{I}$, i.e., an identity matrix. As both $\boldsymbol{K}_T$ and $\boldsymbol{S}_M$ are symmetric, without loss of generality, the eigenvectors $\boldsymbol{\phi}_q$ are $\boldsymbol{S}_M$-orthonormalized, i.e., $\boldsymbol{\phi}_p^T \boldsymbol{S}_M \boldsymbol{\phi}_q = \delta_{pq}$. The construction of the pseudo-mass matrix ($\boldsymbol{S}_M$) is detailed in the following section.

### *4.1 Pseudo-mass matrix construction*

In contrast to the mass matrix interpolation in the eigenfrequency optimization case, the desired pseudo-mass matrix is diagonal and is supposed to degenerate into an identity matrix when all the elements are solid. Under this guidance, the first step is to identify, for each node, whether it belongs to a solid/high-density or void/low-density region. To this end, a nodal pseudo-density ($\varpi_i$, $i$ = 1, …, #nodes) is introduced which is defined as the maximum of the element densities that the node is surrounded by (Figure 1). For the sake of differentiability, this maximum is approximated by a $p$-norm function as

$$\varpi_i = \left(\sum_{r \in \mathcal{B}_i} \rho_r^q\right)^{1/q} \ , \qquad i = 1, \ldots, \#\text{nodes} \tag{18}$$

where $\mathcal{B}_i$ is the set of element IDs that the $i^{\text{th}}$ node is attached to, $q$ is the power of the $p$-norm here chosen as $q$ = 15. For a regular 2D domain discretized with uniform Q4 elements, $\mathcal{B}_i$ can contain 1 (convex corner node), 2 (side node), 3 (concave corner node), or 4 (inner node) element IDs depending on the nodal location, see Figure 2.

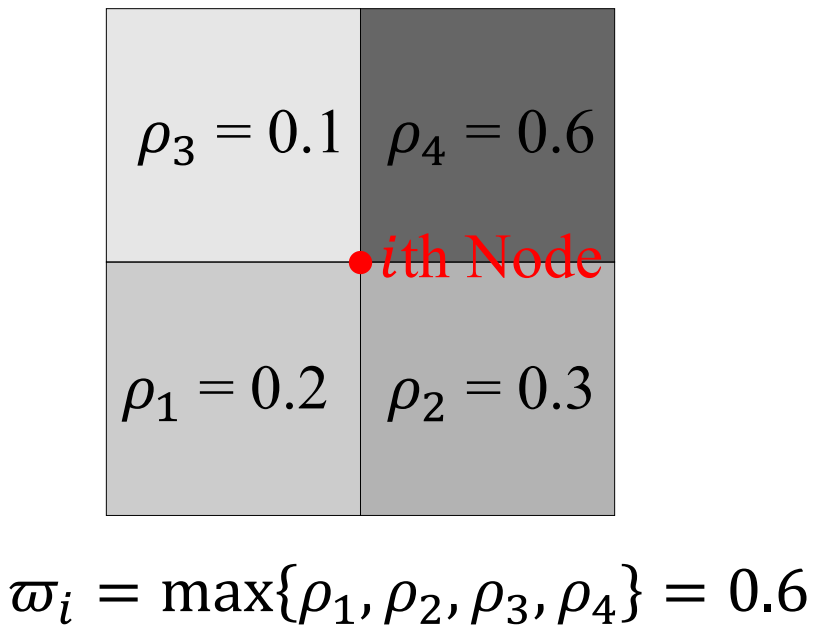

Figure 1. Illustration of the nodal pseudo density.

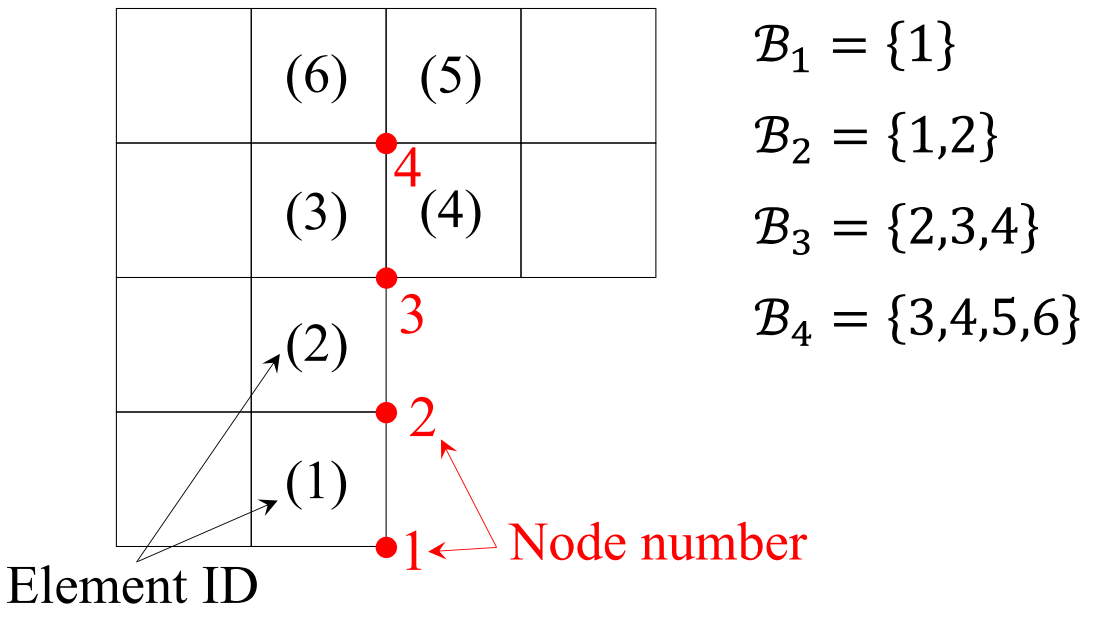

Figure 2. Illustration of the set of element IDs of different types of nodes.

Using the nodal pseudo density $\varpi_i$, the nodal pseudo-mass is interpolated by

$$\tilde{m}_i = \begin{cases} \hat{\epsilon} + (1-\hat{\epsilon})\varpi_i^{p_m} & \text{if } \varpi_i \le \varpi_L \\ a_0 + a_1\varpi_i + a_2\varpi_i^2 + a_3\varpi_i^3 & \text{if } \varpi_L < \varpi_i < \varpi_H \\ 1 & \text{if } \varpi_i \ge \varpi_H \end{cases} \tag{19}$$

where $\hat{\epsilon}$ is a small positive number to avoid singularity, $p_m$ is the penalization power, $\varpi_L$ and $\varpi_H$ are two cut-off values incorporated such that the nodal pseudo-mass is interpolated by $\varpi_i$ in a similar manner as stiffness in Eq. (3) when $\varpi_i \le \varpi_L$, and is equal to one when $\varpi_i \ge \varpi_H$. The transition between the two interpolations is carried out by a cubic polynomial with continuity on both function value and its 1st order derivative at two ends, $\varpi_i = \varpi_L$ and $\varpi_i = \varpi_H$. To this end, the parameters $a_0$, $a_1$, $a_2$ and $a_3$ are determined by the following linear equations

$$\begin{bmatrix} 1 & \varpi_L & \varpi_L^2 & \varpi_L^3 \\ 0 & 1 & 2\varpi_L & 3\varpi_L^2 \\ 1 & \varpi_H & \varpi_H^2 & \varpi_H^3 \\ 0 & 1 & 2\varpi_H & 3\varpi_H^2 \end{bmatrix} \begin{bmatrix} a_0 \\ a_1 \\ a_2 \\ a_3 \end{bmatrix} = \begin{bmatrix} \hat{\epsilon} + (1-\hat{\epsilon})\varpi_L^{p_m} \\ p_m(1-\hat{\epsilon})\varpi_L^{p_m-1} \\ 1 \\ 0 \end{bmatrix} \tag{20}$$

As an illustration, Figure 3 shows the curves of the function $\tilde{m}_i(\varpi_i)$ in Eq. (19) with cut-off values $\varpi_L = 0.1$ and $\varpi_H = 0.2$ and different $p_m$ values. As can be seen, the transitions at the two cut-offs have continuous slope.

Finally, the pseudo-mass matrix $\boldsymbol{S}_M$ is constructed as

$$\boldsymbol{S}_M = \begin{bmatrix} \widetilde{m}_1 \boldsymbol{I}_{d\times d} & \boldsymbol{0} & \boldsymbol{0} \\ \boldsymbol{0} & \ddots & \boldsymbol{0} \\ \boldsymbol{0} & \boldsymbol{0} & \widetilde{m}_{\#\text{nodes}} \boldsymbol{I}_{d\times d} \end{bmatrix} \tag{21}$$

where $\boldsymbol{I}_{d\times d}$ denotes a $d \times d$ identity matrix with $d$ representing dimension ($d$ = 2 for 2D case and $d$ = 3 for 3D case). It should be noted that the boundary conditions should be applied to $\boldsymbol{S}_M$ in the same way, as done for $\boldsymbol{K}_T$ before used in the eigenanalysis in Eq. (17). It is also noted that the construction of the nodal pseudo-mass in this section remains valid for both uniform and *non-uniform* FE meshes. The effectiveness of the stability analysis with a fictitious domain and the proposed pseudo-mass matrix is illustrated with an example in the next section.

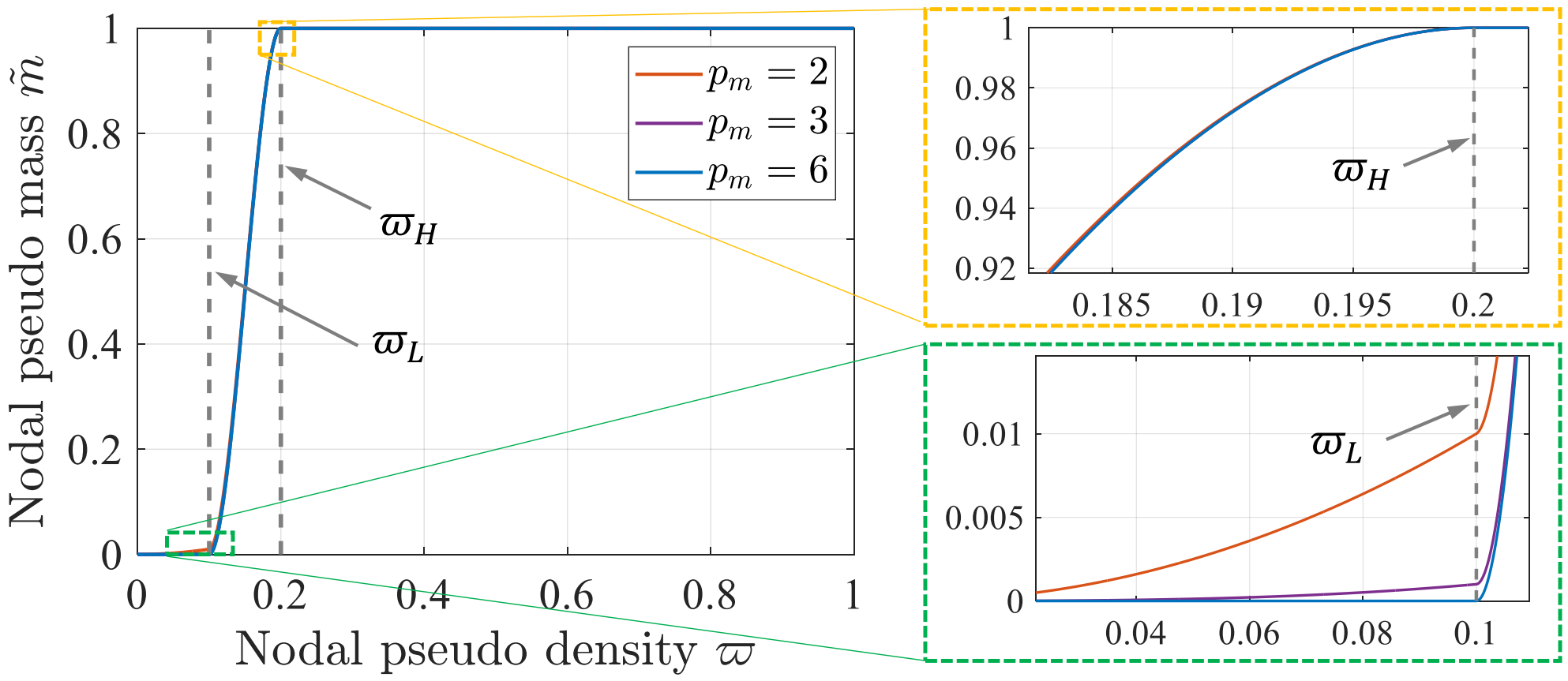

Figure 3. Nodal pseudo-mass $\widetilde{m}$ as a function of nodal pseudo-density $\varpi$ with cut-offs $\varpi_L$ = 0.1 and $\varpi_H$ = 0.2 and different power values $p_m$.

## 4.2 *Example*

In a topology design problem, the buckling modes in the low-density region should be discarded and only the buckling modes in the solid parts or the regions with high-density values should be considered. To illustrate the issue of the spurious mode the column under compressive loads shown in Figure 4a is investigated. The finite element of size 10mm×10mm is used for the discretization of all the FE meshes in Figure 4. The column is immersed in a fictitious domain by adding void elements of negligible stiffness in Figure 4b. Furthermore, the topology of the column is blurred in Figure 4c by considering four columns of solid elements, two columns of $\rho$ = 0.8, $\rho$ = 0.6, $\rho$ = 0.3 and $\rho$ = 0.1 elements, respectively, and $\rho$ = 0.001 for the rest of the elements. The material interpolation scheme in Eq. (3) is used for those intermediate densities with $\epsilon = 10^{-8}$, $p$ = 3 and $p_L$ = 6. The parameters regarding the construction of the pseudo-mass matrix are chosen as: $q$ = 15, $\hat{\epsilon} = 10^{-9}$, $p_m$ = 6, $\varpi_L$ = 0.1 and $\varpi_H$ = 0.2, see Eqns. (18)-(21).

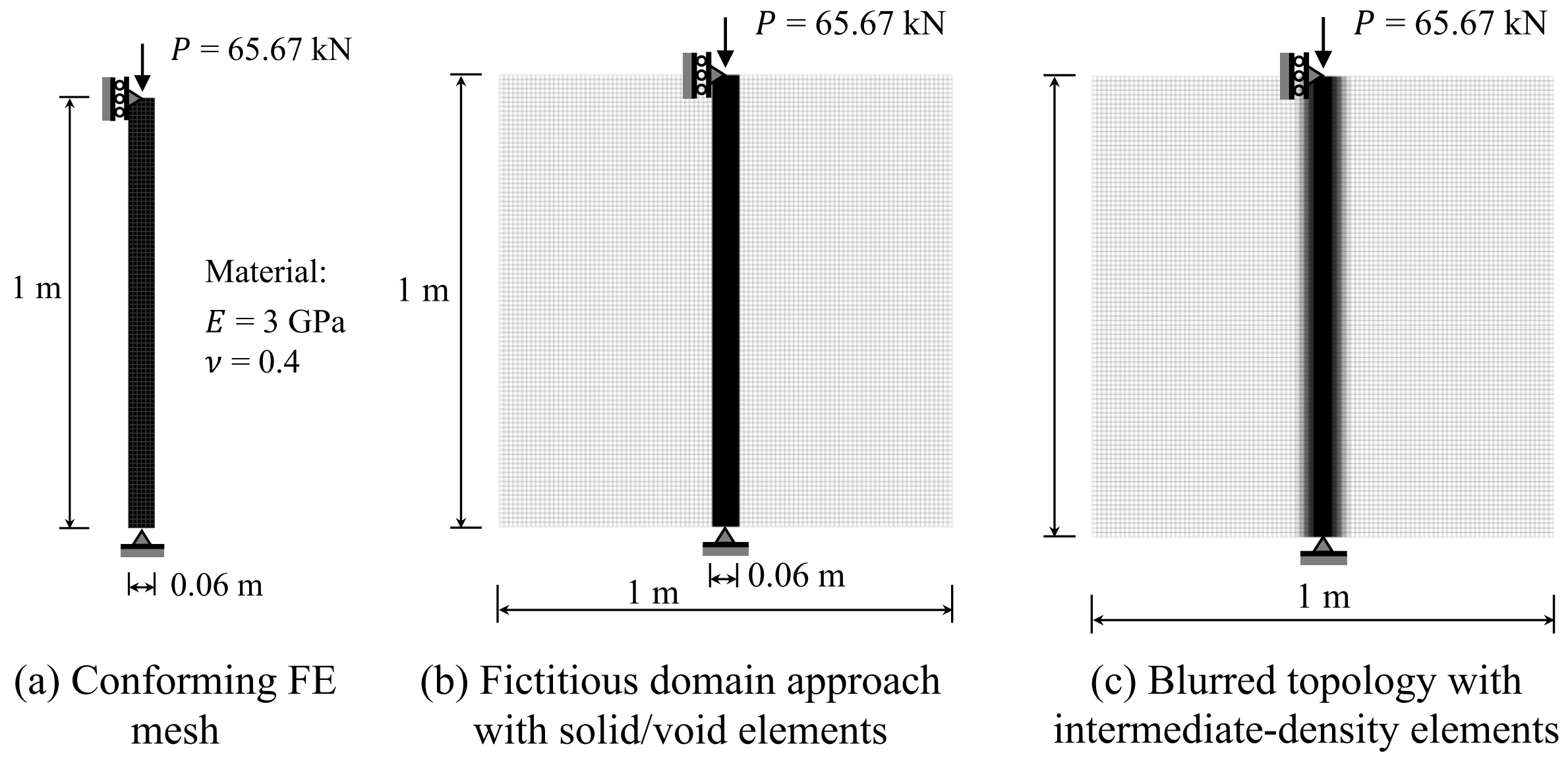

Figure 4. The nonlinear stability analysis of a compressive pinned column with conforming and fictitious domain FE meshes and intermediate-density elements.

The stability analysis is carried out at the final loading step using Eq. (15) for Figure 4 (a)-(c) and Eq. (17) for Figure 4 (b)-(c). The comparison of the first six eigenvalues of the three cases in Figure 4 is given in Table *1*, while the associated eigenmodes are plotted in Figure 5, Figure 6, and Figure 7. The comparison of Figure 5 and Figure 6a illustrates the presence of spurious modes with void elements, whereas the comparison of Figure 5 and Figure 7a demonstrates the effectiveness of the proposed pseudo-mass idea in suppressing the spurious modes. Furthermore, the eigenvalue calculation with the pseudo-mass matrix produces almost the same eigenvalues as compared to the conforming mesh, see Table 1. Due to high penalization ($p$ = 3) on the intermediate densities, the blurred topology gives consistently smaller eigenvalues. However, as shown in Figure 7b, no spurious modes are present with both void and low-density elements.

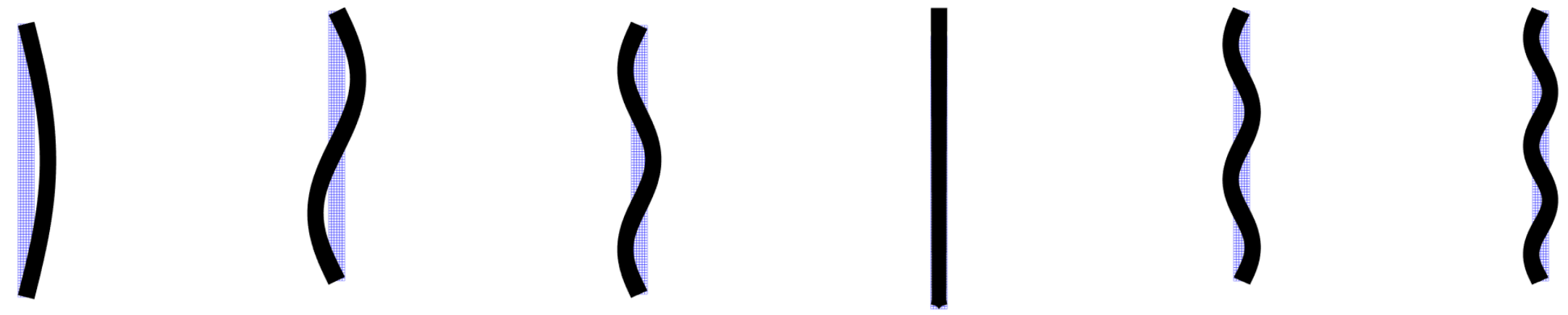

Figure 5. Eigenmodes of the conforming mesh in Figure 4a: from the left to the right are $\boldsymbol{\phi}_1,\ldots,\boldsymbol{\phi}_6$.

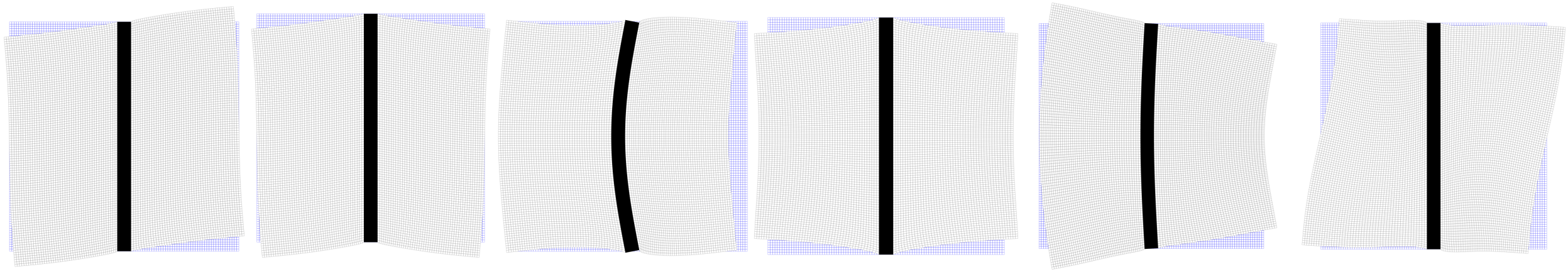

(a) Fictitious domain approach with solid/void elements

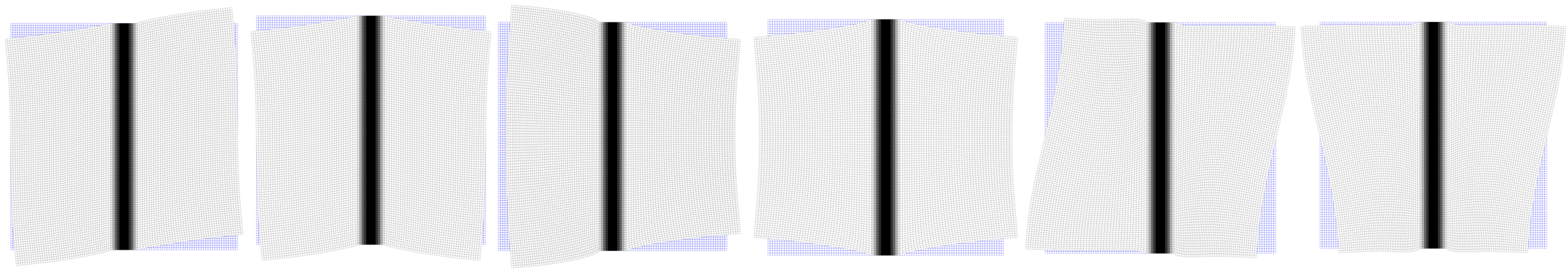

(b) Blurred topology with intermediate-density elements

Figure 6. Eigenmodes of the nonconforming meshes in Figure 4 (b) and (c) *without* pseudo-mass matrix: from the left to the right are $\boldsymbol{\phi}_1,\ldots,\boldsymbol{\phi}_6$.

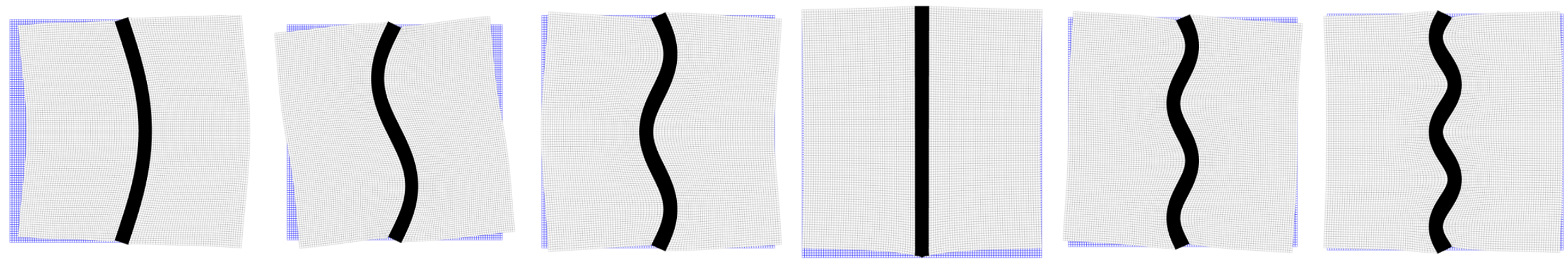

(a) Fictitious domain approach with solid/void elements

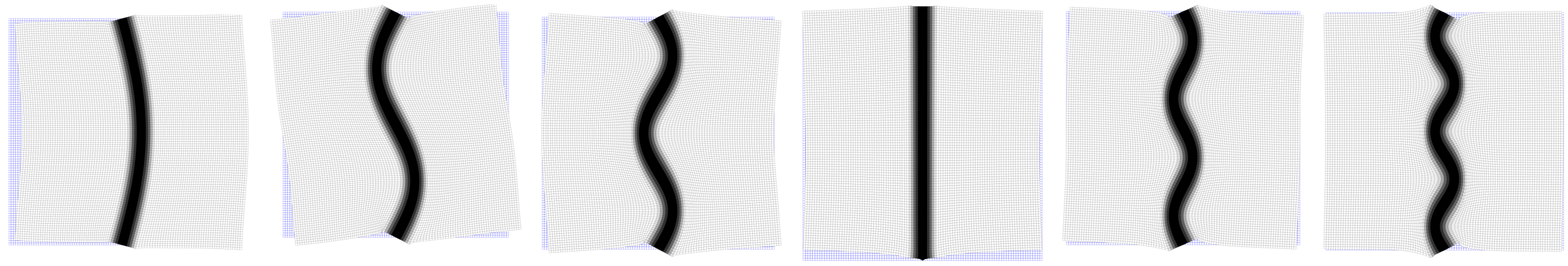

(b) Blurred topology with intermediate-density elements

Figure 7. Eigenmodes of the nonconforming meshes in Figure 4 (b) and (c) *with* pseudo-mass matrix: from the left to the right are $\boldsymbol{\phi}_1$,…, $\boldsymbol{\phi}_6$.

Table 1. Comparison of the eigenvalues by nonlinear stability analysis using Eq. (15) or (17) on conforming or fictitious domain FE meshes

| | Stability analysis with Eq. (*15*) | | | Stability analysis with Eq. (*17*) | |
|---|---|---|---|---|---|
| | Conforming FE mesh in Figure 4a | Fictitious domain with solid-void Figure 4b | Solid-intermediate-void topology in Figure 4c | Fictitious domain with solid-void Figure 4b | Solid-intermediate-void topology in Figure 4c |
| $\begin{bmatrix}\lambda_1\\ \lambda_2\\ \lambda_3\\ \lambda_4\\ \lambda_5\\ \lambda_6\end{bmatrix}$ | $\begin{bmatrix}2.8326\times10^{-8}\\ 0.01052993\\ 0.05912481\\ 0.06507443\\ 0.18120229\\ 0.41148828\end{bmatrix}$ | $\begin{bmatrix}8.9900\times10^{-10}\\ 8.9900\times10^{-10}\\ 2.0254\times10^{-9}\\ 3.1175\times10^{-9}\\ 3.4448\times10^{-9}\\ 4.2193\times10^{-9}\end{bmatrix}$ | $\begin{bmatrix}1.0418\times10^{-9}\\ 1.0418\times10^{-9}\\ 3.3478\times10^{-9}\\ 3.3482\times10^{-9}\\ 4.7736\times10^{-9}\\ 4.7737\times10^{-9}\end{bmatrix}$ | $\begin{bmatrix}3.4671\times10^{-8}\\ 0.01052995\\ 0.05912486\\ 0.06507443\\ 0.18120235\\ 0.41148836\end{bmatrix}$ | $\begin{bmatrix}8.6913\times10^{-6}\\ 0.00677015\\ 0.03760269\\ 0.03838094\\ 0.11421241\\ 0.25709518\end{bmatrix}$ |

## 5 Sensitivity analysis of simple/multiple eigenvalues

This section discusses the calculation of the derivatives of eigenvalues w.r.t the design changes as it is an important ingredient in the gradient-based optimization algorithms in topology optimization. When the eigenvalues are simple, the eigenvalues are differentiable, and the calculation details for this case are given in Section 5.1. When the eigenvalues are multiple (repeated), the eigenvalues are not differentiable and only the directional derivatives exist. In this scenario, the perturbation approach proposed by Seyranian et al. [21] can be applied to compute the directional derivatives of the multiple eigenvalues.

### *5.1 Sensitivity analysis of simple eigenvalues*

The sensitivity of simple eigenvalues w.r.t the density variable $\rho$ for linear elastic structure has been well documented in the literature for various design optimizations, e.g., Ref. [21, 29]. For geometrically nonlinear structures, the calculation of simple eigenvalue derivatives becomes more computationally involved, as the tangent stiffness matrix $\boldsymbol{K}_T$ depends on the density variable $\rho$ explicitly by the material interpolation as well as implicitly through the dependence on the displacement $\boldsymbol{u}$. The extension to geometrically nonlinear structure was carried out by Dalklint et al. [41] with an adjoint formulation which is computationally more efficient when the number of design variables exceeds the number of objective and constraint functions [52]. The adjoint approach is also adopted here, and the derivation is given below.

Suppose the nonlinear structural analysis has been carried out by solving Eq. (7)1 with $\boldsymbol{F}_{int}$ given in Eq. (13), followed by an eigen analysis in Eq. (17), the goal is to compute the derivative of the simple eigenvalue $\lambda_q$, i.e., $d\lambda_q/d\boldsymbol{\rho}$. To this end, consider the total differentiation of the adjoint equation

$$\boldsymbol{\phi}_q^T\left(\boldsymbol{K}_T - \lambda_q \boldsymbol{S}_M\right)\boldsymbol{\phi}_q + \boldsymbol{\eta}^T\boldsymbol{R} = 0 \tag{22}$$

where $\boldsymbol{\eta}$ is the adjoint vector, $\boldsymbol{\phi}_q^T\left(\boldsymbol{K}_T - \lambda_q \boldsymbol{S}_M\right)\boldsymbol{\phi}_q = 0$ and $\boldsymbol{R} = \boldsymbol{0}$ are implied from Eqns. (17) and (7)1 which hold irrespective of the change of the density variables $\boldsymbol{\rho}$. The total differentiation w.r.t. $\boldsymbol{\rho} \equiv [\rho_1 \quad \cdots \quad \rho_{nele}]$ gives

$$\begin{aligned}\frac{d\lambda_q}{d\boldsymbol{\rho}} = &\left.\frac{\partial\left(\boldsymbol{\phi}_q^T\boldsymbol{K}_T\boldsymbol{\phi}_q\right)}{\partial\boldsymbol{\rho}}\right|_{\boldsymbol{\phi}_q\ \text{fixed}} - \lambda_q \left.\frac{\partial\left(\boldsymbol{\phi}_q^T\boldsymbol{S}_M\boldsymbol{\phi}_q\right)}{\partial\boldsymbol{\rho}}\right|_{\boldsymbol{\phi}_q\ \text{fixed}} + \boldsymbol{\eta}^T\frac{\partial\boldsymbol{R}}{\partial\boldsymbol{\rho}} \\ &+ \left[\left.\frac{\partial\left(\boldsymbol{\phi}_q^T\boldsymbol{K}_T\boldsymbol{\phi}_q\right)}{\partial\boldsymbol{u}}\right|_{\boldsymbol{\phi}_q\ \text{fixed}} + \boldsymbol{\eta}^T\frac{\partial\boldsymbol{R}}{\partial\boldsymbol{u}}\right]\frac{\partial\boldsymbol{u}}{\partial\boldsymbol{\rho}}\end{aligned} \tag{23}$$

where the condition $\boldsymbol{\phi}_q\boldsymbol{S}_M\boldsymbol{\phi}_q = 1$ implied from the $\boldsymbol{S}_M$-orthonormalization and Eq. (17) are used for the simplifications. Finally, the sensitivities $d\lambda_q/d\boldsymbol{\rho}$ are given by

$$\frac{d\lambda_q}{d\boldsymbol{\rho}} = \left.\frac{\partial\left(\boldsymbol{\phi}_q^T \boldsymbol{K}_T \boldsymbol{\phi}_q\right)}{\partial \boldsymbol{\rho}}\right|_{\boldsymbol{\phi}_q \text{ fixed}} - \lambda_q \left.\frac{\partial\left(\boldsymbol{\phi}_q^T \boldsymbol{S}_M \boldsymbol{\phi}_q\right)}{\partial \boldsymbol{\rho}}\right|_{\boldsymbol{\phi}_q \text{ fixed}} + \boldsymbol{\eta}^T \frac{\partial \boldsymbol{R}}{\partial \boldsymbol{\rho}}$$

$$\text{with } \boldsymbol{\eta}^T = -\left.\frac{\partial\left(\boldsymbol{\phi}_q^T \boldsymbol{K}_T \boldsymbol{\phi}_q\right)}{\partial \boldsymbol{u}}\right|_{\boldsymbol{\phi}_q \text{ fixed}} \left[\frac{\partial \boldsymbol{R}}{\partial \boldsymbol{u}}\right]^{-1} \tag{24}$$

where the following derivatives are needed

$$\left.\frac{\partial\left(\boldsymbol{\phi}_q^T \boldsymbol{K}_T \boldsymbol{\phi}_q\right)}{\partial \boldsymbol{\rho}}\right|_{\boldsymbol{\phi}_q \text{ fixed}}, \quad \left.\frac{\partial\left(\boldsymbol{\phi}_q^T \boldsymbol{K}_T \boldsymbol{\phi}_q\right)}{\partial \boldsymbol{u}}\right|_{\boldsymbol{\phi}_q \text{ fixed}}, \quad \left.\frac{\partial\left(\boldsymbol{\phi}_q^T \boldsymbol{S}_M \boldsymbol{\phi}_q\right)}{\partial \boldsymbol{\rho}}\right|_{\boldsymbol{\phi}_q \text{ fixed}}, \quad \frac{\partial \boldsymbol{R}}{\partial \boldsymbol{\rho}}, \quad \frac{\partial \boldsymbol{R}}{\partial \boldsymbol{u}} \tag{25}$$

which are given in Appendix A.

### *5.2 Sensitivity analysis of multiple eigenvalues*

When the eigenvalues are repeated, e.g., $N$-fold multiple eigenvalues $\lambda_1 = \lambda_2 = \cdots = \lambda_N$, the associated eigenvectors are not uniquely defined, and the multiple eigenvalues are not differentiable. However, the *directional derivatives* of multiple eigenvalues can be obtained and used together with the constraints on the design change directions in the topology optimization, see Section 6. To this end, consider an $N$-fold multiple eigenvalues with value $\tilde{\lambda}$, i.e.,

$$\lambda_1 = \lambda_2 = \cdots = \lambda_N = \tilde{\lambda} \tag{26}$$

and a set of $\boldsymbol{S}_M$-orthonormalized eigenvectors $\boldsymbol{\phi}_r$, $r = 1, \dots, N$. In this case, these eigenvectors are not unique, and any orthogonal transformation of these vectors also yield $\boldsymbol{S}_M$-orthonormalized eigenvectors. For a given design perturbation $\boldsymbol{\rho} \rightarrow \boldsymbol{\rho} + \Delta\boldsymbol{\rho}$, there exist eigenvectors $\widetilde{\boldsymbol{\phi}}_r$ $(r = 1, \dots, N)$ that are continuous w.r.t the design changes. However, the determination of these continuous eigenvectors is not trivial and may require higher-order derivatives, see Friswell [23] for more details. Following the work of Seyranian et al. [21] (see also Ref. [53], Ch. 5-13), the attention is focused on directional derivatives where the perturbation of the density $\boldsymbol{\rho}$ in the direction $\Delta\boldsymbol{\rho}$ is considered, i.e., $\boldsymbol{\rho} \rightarrow \boldsymbol{\rho} + \varepsilon\Delta\boldsymbol{\rho}$, where $\varepsilon$ is the perturbation parameter. Let $\widetilde{\boldsymbol{\phi}}_r$ be the basis that is continuous w.r.t this perturbation, then there exists an orthogonal matrix $\boldsymbol{Q}$ such that $\widetilde{\boldsymbol{\phi}}_r$ can be expressed as

$$\widetilde{\boldsymbol{\phi}}_r = \sum_{k=1}^{N} Q_{rk}\boldsymbol{\phi}_k \quad , \quad r = 1, \dots, N \tag{27}$$

where $Q_{rk}$ are the unknown coefficients of the orthogonal matrix $\boldsymbol{Q}$. The resulting perturbation expansion of the multiple eigenvalues $\lambda_r$ and eigenvectors $\widetilde{\boldsymbol{\phi}}_r$ are given by

$$\begin{aligned} \lambda_r(\boldsymbol{\rho} + \varepsilon\Delta\boldsymbol{\rho}) &= \tilde{\lambda} + \varepsilon\Delta\lambda_r + \mathcal{O}(\varepsilon^2), \qquad r = 1, \dots, N \\ \widetilde{\boldsymbol{\phi}}_r(\boldsymbol{\rho} + \varepsilon\Delta\boldsymbol{\rho}) &= \widetilde{\boldsymbol{\phi}}_r + \varepsilon\Delta\widetilde{\boldsymbol{\phi}}_r + \mathcal{O}(\varepsilon^2), \;\; r = 1, \dots, N \end{aligned} \tag{28}$$

where $\tilde{\lambda} = \tilde{\lambda}(\boldsymbol{\rho})$, $\widetilde{\boldsymbol{\phi}}_r = \widetilde{\boldsymbol{\phi}}_r(\boldsymbol{\rho})$, and $\Delta\lambda_r$ and $\Delta\widetilde{\boldsymbol{\phi}}_r$ are the first-order corrections.

Considering up to the first order perturbation $\mathcal{O}(\varepsilon)$ terms, the eigenequation (17) becomes

$$\left(\boldsymbol{K}_T + \varepsilon\frac{d\boldsymbol{K}_T}{d\boldsymbol{\rho}}\Delta\boldsymbol{\rho}\right)\left(\widetilde{\boldsymbol{\phi}}_r + \varepsilon\Delta\widetilde{\boldsymbol{\phi}}_r\right) = \left(\tilde{\lambda} + \varepsilon\Delta\lambda_r\right)\left(\boldsymbol{S}_M + \varepsilon\frac{d\boldsymbol{S}_M}{d\boldsymbol{\rho}}\Delta\boldsymbol{\rho}\right)\left(\widetilde{\boldsymbol{\phi}}_r + \varepsilon\Delta\widetilde{\boldsymbol{\phi}}_r\right) \tag{29}$$

which, by expansion and dropping higher order $\mathcal{O}(\varepsilon^2)$ terms give

$$\left(\boldsymbol{K}_T - \tilde{\lambda}\boldsymbol{S}_M\right)\Delta\widetilde{\boldsymbol{\phi}}_r + \left(\frac{d\boldsymbol{K}_T}{d\boldsymbol{\rho}}\Delta\boldsymbol{\rho} - \Delta\lambda_r\boldsymbol{S}_M - \tilde{\lambda}\frac{d\boldsymbol{S}_M}{d\boldsymbol{\rho}}\Delta\boldsymbol{\rho}\right)\widetilde{\boldsymbol{\phi}}_r = \boldsymbol{0} \tag{30}$$

Next, pre-multiplying Eq. (30) with $\boldsymbol{\phi}_s^T$ $(s = 1, \dots, N)$ gives

$$\boldsymbol{\phi}_s^T\left(\frac{d\boldsymbol{K}_T}{d\boldsymbol{\rho}}\Delta\boldsymbol{\rho} - \Delta\lambda_r\boldsymbol{S}_M - \tilde{\lambda}\frac{d\boldsymbol{S}_M}{d\boldsymbol{\rho}}\Delta\boldsymbol{\rho}\right)\widetilde{\boldsymbol{\phi}}_r = \boldsymbol{0}\,, \qquad s = 1, \dots, N \tag{31}$$

which can be further rephrased by expanding $\widetilde{\boldsymbol{\phi}}_r$ using $\boldsymbol{\phi}_k$ in Eq. (*27*) as

$$\sum_{k=1}^{N} Q_{rk}\left[\boldsymbol{\phi}_s^T\left(\frac{d\boldsymbol{K}_T}{d\boldsymbol{\rho}}\Delta\boldsymbol{\rho} - \tilde{\lambda}\frac{d\boldsymbol{S}_M}{d\boldsymbol{\rho}}\Delta\boldsymbol{\rho}\right)\boldsymbol{\phi}_k - \Delta\lambda_r\delta_{sk}\right] = 0, \quad s = 1, \dots, N \tag{32}$$

Eq. (*32*) can be seen as a system of $N$ linear equations $\boldsymbol{Ax} = \boldsymbol{0}$ with $\boldsymbol{A} \in \mathbb{R}^{N\times N}$, $\boldsymbol{x} \in \mathbb{R}^{N\times 1}$, $A_{sk} = \boldsymbol{\phi}_s^T \left(\frac{d\boldsymbol{K}_T}{d\boldsymbol{\rho}}\Delta\boldsymbol{\rho} - \tilde{\lambda}\frac{d\boldsymbol{S}_M}{d\boldsymbol{\rho}}\Delta\boldsymbol{\rho}\right)\boldsymbol{\phi}_k - \Delta\lambda_r\delta_{sk}$, and $x_k = Q_{rk}$. Therefore, for $\boldsymbol{x}$ to have non-trivial solution it requires that

$$\det\left[\boldsymbol{\phi}_s^T \left(\frac{d\boldsymbol{K}_T}{d\boldsymbol{\rho}}\Delta\boldsymbol{\rho} - \tilde{\lambda}\frac{d\boldsymbol{S}_M}{d\boldsymbol{\rho}}\Delta\boldsymbol{\rho}\right)\boldsymbol{\phi}_k - \Delta\lambda_r\delta_{sk}\right] = 0\,, \quad s,k = 1,\dots,N \tag{33}$$

By introducing a series of vectors $\boldsymbol{z}_{sk} \in \mathbb{R}^{n_{ele}\times 1}$ $(s,k = 1,\dots,N)$ with

$$\boldsymbol{z}_{sk} = \begin{bmatrix}\boldsymbol{\phi}_s^T \left(\frac{d\boldsymbol{K}_T}{d\rho_1} - \tilde{\lambda}\frac{d\boldsymbol{S}_M}{d\rho_1}\right)\boldsymbol{\phi}_k & \dots & \boldsymbol{\phi}_s^T \left(\frac{d\boldsymbol{K}_T}{d\rho_{n_{ele}}} - \tilde{\lambda}\frac{d\boldsymbol{S}_M}{d\rho_{n_{ele}}}\right)\boldsymbol{\phi}_k\end{bmatrix}_{n_{ele}\times 1} \tag{34}$$

Eq. (33) can be rewritten as

$$\det[\boldsymbol{T} - \Delta\lambda_r\boldsymbol{I}_N] = 0\,, \quad s,k = 1,\dots,N \tag{35}$$

where the components of matrix $\boldsymbol{T}$ are $T_{sk} = \boldsymbol{z}_{sk}^T\Delta\boldsymbol{\rho}$, $T_{sk} = T_{ks}$, and $\boldsymbol{I}_N$ is an $N\times N$ identity matrix. Thus, the first-order corrections $\Delta\lambda_r$ in Eq. (35) are obtained via eigenanalysis on matrix $\boldsymbol{T}$. Moreover, in case if $\Delta\lambda_r \neq \Delta\lambda_s$ for $r \neq s$, the eigenvectors that are continuous w.r.t the design change can be uniquely determined from Eqns. (*32*) and (*27*), otherwise higher order corrections are required [23]. Nevertheless, the first-order corrections $\Delta\lambda_r$ are still valid in this case.

Further simplifications can be made if the perturbation vector $\Delta\boldsymbol{\rho}$ is chosen such that all the off-diagonal terms in $\boldsymbol{T}$ vanishes, i.e., $T_{sk} = \boldsymbol{z}_{sk}^T\Delta\boldsymbol{\rho} = 0$ for $r \neq s$. In this case, consider the directional derivative operator $D_{\Delta\boldsymbol{\rho}}[*]$ is defined by

$$D_{\Delta\boldsymbol{\rho}}[f] \coloneqq \lim_{h\to 0^+}\frac{f(\boldsymbol{\rho} + h\Delta\boldsymbol{\rho}) - f(\boldsymbol{\rho})}{h} \tag{36}$$

where $f(*)$ is a function of $\boldsymbol{\rho}$ that is continuous along the direction $\Delta\boldsymbol{\rho}$ with finite value in Eq. (36). Then, $D_{\Delta\boldsymbol{\rho}}[\lambda_r] = \Delta\lambda_r = \boldsymbol{z}_{rr}^T\Delta\boldsymbol{\rho}$, $(r = 1,\dots,N)$ which is a *linear function* of the direction $\Delta\boldsymbol{\rho}$.

Therefore, for this special choice of direction $\Delta\boldsymbol{\rho}$, the directional derivative of the repeated eigenvalues can be computed, and further employed in a gradient-based optimization algorithm.

As the total differentiation of $\boldsymbol{K}_T$ leads to $d\boldsymbol{K}_T/d\boldsymbol{\rho} = \partial\boldsymbol{K}_T/\partial\boldsymbol{\rho} + (\partial\boldsymbol{K}_T/\partial\boldsymbol{u})(\partial\boldsymbol{u}/\partial\boldsymbol{\rho})$ and the implicit derivative $\partial\boldsymbol{u}/\partial\boldsymbol{\rho}$ can be evaluated by the total differentiation of Eq. (7)$_1$ as $\partial\boldsymbol{u}/\partial\boldsymbol{\rho} = -[\partial\boldsymbol{R}/\partial\boldsymbol{u}]^{-1}(\partial\boldsymbol{R}/\partial\boldsymbol{\rho})$, the component of $\boldsymbol{z}_{sk}$ in Eq. (34) can be computed more efficiently by

$$[\boldsymbol{z}_{sk}]_r = \left.\frac{\partial(\boldsymbol{\phi}_s^T\boldsymbol{K}_T\boldsymbol{\phi}_k)}{\partial\rho_r}\right|_{\boldsymbol{\phi}_s,\boldsymbol{\phi}_k \text{ fixed}} - \tilde{\lambda}\left.\frac{\partial(\boldsymbol{\phi}_s^T\boldsymbol{S}_M\boldsymbol{\phi}_k)}{\partial\rho_r}\right|_{\boldsymbol{\phi}_s,\boldsymbol{\phi}_k \text{ fixed}} + \widehat{\boldsymbol{\eta}}_{sk}^T\frac{\partial\boldsymbol{R}}{\partial\rho_r}\ ,\ \ r = 1,\dots,n_{ele} \tag{37}$$

with

$$\widehat{\boldsymbol{\eta}}_{sk}^T = -\left.\frac{\partial(\boldsymbol{\phi}_s^T\boldsymbol{K}_T\boldsymbol{\phi}_k)}{\partial\boldsymbol{u}}\right|_{\boldsymbol{\phi}_s,\boldsymbol{\phi}_k \text{ fixed}}\left[\frac{\partial\boldsymbol{R}}{\partial\boldsymbol{u}}\right]^{-1} \tag{38}$$

So, the terms that remain to be calculated are

$$\left.\frac{\partial(\boldsymbol{\phi}_s^T\boldsymbol{K}_T\boldsymbol{\phi}_k)}{\partial\boldsymbol{\rho}}\right|_{\boldsymbol{\phi}_s,\boldsymbol{\phi}_k \text{ fixed}},\quad \left.\frac{\partial(\boldsymbol{\phi}_s^T\boldsymbol{K}_T\boldsymbol{\phi}_k)}{\partial\boldsymbol{u}}\right|_{\boldsymbol{\phi}_s,\boldsymbol{\phi}_k \text{ fixed}},\quad \left.\frac{\partial(\boldsymbol{\phi}_s^T\boldsymbol{S}_M\boldsymbol{\phi}_k)}{\partial\boldsymbol{\rho}}\right|_{\boldsymbol{\phi}_s,\boldsymbol{\phi}_k \text{ fixed}} \tag{39}$$

$s, k = 1, \dots, N$

and are given in Appendix A.

## 6 Topology optimization

The goal is to design structures with minimum end compliance for a given amount of material and specified stability objective under the given loads. The stability objective is specified in terms of the constraints on the first (smallest) $m$ *different* eigenvalues of the tangent stiffness matrix at the final deformation stage. It is noted that $m$ should be large enough to incorporate all the relevant buckling modes for handling mode switching. If the first $m$ eigenvalues are always simple for all the optimization iterations, gradient-based optimization procedures can be employed for optimization, as the design sensitivities can be computed (Section 5.1). However, if some of the eigenvalues are multiple, then only directional derivatives are available (Section 5.2) and appropriate optimization formulations will be required that can employ these directional derivatives.

### *6.1 Optimization formulations*

Inspired by [40], a new optimization procedure is developed to handle the presence of multiple eigenvalues during optimization. The overall idea is to optimize the design variables directly when all the eigenvalues are simple and to optimize design variables increment when there are multiple eigenvalues via a sub-optimization problem. The two optimization formulations are described next.

*Case (a): Simple Eigenvalues*

For optimization iterations that involve only *simple* eigenvalues of $\boldsymbol{K}_T$, i.e., $\lambda_1 < \lambda_2 < \cdots < \lambda_m < \lambda_{m+1}$, the optimization problem is formulated as

$$\begin{aligned} \min_{\boldsymbol{x}} \quad & f_0(\boldsymbol{x}) = \boldsymbol{F}_{ext}^T \boldsymbol{u} \\ \text{s.t.} \quad & f_1(\boldsymbol{x}) = V(\boldsymbol{x})/V_f - 1 \leq 0 \\ & f_{q+1}(\boldsymbol{x}) = 1 - \lambda_q/\hat{\lambda} \leq 0\,, \qquad q = 1, \dots m \\ & \boldsymbol{0} \leq \boldsymbol{x} \leq \boldsymbol{1} \end{aligned} \tag{40}$$

*Case (b): Multiple Eigenvalues*

For optimization iterations that involve *multiple* eigenvalues of $\boldsymbol{K}_T$ (i.e., $\lambda_{k-1} = \lambda_k$ for some $k$), say at $n^{\text{th}}$ iteration, a sub-optimization problem is formulated in terms of design increments. Let $\boldsymbol{x}_n$ be the design variable at $n^{\text{th}}$ iteration, the goal is then to optimize the incremental design update $\Delta\boldsymbol{x}$ where $\boldsymbol{x}_{n+1} = \boldsymbol{x}_n + \Delta\boldsymbol{x}$. To this end, the sub-optimization problem is constructed using linear approximations of compliance, volume, and stability constraints at $\boldsymbol{x} = \boldsymbol{x}_n$, and the eigenvalue constraints are applied to the first $m$ clusters of different eigenvalues, e.g., $\underbrace{\lambda_1 = \cdots = \lambda_{N_1}}_{\text{cluster-1}} < \underbrace{\lambda_{N_1+1} = \cdots = \lambda_{N_1+N_2}}_{\text{cluster-2}} < \cdots < \underbrace{\lambda_{N_1+\cdots+N_{m-1}+1} = \cdots = \lambda_{N_1+\cdots+N_m}}_{\text{cluster-}m}$ with $\bar{m} \coloneqq N_1 + \cdots + N_m$ . The sub-optimization problem is formulated as

$$\begin{aligned}
&\min_{\Delta x} \ \check{f}_0(\Delta \boldsymbol{x}) = \left.\frac{df_0}{d\boldsymbol{x}}\right|_{\boldsymbol{x}_n} \Delta \boldsymbol{x} \\
&\text{s.t.} \ \ \check{f}_1(\Delta \boldsymbol{x}) = V(\boldsymbol{x}_n + \Delta \boldsymbol{x})/V_f - 1 \le 0 \\
&\qquad \check{f}_{q+1}(\Delta \boldsymbol{x}) = 1 - \left(\lambda_q + \Delta\lambda_q\right)/\hat{\lambda} \le 0, \qquad q = 1, \dots \bar{m} \\
&\qquad \check{f}_q(\Delta \boldsymbol{x}) = g_{sk}(\Delta \boldsymbol{x}) = 0, \qquad j = (\bar{m}+2), \dots, (\bar{m}+1+c_{nt}) \\
&\qquad (\text{with } \ g_{sk}(\Delta \boldsymbol{x}) = \boldsymbol{z}_{sk}^T \Delta \boldsymbol{\rho}, \ \ s > k, \ \ \text{for multiple eigenvalues}) \\
&\qquad \max\{-\boldsymbol{x}_n, -\theta \boldsymbol{1}\} \le \Delta \boldsymbol{x} \le \min\{(\boldsymbol{1} - \boldsymbol{x}_n), \theta \boldsymbol{1}\}
\end{aligned} \tag{41}$$

In the above formulations, for addressing mesh-dependency and checkerboard issues [54], a density filter [7, 55] is used that maps the design variables $\boldsymbol{x}$ to the density variables $\boldsymbol{\rho}$ by

$$\boldsymbol{\rho} = \boldsymbol{W}\boldsymbol{x} \quad \text{with} \quad W_{pq} = \frac{w_{pq} v_q}{\sum_{q=1}^{n_{ele}} w_{pq} v_q} \quad \text{and} \quad w_{pq} = \max\left(r_{min} - \left\|\boldsymbol{X}_p - \boldsymbol{X}_q\right\|, 0\right) \tag{42}$$

in which $r_{min}$ is the filter radius, $v_q$ the volume of $q^{\text{th}}$ element and $\boldsymbol{X}_p$ the coordinates of the centroid of the $p^{\text{th}}$ element. This filter is also applied to the density increments, i.e., $\Delta\boldsymbol{\rho} = \boldsymbol{W}\Delta\boldsymbol{x}$, in the sub-optimization problem, i.e., case (b) above.

In Eq. (40) and Eq. (41), $\hat{\lambda} > 0$ is a user-defined threshold value on the eigenvalues of $\boldsymbol{K}_T$ that are enforced for the structural stability, $V_f$ denotes the allowable material volume fraction, and the material volume of the design, $V(\boldsymbol{x})$, is calculated by

$$V(\boldsymbol{x}) = \frac{1}{V_s} \boldsymbol{\rho}^T \widetilde{\boldsymbol{v}} \tag{43}$$

where $\widetilde{\boldsymbol{v}}$ is a vector of element volumes and $V_s$ is the total volume of the design domain $\Omega_0$.

In Eq. (41), the constraints $g_{sk}(\Delta\boldsymbol{x}) = 0$ are used to achieve vanishing off-diagonal terms in the matrix $[T_{sk}]$ where $T_{sk} = \boldsymbol{z}_{sk}^T\Delta\boldsymbol{\rho}$, $(s, k = 1, \dots, N)$, for some $N$-fold multiple eigenvalues, such that the increments of the multiple eigenvalues are $\Delta\lambda_r = \boldsymbol{z}_{rr}^T\Delta\boldsymbol{\rho}$ $(r = 1, \dots, N)$, which implies a linear dependence of $\Delta\lambda_r$ on $\Delta\boldsymbol{\rho}$. Suppose there are $k$ multiple eigenvalues of $N_1$, …, $N_k$ multiplicities in the first $m$ clusters of eigenvalues, then the number of this set of constraints, $c_{nt}$, can be computed as $c_{nt} = \sum_{r=1}^{k} N_r(N_r - 1)/2$ which is due to the fact that $\boldsymbol{z}_{sk} = \boldsymbol{z}_{ks}$. For example, considering the first six clusters of eigenvalues ($m$ = 6) and assuming that $\underbrace{\lambda_1}_{\text{cluster-1}} < \underbrace{\lambda_2 = \lambda_3}_{\text{cluster-2}} < \underbrace{\lambda_4 = \lambda_5 = \lambda_6}_{\text{cluster-3}} < \underbrace{\lambda_7}_{\text{cluster-4}} < \underbrace{\lambda_8 = \lambda_9}_{\text{cluster-5}} < \underbrace{\lambda_{10}}_{\text{cluster-6}}$ ($N_1$ = 2, $N_2$ = 3 and $N_3$ = 2), $c_{nt}$ = 1 + 3 + 1 = 5 and Eq. $(41)_4$ has to be considered separately for $N_1$-fold eigenvalues $\lambda_2 = \lambda_3$, $N_2$-fold eigenvalues $\lambda_4 = \lambda_5 = \lambda_6$ and $N_3$-fold eigenvalues $\lambda_8 = \lambda_9$.

Finally, in the box constraint in Eqns. (40) and (41), $\mathbf{1}$ is a vector of ones, and the *min* and *max* operators are applied in an element-wise manner. As the sub-optimization formulation for the design increment $\Delta\boldsymbol{x}$ in Eq. (41) consider linearized approximations at the last converged design $\boldsymbol{x}_n$, the overall design change should be confined to a small neighborhood of $\boldsymbol{x}_n$. Therefore, a move limit parameter $\theta$ is introduced to control the step size of the design increments $\Delta\boldsymbol{x}$. In this study, a move limit value $\theta \leq 0.1$ is used. A flowchart of the overall topology optimization process is given in Figure 8. Detailed descriptions related to the optimization algorithms and associated parameter settings are given in the following subsection.

*Remarks:*

1. The equality (equilibrium and eigenvalue) constraints in Eqns. $(7)_1$ and (17) that specify the dependence of displacement $\boldsymbol{u}$ and eigenpairs $(\lambda_r, \boldsymbol{\phi}_r)$ on the design variables $\boldsymbol{x}$ are not enforced by the optimization algorithm but through finite element analysis and eigenanalysis, respectively.

2. The number of constraints in Eq. (41) may change during optimization iterations due to the change in the multiplicity of the eigenvalues. However, this creates no difficulties as the solution of $\Delta\boldsymbol{x}$ from Eq. (41) is an independent optimization problem.

3. In the design phase, the stability constraints are only enforced at the final load step, and it is assumed that the system is stable below this load. This is indeed the case for all the numerical examples considered in this study.

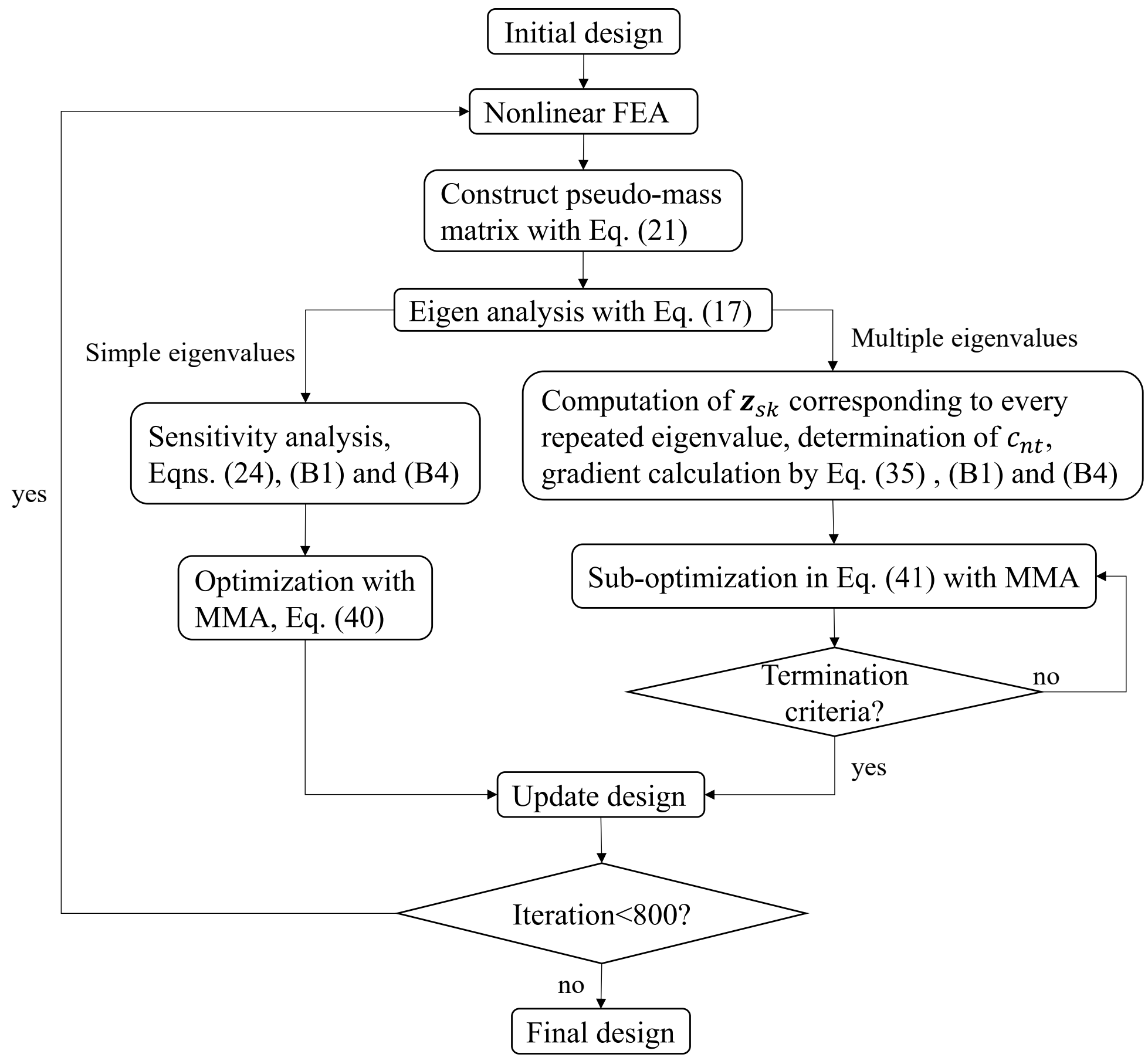

Figure 8. Flow chart of the finite strain topology optimization with nonlinear buckling constraints.

## 6.2 Optimization algorithms

The method of moving asymptotes (MMA) [56] is used for both the optimization problem in Eq. (40) as well as the sub-optimization problem in Eq. (41), wherein the default parameter values are used unless otherwise specified. The optimization process sketched in Figure 8 consists of two main loops for solving Eq. (40) and Eq. (41), which are referred to as outer-loop and inner-loop, respectively. The inner loop is only present when there are repeated eigenvalues in the eigenvalue clusters incorporated. The *inner loop* refers to the fact that a full sub-optimization problem (Eq. (41)) is solved to update the design variable at a single iteration of the outer loop.

### *6.2.1 Outer loop*

In the outer loop of the optimization, there are optimization parameters, i.e., moving upper and lower asymptotes in the MMA algorithm that are iteratively updated by the MMA optimizer. If there are optimization iterations that have multiple eigenvalues, then sub-optimizations problems have to be formulated using Eq. (41) and solved for each of these iterations with multiple eigenvalues. In this case, the upper and lower asymptotes in the outer loop MMA algorithm are reinitiated to their initial values after the inner loop sub-optimization is completed. The criterion for outer-loop termination is based on the total number of iterations, which is chosen as 800 in this study.

### *6.2.2 Inner loop*

If the considered clusters of eigenvalues of some intermediate design include multiple eigenvalues, the sub-optimization problem in Eq. (41) becomes activated and is again solved using the MMA optimizer. The criteria for terminating the sub-optimization iterations are based on: (a) the change of two successive (linear) objective function values normalized by its initial value, i.e., $\left|\check{f}_0^{(I+1)} - \check{f}_0^{(I)}\right|/|f_0| \leq 10^{-8}$, or (b) the 2-norm of two successive design variables difference, i.e., $\left\|\Delta\boldsymbol{x}^{(I+1)} - \Delta\boldsymbol{x}^{I}\right\|_2 \leq 10^{-4}$, or (c) the total number of inner loop iterations, i.e., $I > 100$, where $I$ denotes the sub-optimization iteration number.

## 7 Numerical examples

Before topology optimization, the correctness of the sensitivity analysis is verified. With the derivations given in Section 5, Appendices A and B, the verification of the sensitivity analysis by the central difference method is given in Appendix C. In the topology optimization, the optimization starts with an initial design consisting of a homogeneous distribution of the target volume fraction values (i.e., $\rho = V_f$, where $V_f$ is the threshold value in the volume fraction constraint in Eqns. (40) and (41)) and is terminated after 800 iterations (Section 6.2.1) when a discrete topology has emerged. Except for the pinned column example in Section 7.2 where $m$ = 4 clusters of different eigenvalues are included, in rest of the examples $m$ = 6 clusters of eigenvalues are included in the nonlinear stability constraints in Eqns. (40) and (41). In the numerical implementation, the two eigenvalues are considered to be equal if their absolute difference is less than 1.0E-8. In the following examples, the move limit for the sub-optimization in Eq. (41) is chosen as $\theta$ = 0.04, except for the symmetric compressive block example in Section 7.4 where $\theta$ = 0.1. A continuation scheme is used to slowly increase the penalization powers $p$ [Eq. (3)] and $p_L$ [see descriptions following Eq. (*13*)] to avoid analysis failure during early optimization iterations due to large deformations, and to relax the nonconvexity of the optimization problem. Specifically, $p$ is raised from 1 to 3 @ 0.1 per 5 iterations; $p_L$ is increased from 4 to 6 @ 0.1 per 5 iterations. Besides, the penalization power $p_m$ in the nodal pseudo-mass construction [Eq.

($19$)] is also increased from 1 to 6 @ 0.1 per 5 iterations, while all other parameters related to the pseudo-mass matrix are chosen the same values as those in Section 4.2, i.e., $q$ = 15, $\hat{\epsilon}$ = $10^{-9}$, $\varpi_L$ = 0.1 and $\varpi_H$ = 0.2. All the optimization examples consider load control and the solver in the finite element analysis uses the Newton–Raphson scheme with an adaptive step-size strategy and convergence is assumed when the energy residual drops below 1.0E-12 [57].

The B-spline fitting of the optimized designs for post-performance analyses is carried out using Rhino [58] with a level-set value of 0.5 unless otherwise stated. For the post-buckling analysis, the branch-switching method described in Appendix D is used that can effectively capture multiple secondary branches emanating from the bifurcation point, while the cylindrical arc-length method [57] is adopted to trace primary/secondary solution paths. All the numerical computations are carried out in a Matlab-based in-house finite element library CPSSL-FEA developed at the University of Notre Dame.

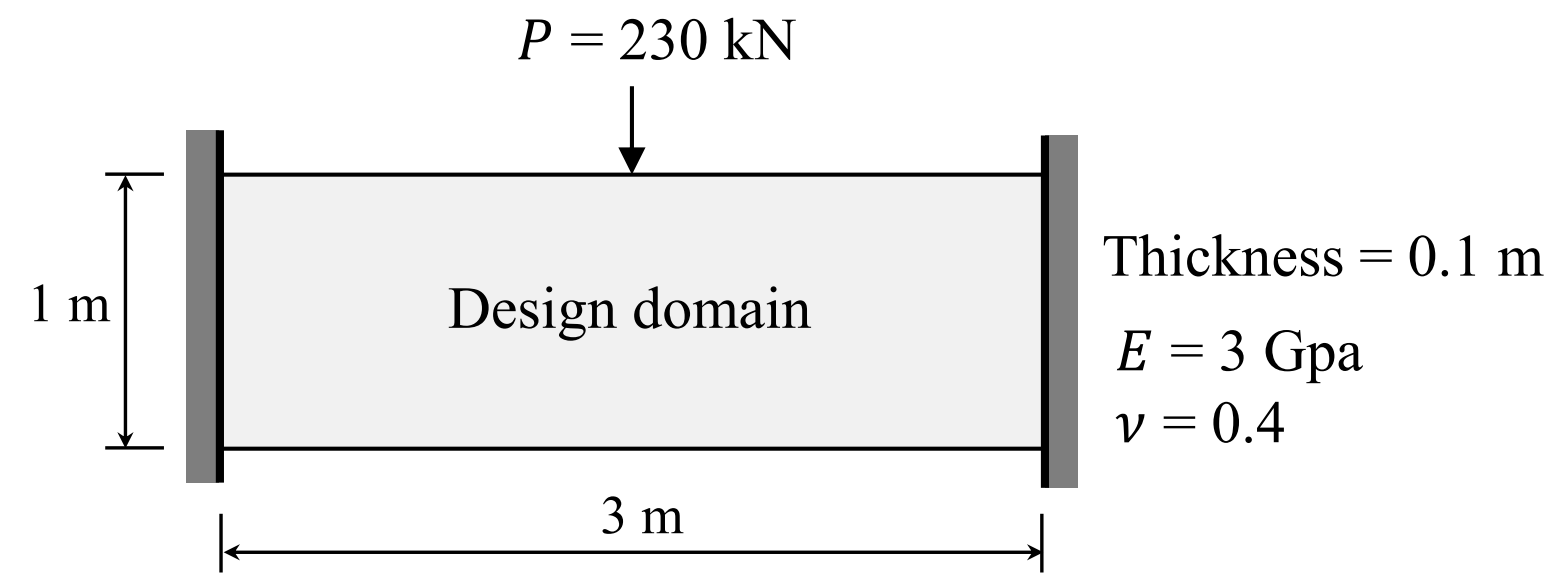

Figure 9. Sketch of the double-clamped beam problem.

### *7.1 Double clamped beam*

The first example considers the double-clamped beam problem sketched in Figure 9 with the design domain discretized by a 120×40 FE mesh. The volume constraint is considered with $V_f$ = 0.1, and the density filter radius $r_{min}$ = 50 mm. Due to the symmetry of the problem, the optimization space considers only half of the design domain, while the full domain is included in the nonlinear FEA to capture asymmetric buckling modes. This example has been examined in various works [6, 8, 38, 48] and is challenging for two reasons: (a) There is a linear-type design featured by a flipped V-shape [38] and a nonlinear-type design that has features shown in Figure 10a, and these two designs compete with each other during the optimization process. The linear design cannot support the load as the penalization power $p$ is raised to a high value and this leads to FEA failure. (b) Even if the design with nonlinear features emerges during the optimization process, the FEA may still fail as the design can be unstable depending on the overall depth of the design and other unstable features.

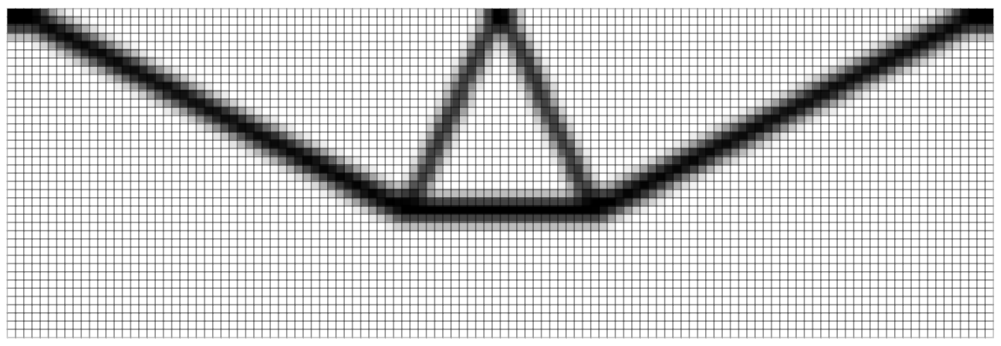

(a) Optimized topology at iteration #102 for which FEA fails

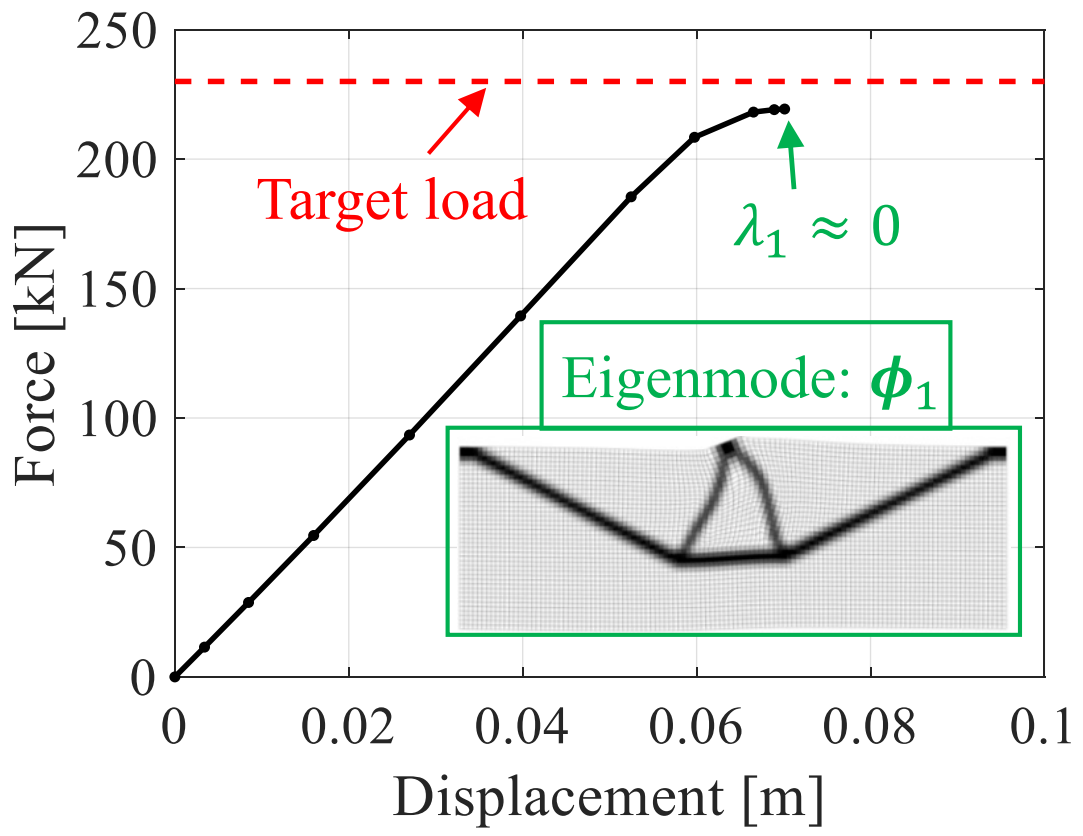

(b) Load-displacement curve of the optimized topology reaching a critical point before the target load

Figure 10. Topology optimization of the double-clamped beam without buckling constraint.

The first difficulty can be overcome by using a relatively fast continuation scheme [38] or starting with a higher penalization power $p$, say 2 as done in [59], or by choosing an initial design that is perturbed to favor nonlinear features [8]. The second issue, however, cannot be remedied without stability constraints. To demonstrate this, Figure 10a shows an optimized topology without stability constraints, i.e., Eq. (40) with $f_q$ $(q = 2, \ldots, m + 1)$ being removed. The optimization is initiated by a homogeneous design and runs with a continuation scheme: $p$ starts with 1.6 and increased by 0.2 every 25 iterations. The optimization is interrupted at the 102$^{th}$ iteration by a FEA failure wherein a critical point is reached before the target load, see the plotted load-displacement curve in Figure 10b where the displacement is measured at the loading point, i.e., the middle node on the top surface.

Table 2. Four cases of topology optimization of the double-clamped beam with different stability threshold values $\hat{\lambda}$

| Case number | Stability threshold $\hat{\lambda}$ | Minimized objective value $f_0$ [kN·m] | First six eigenvalues of the optimized topology |
|---|---|---|---|
| Case-1 | 0.002 | 18.21 | $\begin{bmatrix}\lambda_1\\\lambda_2\\\lambda_3\\\lambda_4\\\lambda_5\\\lambda_6\end{bmatrix}=\begin{bmatrix}0.00200001\\0.00326693\\0.00597627\\0.01280943\\0.01591148\\0.03901548\end{bmatrix}$ |
| Case-2 | 0.004 | 21.94 | $\begin{bmatrix}\lambda_1\\\lambda_2\\\lambda_3\\\lambda_4\\\lambda_5\\\lambda_6\end{bmatrix}=\begin{bmatrix}0.00400000\\0.00606846\\0.01731487\\0.01828198\\0.02475417\\0.05668777\end{bmatrix}$ |
| Case-3 | 0.006 | 27.31 | $\begin{bmatrix}\lambda_1\\\lambda_2\\\lambda_3\\\lambda_4\\\lambda_5\\\lambda_6\end{bmatrix}=\begin{bmatrix}0.00599953\\0.00599969\\0.02345340\\0.02457844\\0.03094318\\0.06674226\end{bmatrix}$ |
| Case-4 | 0.008 | 46.38 | $\begin{bmatrix}\lambda_1\\\lambda_2\\\lambda_3\\\lambda_4\\\lambda_5\\\lambda_6\end{bmatrix}=\begin{bmatrix}0.00800098\\0.00919497\\0.02950062\\0.03670040\\0.05752595\\0.05953313\end{bmatrix}$ |

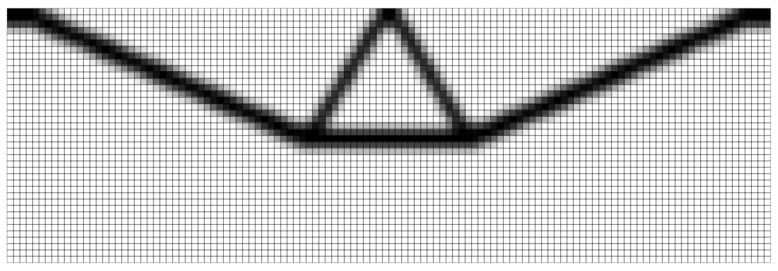

(a) Case-1

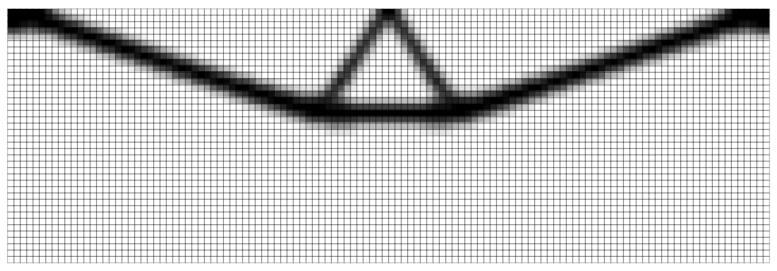

(b) Case-2

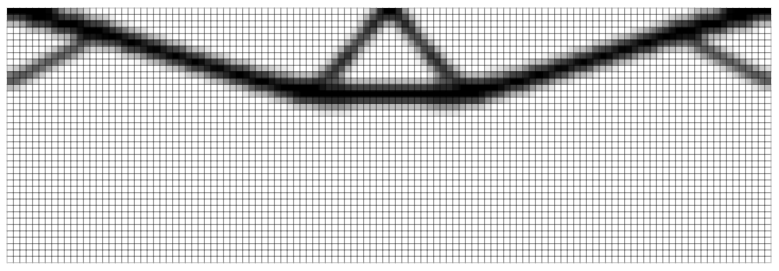

(c) Case-3

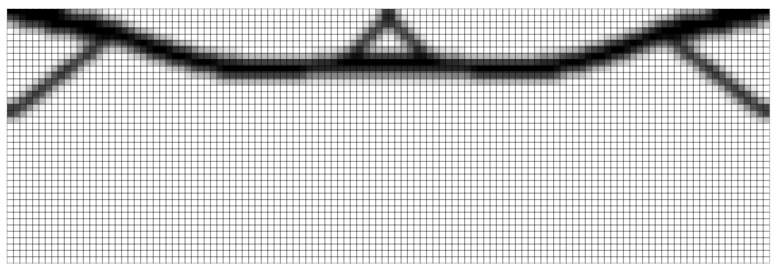

(d) Case-4

Figure 11. Optimized topologies for the four cases listed in Table *2*.

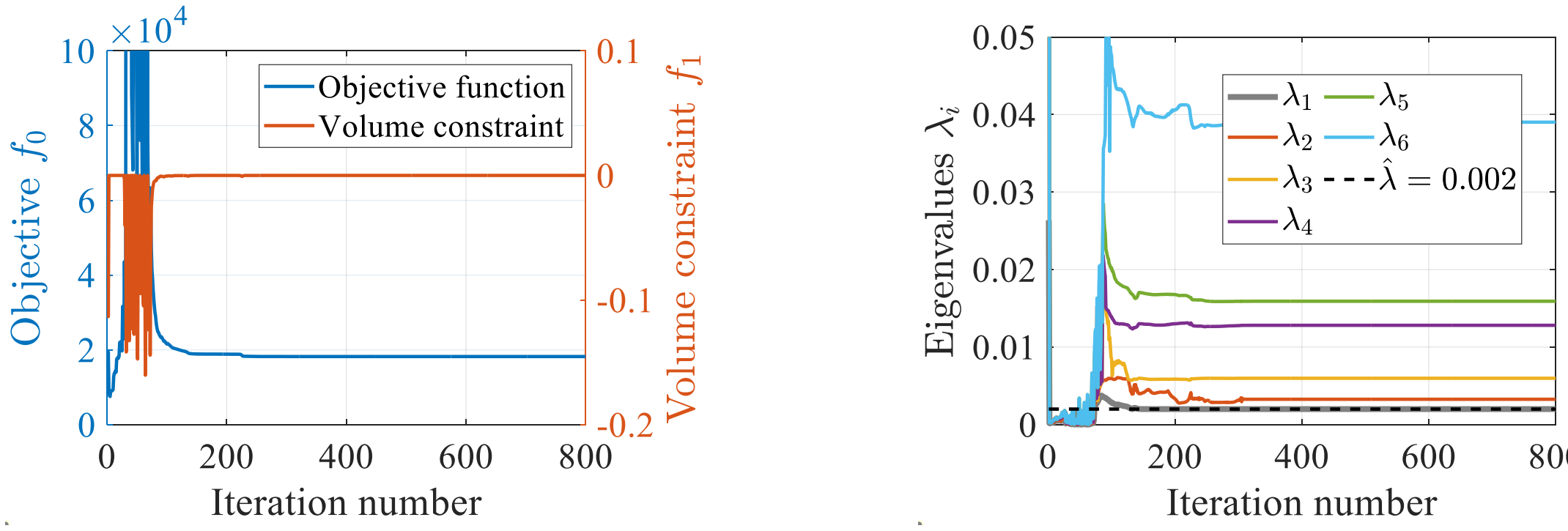

(a) Case-1

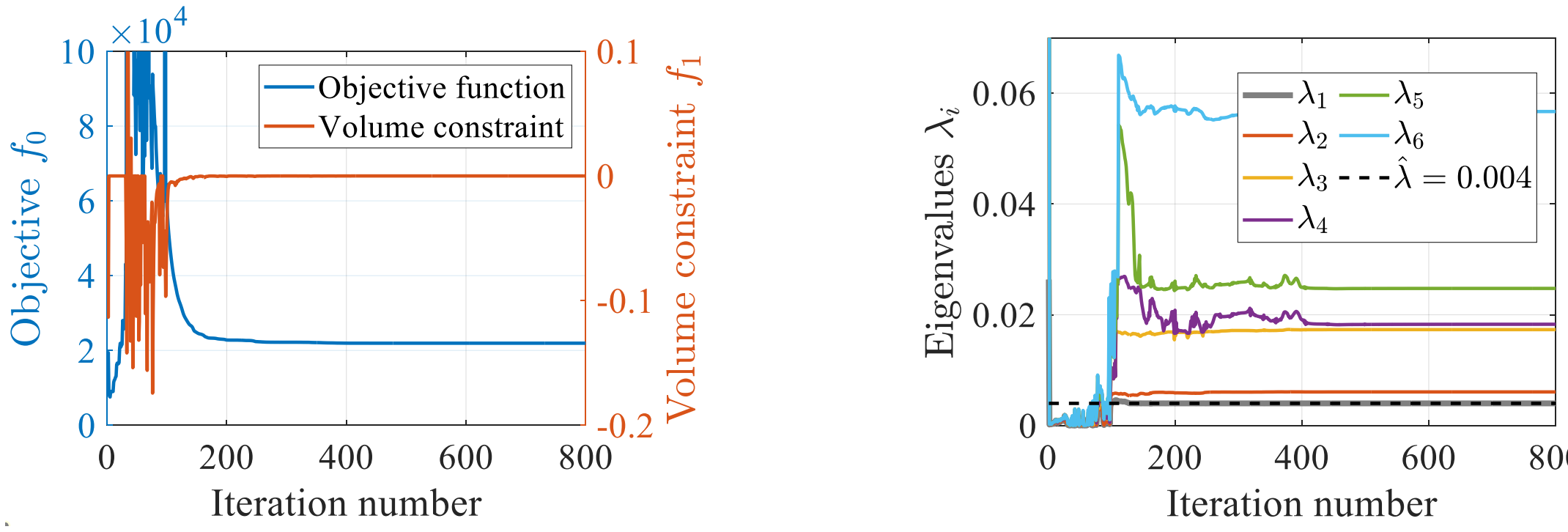

(b) Case-2

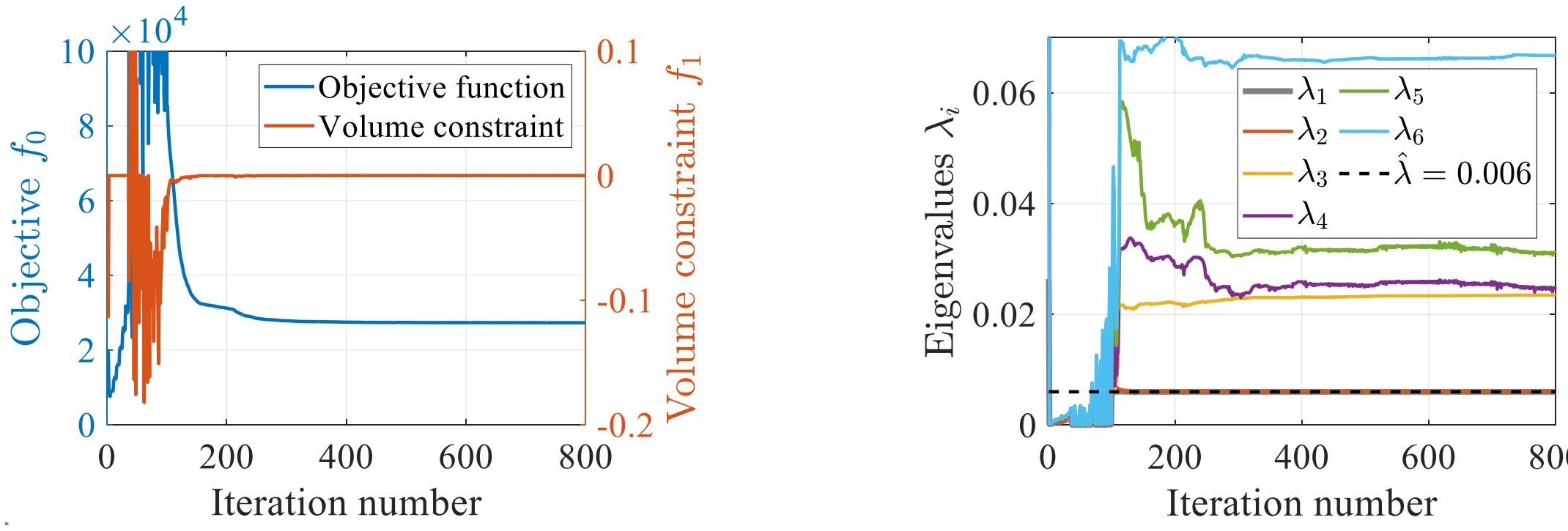

(c) Case-3

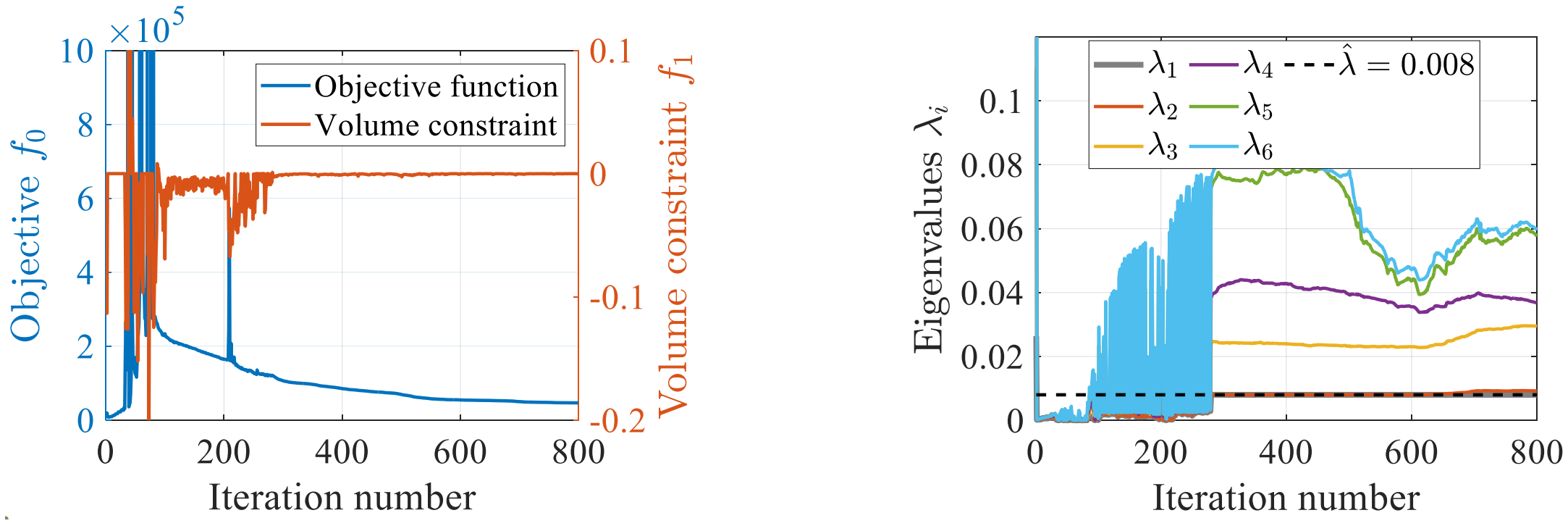

(d) Case-4

Figure 12. Optimization histories of objective, material volume constraint function values, as well as the first six eigenvalues of the tangent stiffness matrix of the designs at the target load for the four cases in Table *2*.

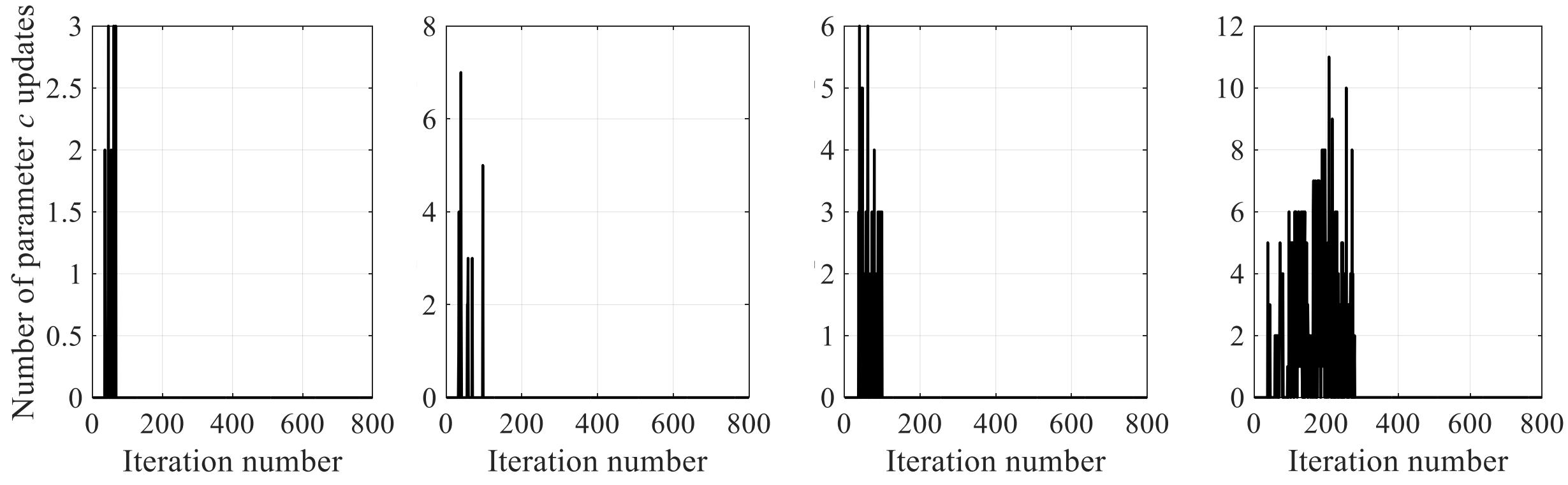

(a) Case-1 (b) Case-2 (c) Case-3 (d) Case-4

Figure 13. History of cutoff parameter *c* updates in the adaptive linear energy interpolation during the optimization for the four cases listed in Table *2*.

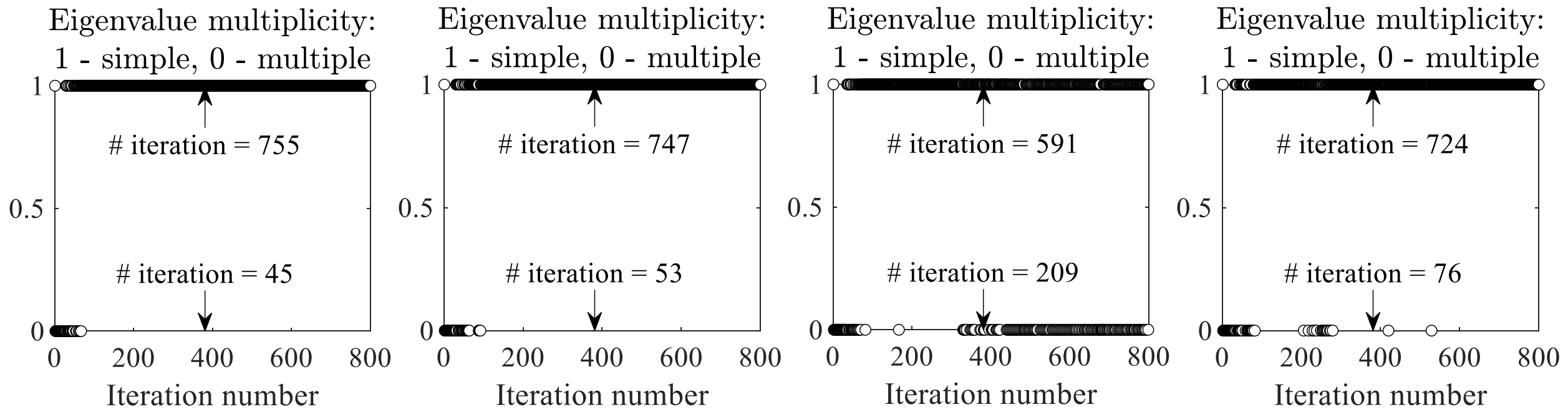

(a) Case-1 (b) Case-2 (c) Case-3 (d) Case-4

Figure 14. Multiplicities of the included first $m$ (= 6) clusters of eigenvalues of the designs at the target load during the optimization for the four cases listed in Table *2*.

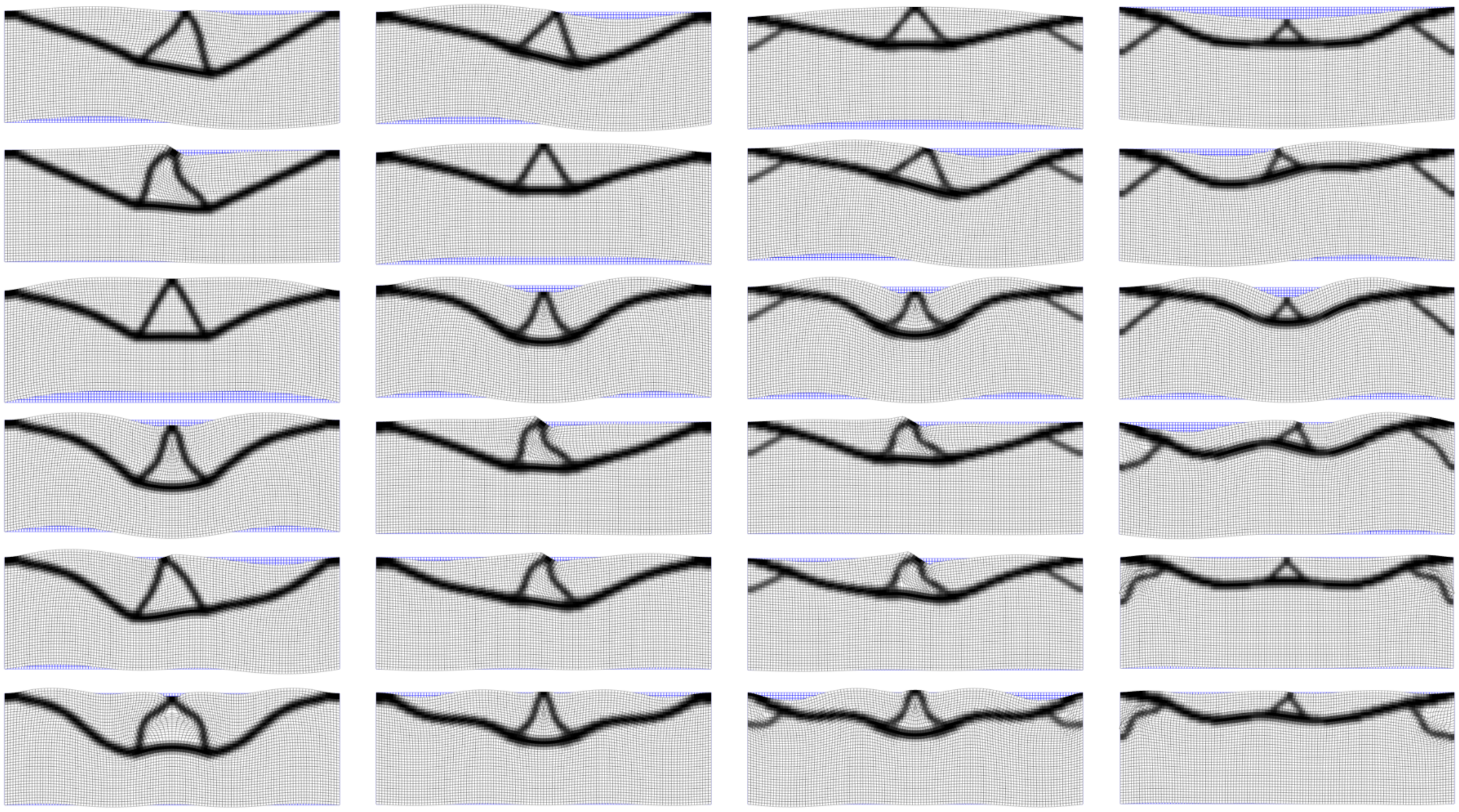

(a) Case-1 (b) Case-2 (c) Case-3 (d) Case-4

Figure 15. Eigenmodes of the optimized designs for the four cases listed in Table *2* at the target load: from the top to the bottom are $\boldsymbol{\phi}_1$,…, $\boldsymbol{\phi}_6$. (Note: appropriate scales are applied to the eigenvectors for visualization purposes)

By considering buckling constraints in optimization formulations, i.e., Eqns. (*40*) and (*41*), the second difficulty can be well addressed. To show this, we consider the nonlinear buckling constraints with different threshold values $\hat{\lambda}$ in Eqns. (40) and (41), which can be intuitively perceived as targeting different levels of safety margins. To avoid large jumps in design updates during optimization iterations, the *move* parameter in the MMA algorithm is set to 0.3. Four cases with different $\hat{\lambda}$ values are given in Table *2*. The optimized topologies of these cases are given in Figure 11 with the optimization histories of objective and constraint values shown in Figure 12. It is noted that all the constraints are satisfied at the final optimization stage and the objective function is decreasing smoothly after the desired nonlinear features appear. In addition, from Figure 12, during the optimization process, only the first stability constraint related to $\lambda_1$ is activated while other eigenvalues are well above the threshold $\hat{\lambda}$. With the increase of the stability thresholds, the optimized topology tends to become shallower. When compared to the topology in Figure 10, it is clear that the shallower topology is to stabilize the two bars under compression by shortening their lengths. Besides, from Table 2 it can be seen that as the stability threshold ($\hat{\lambda}$) increases, the objective value tends to increase, meaning that these two targets are competitive, as expected. In Figure 13, the number of updates of $c$ value in the adaptive linear energy interpolation (Section 3.2) at each optimization iteration is plotted for all the cases. As can be seen, the adaptive scheme is necessary for guaranteeing the convergence of FEA during early stages where there are large amounts of the gray region for geometrically nonlinear topology optimization, which is especially true for Case-4 as both the objective and constraint values are experiencing lots of oscillations (Figure 12d) when the optimizer is looking for feasible solutions that satisfy the stringent stability constraints. Figure 14 shows the proportions of the simple- and multiple-eigenvalue scenarios during the optimization process, where it can be seen that for a large proportion of the iterations the first $m$ (= 6) eigenvalues are simple and more importantly at the final optimization stage the eigenvalues are simple (Table 2). Finally, the six eigenmodes of the optimized structures of the four cases are plotted in Figure 15, where it can be confirmed that all the buckling constraints are enforced on the "real" buckling modes.

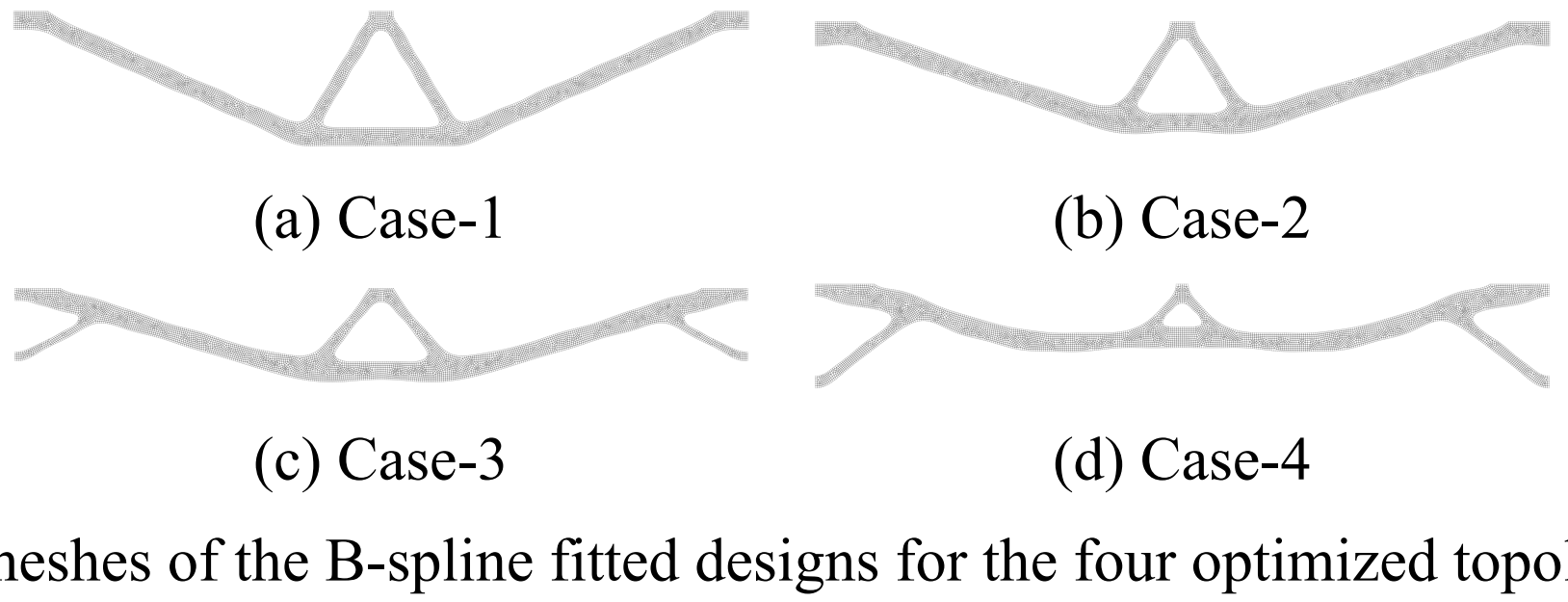

(a) Case-1 (b) Case-2

(c) Case-3 (d) Case-4

Figure 16. FE meshes of the B-spline fitted designs for the four optimized topologies in Figure 11.

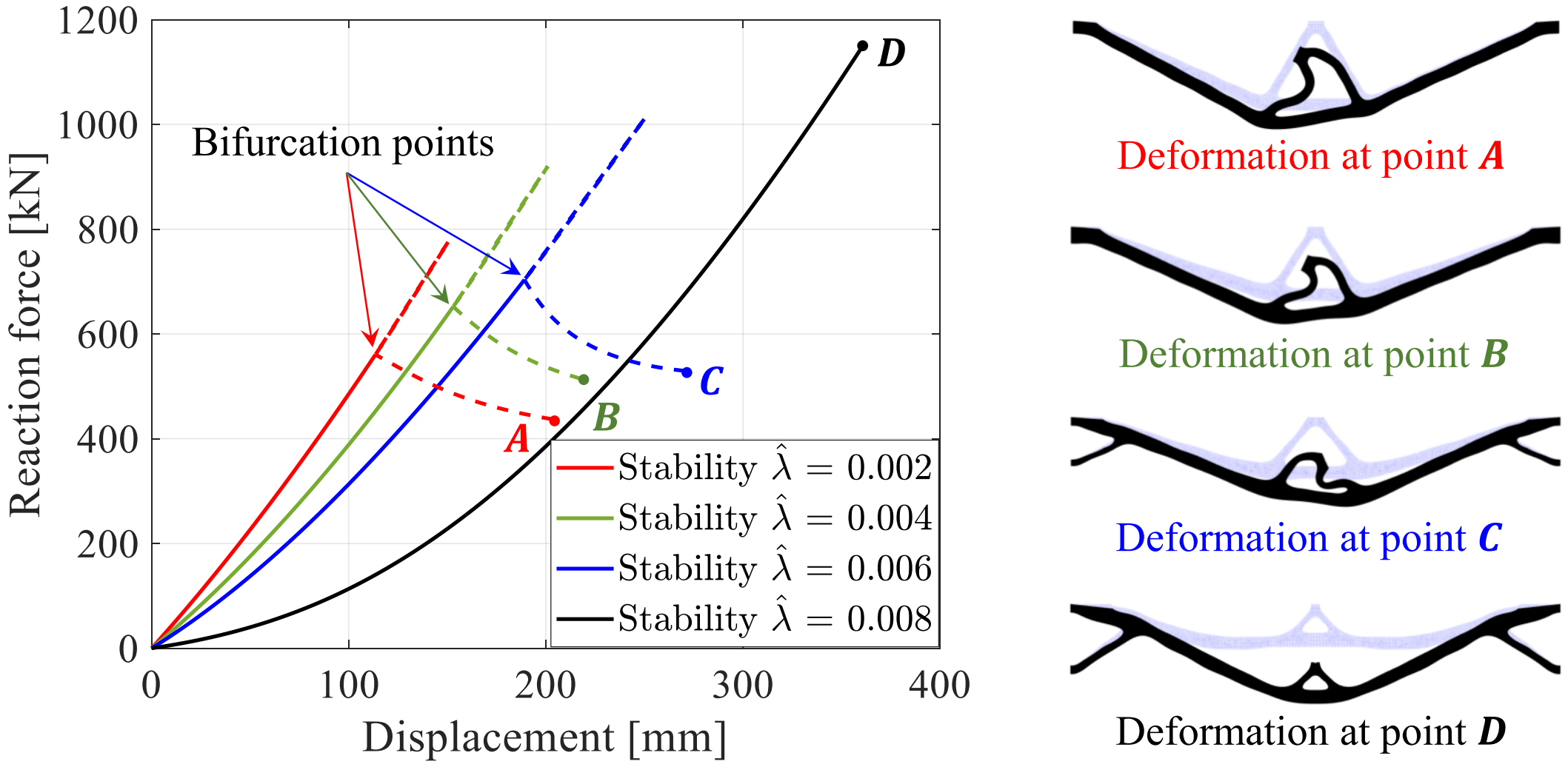

Figure 17. Load-displacement curves of the B-spline fitted optimized double-clamped beam topologies corresponding to different stability constraints and their deformed shapes (solid line corresponds to stable region, while dashed line corresponds to unstable region).

To study the stability performance of the optimized topologies under the same load pattern, the optimized designs in Figure 11 are fitted using B-splines to get rid of the void and intermediate densities, and the corresponding FE models are shown in Figure 16. The fitted designs are then analyzed under the same load pattern using arc-length control together with the branch switching technique (Appendix D1) for tracing secondary branches. The load-displacement curves of the fitted designs together with the deformed shapes are shown in Figure 17. As expected, the designs with higher stability thresholds ($\hat{\lambda}$) have higher critical load at the expense of stiffness. Moreover, for the optimized designs with $\hat{\lambda}$ = 0.002, 0.004 and 0.006, both the principal and bifurcated branches are unstable after the critical (bifurcation) point.

## *7.2 Pinned column*

The second example considers a pinned column design problem with dimension, load and boundary condition, and material properties specified in Figure 18. The design domain is discretized by a 60×180 FE mesh, and the pin and roller boundaries are applied to the middle three nodes on the bottom and top sides, respectively. The total load is equally distributed to the three middle nodes. The volume fraction constraint is considered with $V_f$ = 0.3, and the density filter radius $r_{min}$ = 40mm. Due to the structural symmetry, the optimization space considers only a quarter of the design domain. However, to simulate asymmetric buckling modes, the full domain is included in the nonlinear FEA. Similar to the previous example, the *move* parameter in the MMA algorithm is set to 0.3 for this case as well.

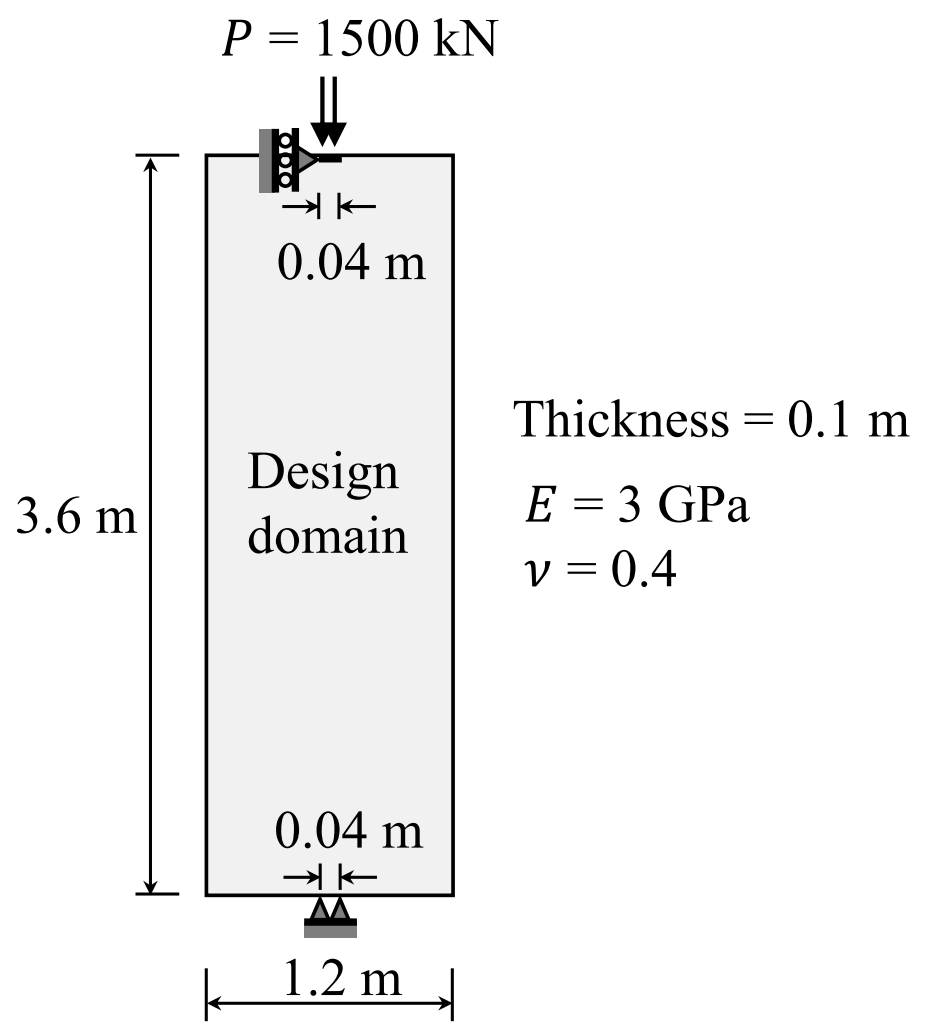

Figure 18. Sketch of the pinned column problem.

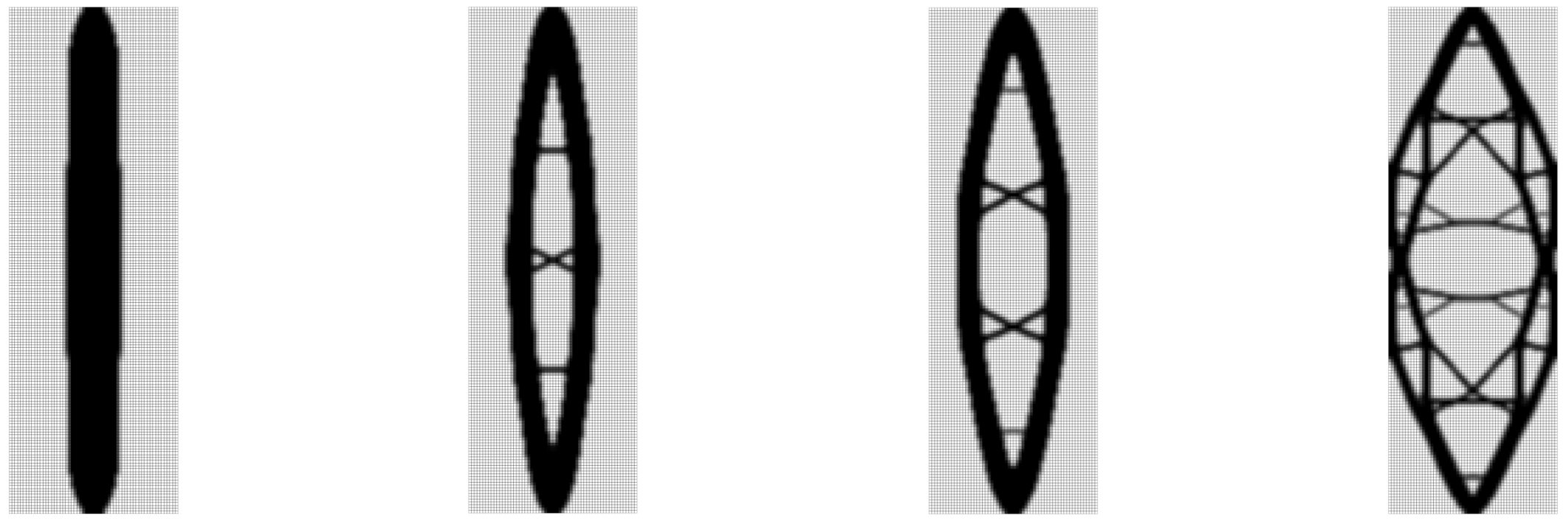

(a) No stability constraint

(b) Stability constraint $\hat{\lambda}$ = 0.001

(c) Stability constraint $\hat{\lambda}$ = 0.002

(d) Stability constraint $\hat{\lambda}$ = 0.003

Figure 19. Optimized topologies of the pinned column problem for different stability constraints.

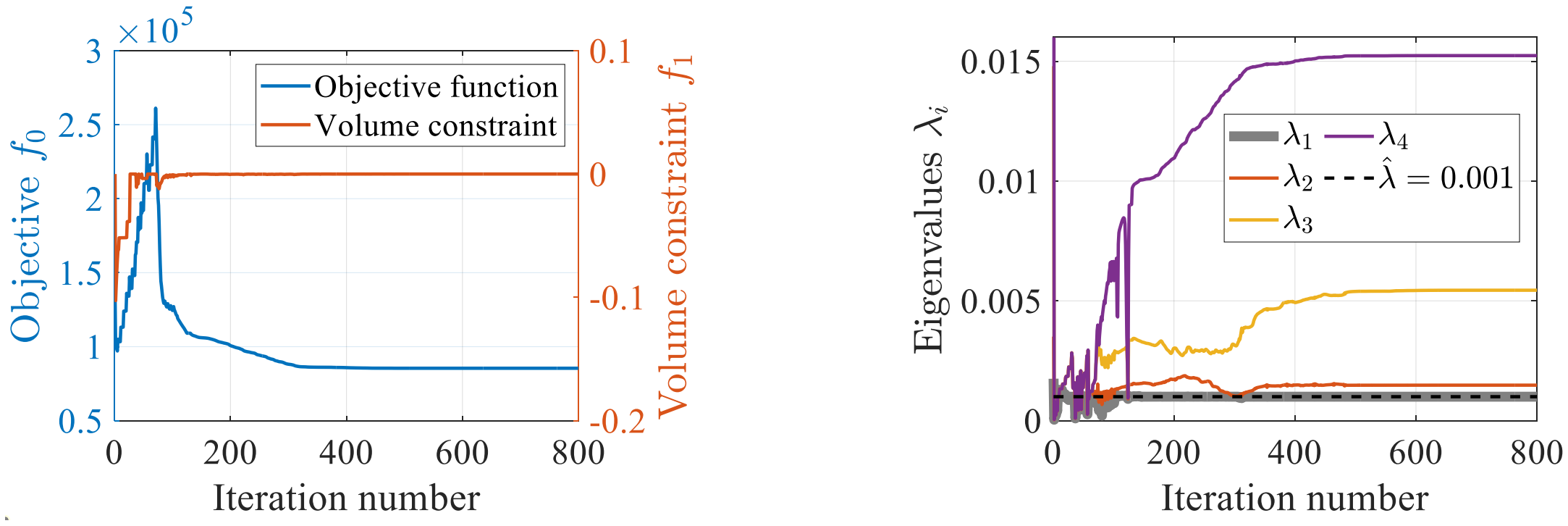

(a) Stability constraint $\hat{\lambda} = 0.001$

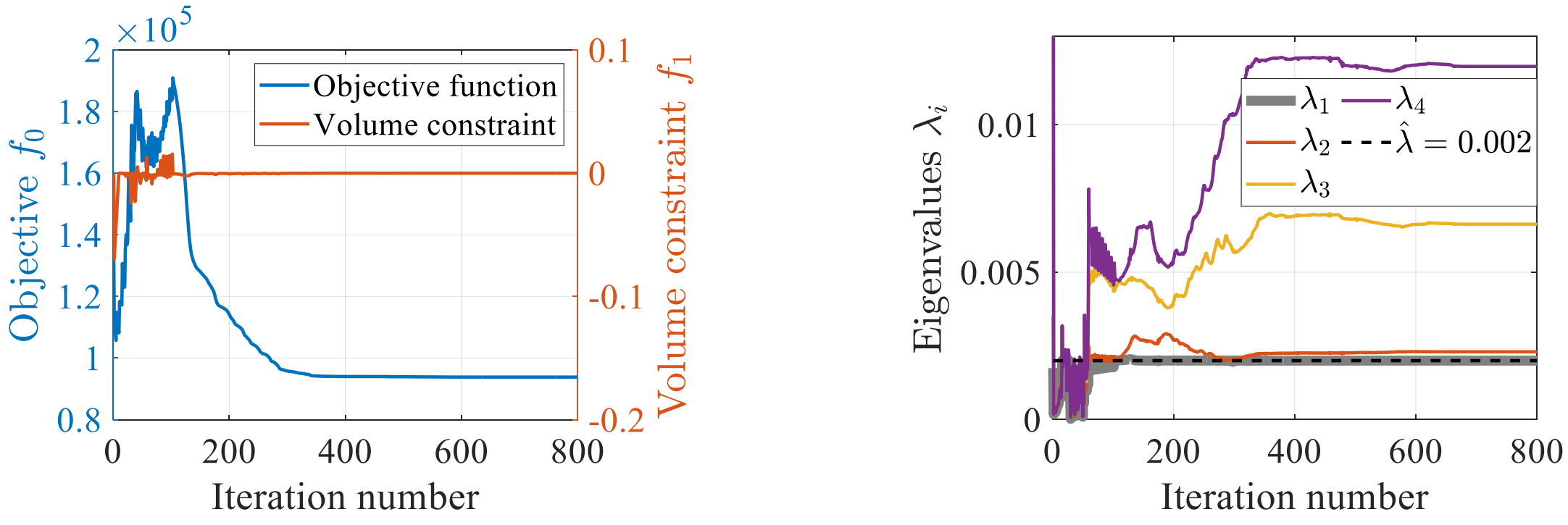

(b) Stability constraint $\hat{\lambda} = 0.002$

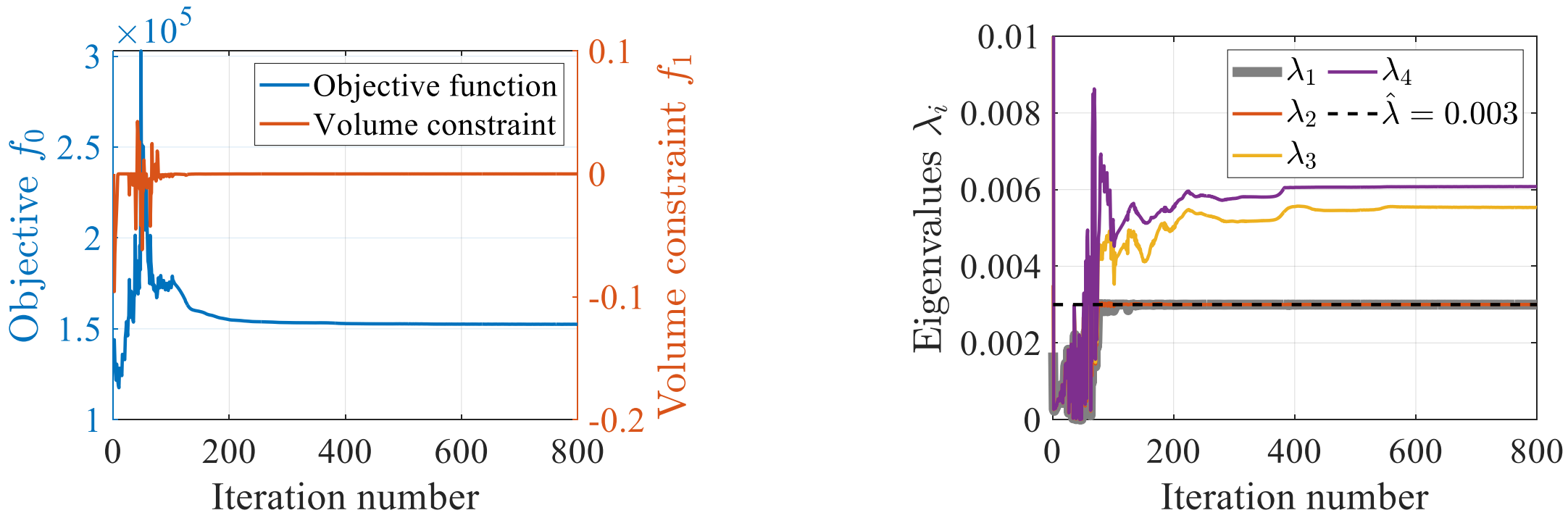

(c) Stability constraint $\hat{\lambda} = 0.003$

Figure 20. Optimization histories of objective, material volume constraint function values, as well as the first four eigenvalues of the tangent stiffness matrix of the designs at the target load for the different stability threshold values in the pinned column problem.

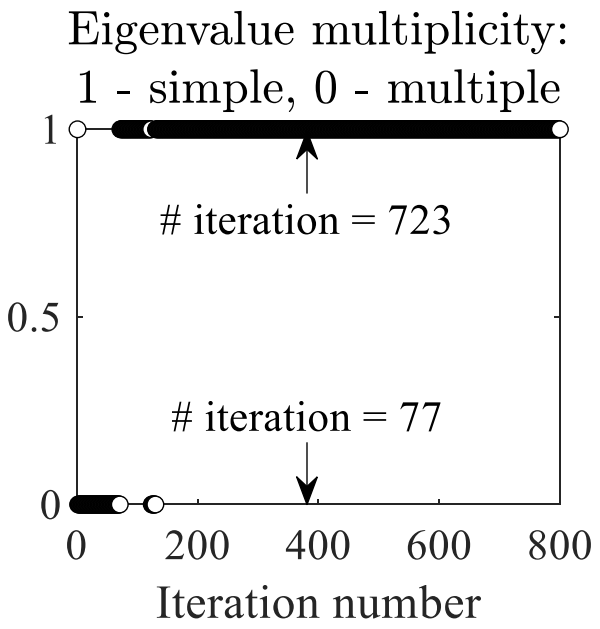

(a) $\hat{\lambda} = 0.001$

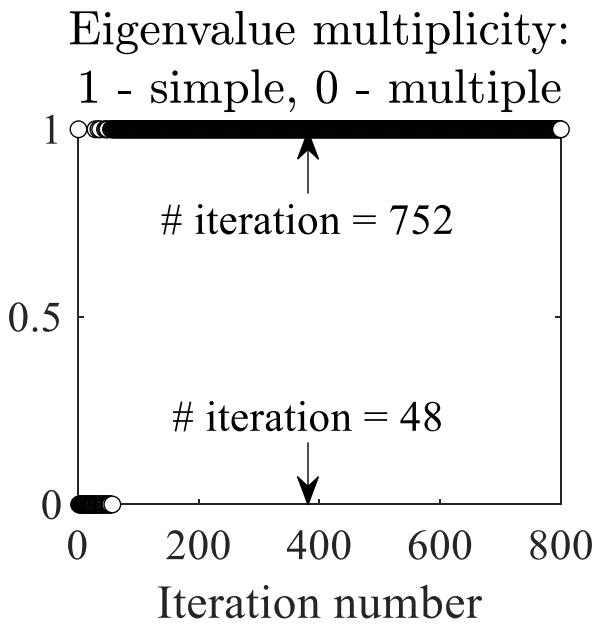

(b) $\hat{\lambda} = 0.002$

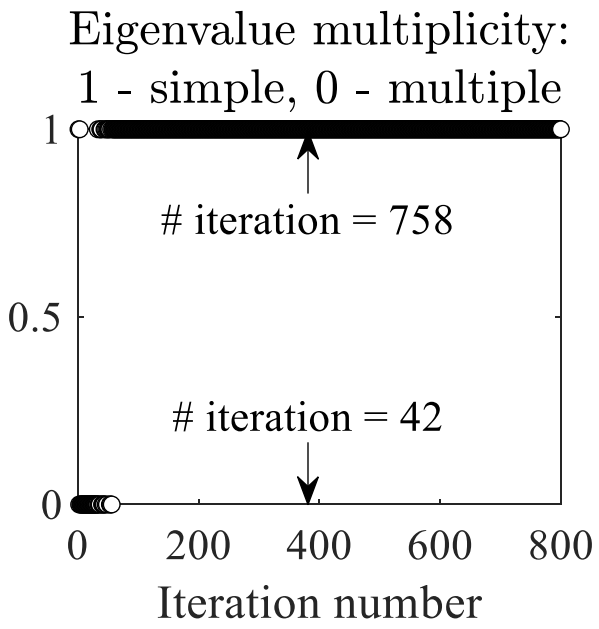

(c) $\hat{\lambda} = 0.003$

Figure 21. Eigenvalue multiplicity during optimization for different stability thresholds in the pinned column problem.

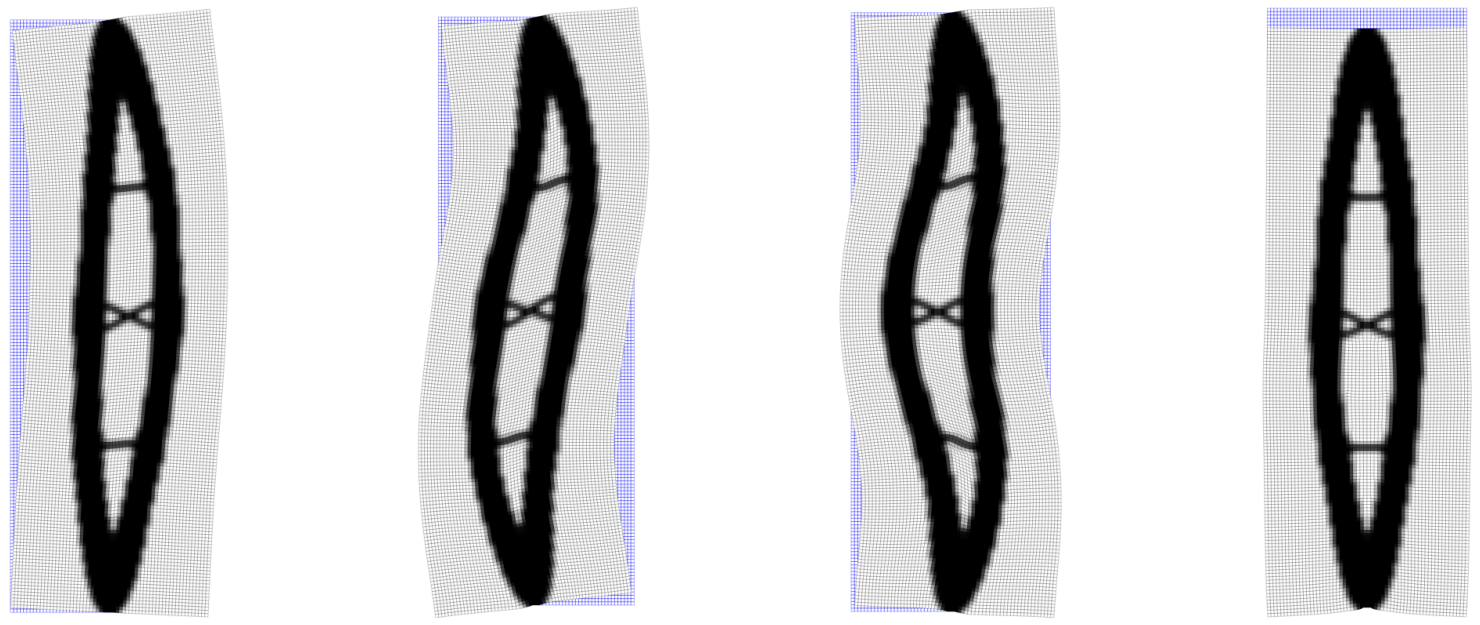

(a) Stability constraint $\hat{\lambda} = 0.001$

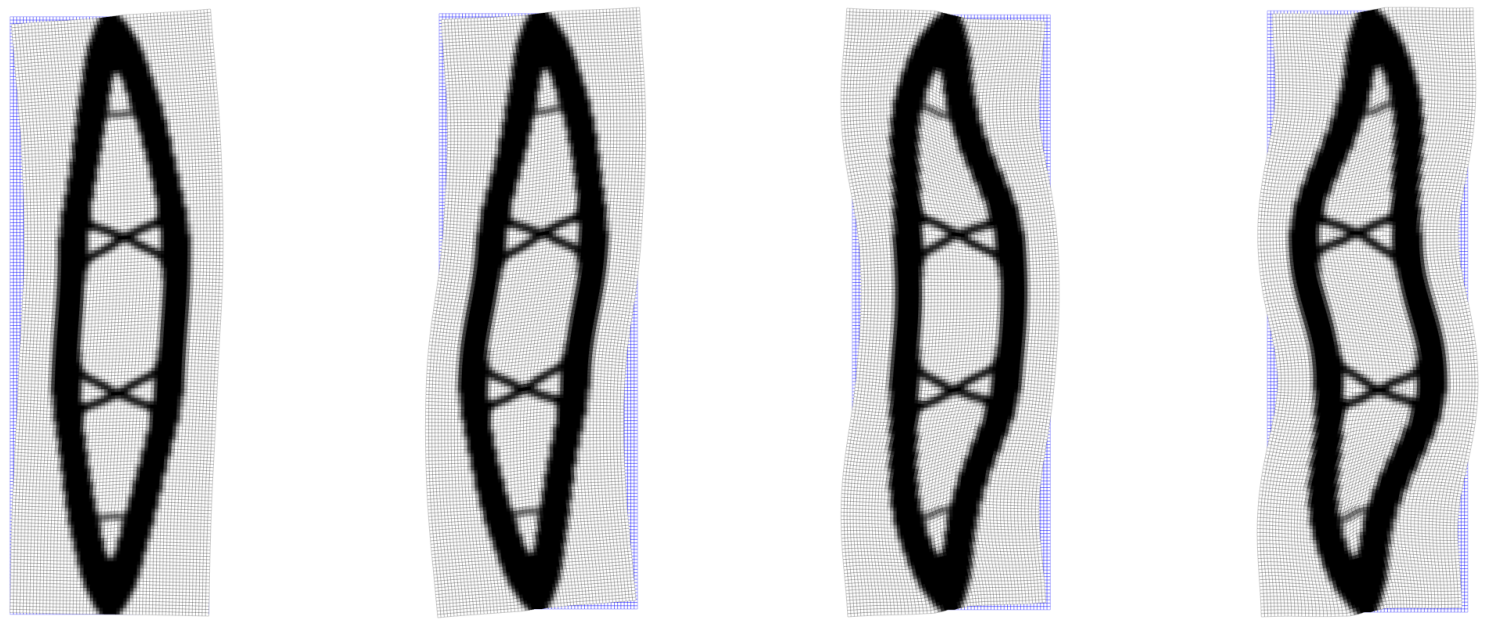

(b) Stability constraint $\hat{\lambda} = 0.002$

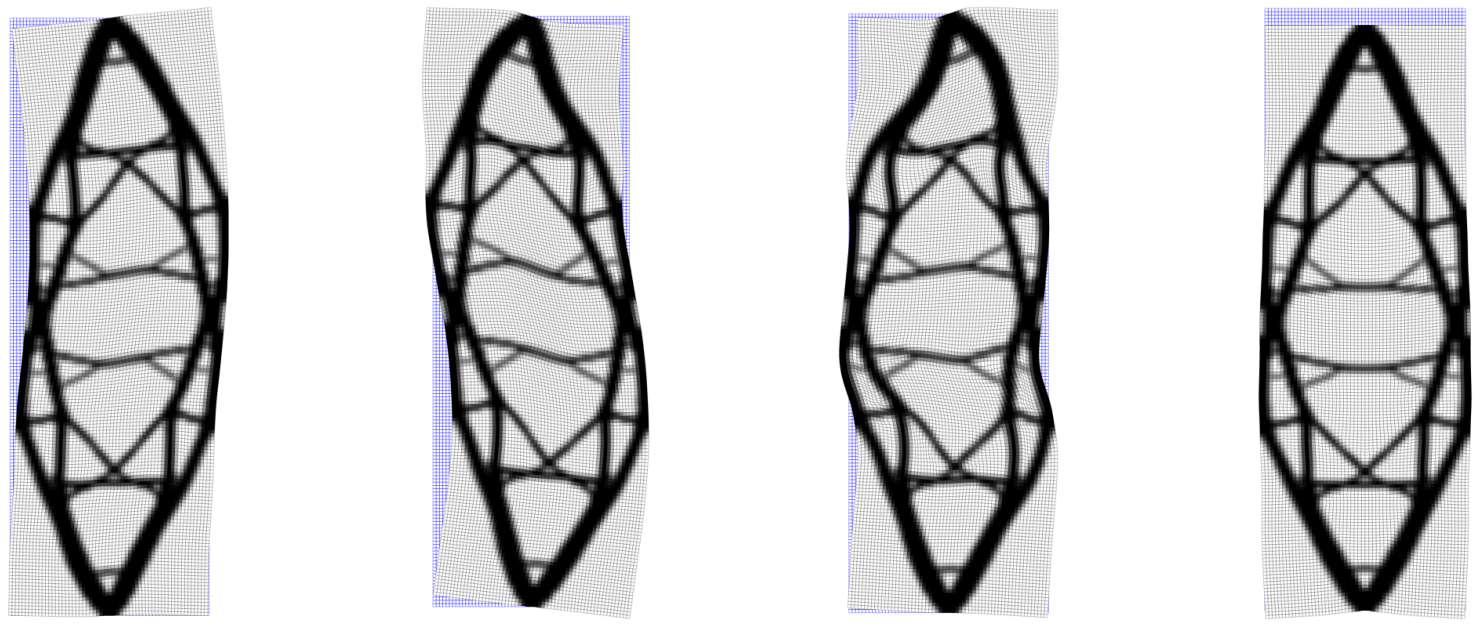

(c) Stability constraint $\hat{\lambda} = 0.003$

Figure 22. Eigenmodes of the optimized stable designs at the target load: from the left to the right are $\boldsymbol{\phi}_1, \ldots, \boldsymbol{\phi}_4$. (Note: appropriate scales are applied to the eigenvectors for visualization purposes)

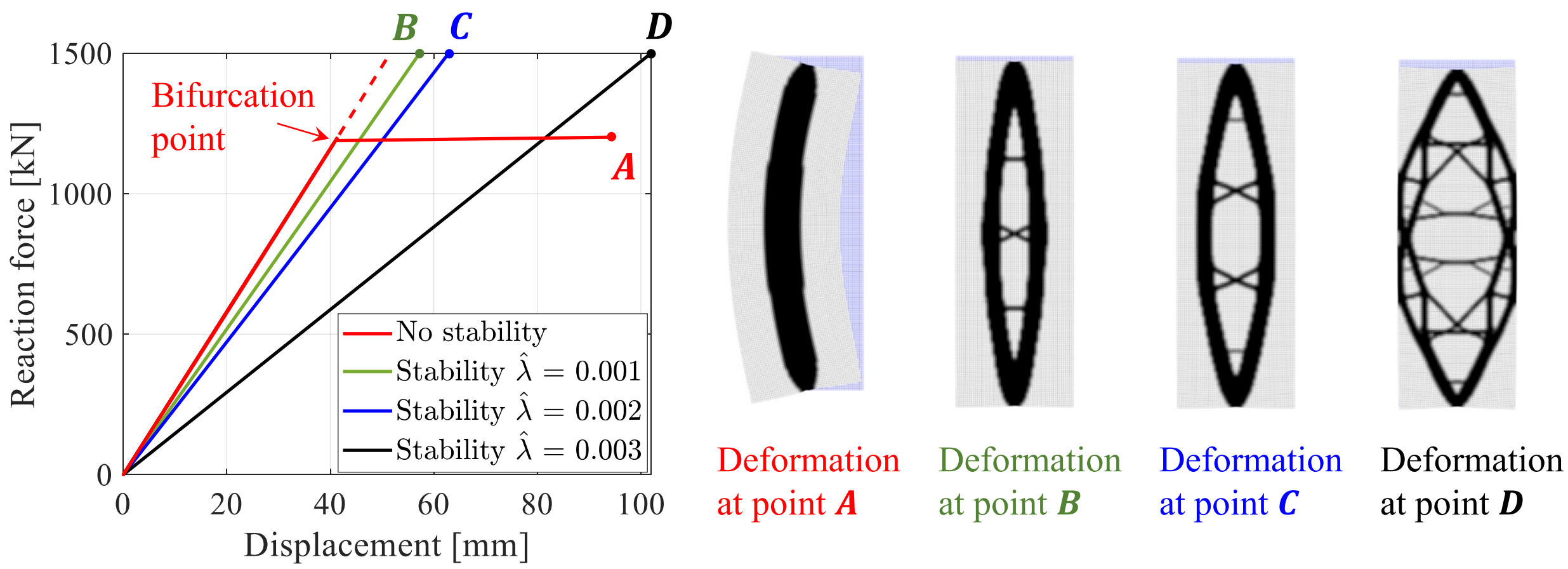

Figure 23. Load-displacement curves of the optimized pinned column topologies corresponding to different stability constraints and their deformed shapes at the target load or buckling state (solid line corresponds to stable region, while dashed line corresponds to unstable region).

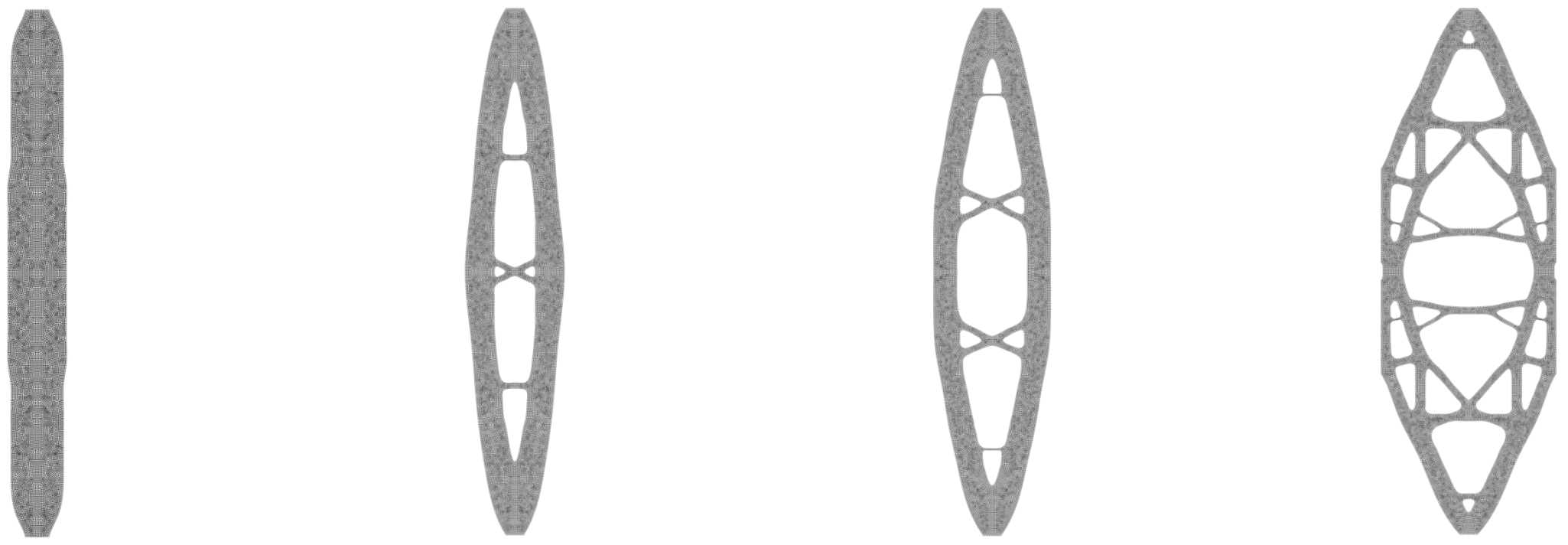

(a) No stability constraint

(b) Stability constraint $\hat{\lambda} = 0.001$

(c) Stability constraint $\hat{\lambda} = 0.002$

(d) Stability constraint $\hat{\lambda} = 0.003$

Figure 24. FE meshes of the B-spline fitted designs for the four optimized topologies in Figure 19.

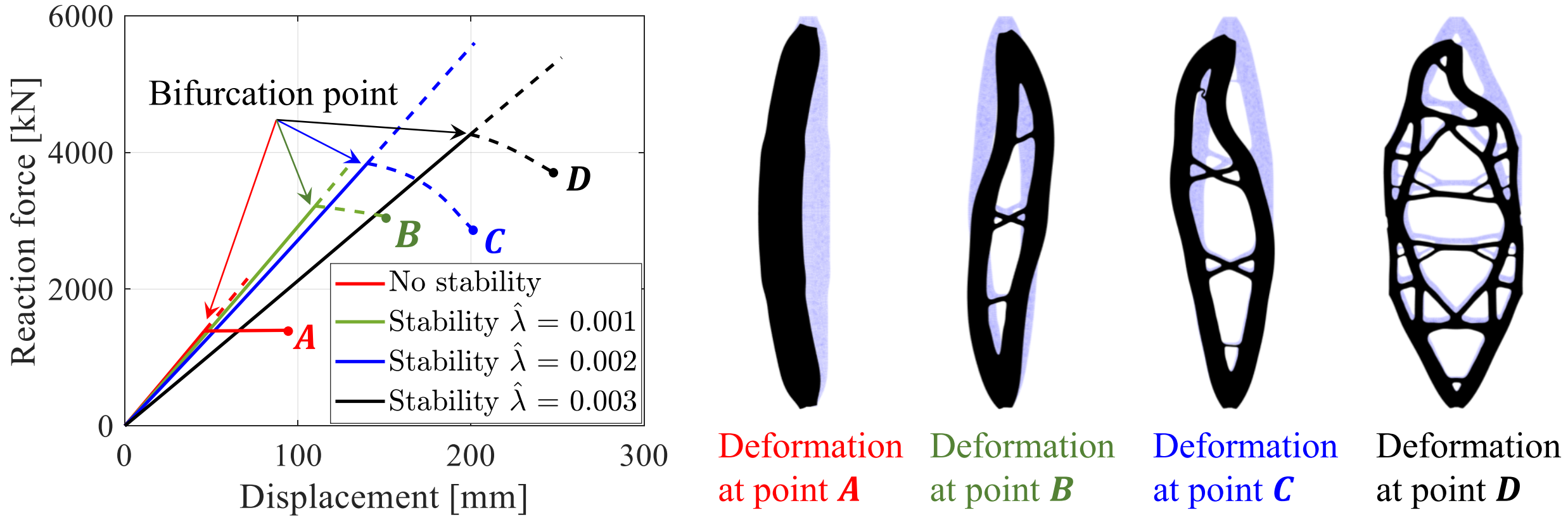

Figure 25. Load-displacement curves of the four B-spline fitted optimized pinned column topologies and their deformed shapes ((solid line corresponds to stable region, while dashed line corresponds to unstable region).

The optimized topology without stability constraint is shown in Figure 19a which is a straight bar. A stability analysis using Eq. (17) shows that the bar will buckle before reaching the load as the minimum eigenvalue is -2×10$^{-4}$ at the target load and the buckling mode is similar to the first eigenmode in Figure 5. To achieve a stable design, the stability constraints are then included and the optimized topologies corresponding to three different stability thresholds ($\hat{\lambda}$) are shown in Figure 19 (b), (c) and (d). With the increase in the stability thresholds, the topology tends to expand horizontally with an increasing number of supporting bars inside. The optimization histories of the objective and constraints are given in Figure 20, where a smooth decrease of the objective after the emergence of desired features can be seen, and all the constraints are well satisfied at the final optimization stage. It should be noted that in Figure 20c the first two eigenvalues ($\lambda_1$ and $\lambda_2$) are, although, visually indistinguishable, their difference is greater than the set threshold (1.0E-8). This is further confirmed in Figure 21, which shows the multiplicity of the eigenvalues of the designs during the optimization. Roughly, only the first hundred iterations have repeated eigenvalues. Figure 22 plots the controlled eigenmodes of the final optimized designs at the target load. Again, with the proposed pseudo-mass strategy there are no spurious modes, as desired. The load-displacement curves of these optimized designs up to the target load with stability examined are shown in Figure 23, where the displacement is measured at the middle node on the top surface. As can be seen, all the designs with stability constraints are stable at the target load of 1500 kN, while the one without stability constraints loses stability at a load of 1188 kN.

To investigate the stability performance of these optimized designs, the B-spline fitting is carried out, and the FE meshes are shown in Figure 24. To preserve all the features in the optimized topologies, the level set is chosen as 0.4 for Figure 24d. The fitted designs are then analyzed using arc-length control under the same load pattern. The displacement-load curves and the corresponding deformations at the marked loading stages are shown in Figure 25. It can be seen that with the incorporation of stability constraints the stability of the structure is significantly enhanced, and this improvement can be controlled by specifying different threshold values ($\hat{\lambda}$) of the constraints. It is remarked that for the designs in Figure 24, the principal branch above the 1st critical point may have multiple critical (limit and bifurcation) points that lead to complex responses which are not investigated here.

## *7.3 L-bracket*

The third example considers an L-shape problem shown in Figure 26 discretized with a uniform FE mesh of element size 0.02 m × 0.02 m, where a total load of 300 kN is applied to the rightmost four nodes distributed as 50kN, 100kN, 100kN, and 50kN. The allowable volume fraction is $V_f$ = 0.4, and the density filter radius $r_{min}$ = 40 mm. The optimized topologies with and without stability constraints are shown in Figure 27. It can be observed that the incorporation of stability constraints leads to the emergence of supporting bars that connect the two vertical bars on the left side. These connecting bars become thicker as the stability threshold ($\hat{\lambda}$) increases. Besides, the peripheral curved bar that undergoes compression becomes thicker and shorter as the stability threshold ($\hat{\lambda}$) increases. The stabilization effect of these features will be discussed in the stability analysis. The optimization histories of the objective, volume constraint and the constrained first six eigenvalues are plotted in Figure 28. During optimization, the first six eigenvalues of all the designs are simple, i.e., $\lambda_a \neq \lambda_b$ for $a \neq b$, $a, b = 1, \dots, 6$. As is shown, both the material volume constraint and the stability constraints are satisfied at the final optimization iteration and the objective function is decreasing smoothly after a few oscillations. The corresponding six eigenmodes of the optimized structures at the target load are plotted in Figure 29. Again, all the eigenmodes that are constrained are "real" eigenmodes. The load-displacement curves of the three optimized designs are compared in Figure 30, where the displacement is measured as the vertical displacement at the right corner node of the top surface on which the load is applied. It can be seen that all the designs including the one without stability constraint are stable up to the target load and increasing stability threshold ($\hat{\lambda}$) leads to structures with reduced stiffness.

To investigate the stability performance of the optimized topologies, the designs are fitted using B-splines with FE meshes shown in Figure 31, which are then analyzed under the same load pattern with arc-length control. The obtained load-displacement curves are shown in Figure 32 together with deformations at different loading stages. As shown, the design without stability constraint encounters a 1st critical (limit) point at the load of 418.07 kN, where the left vertical bar buckles under compression. Incorporating stability constraints with threshold $\hat{\lambda}$ = 0.001 helps to postpone the critical point to the load of 820.59 kN. After reaching the critical load, the left vertical bar, although supported by the two connecting bars, still buckles. With the stability constraints using threshold $\hat{\lambda}$ = 0.002, the optimized design does not encounter the critical point before the load of 834.00 kN. The compressive peripheral curved bar remains stable due to the thicker supporting bars, the increase in its width, and the reduction of its length.

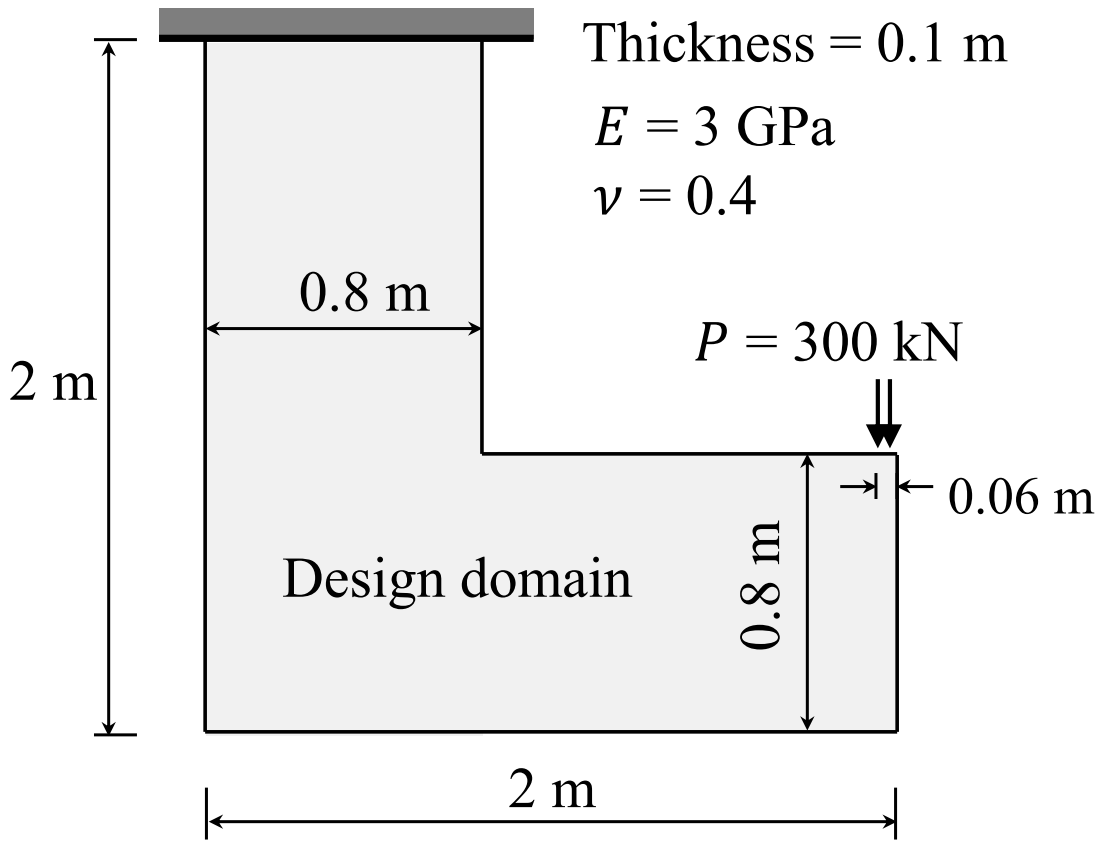

Figure 26. Sketch of the L-shape problem.

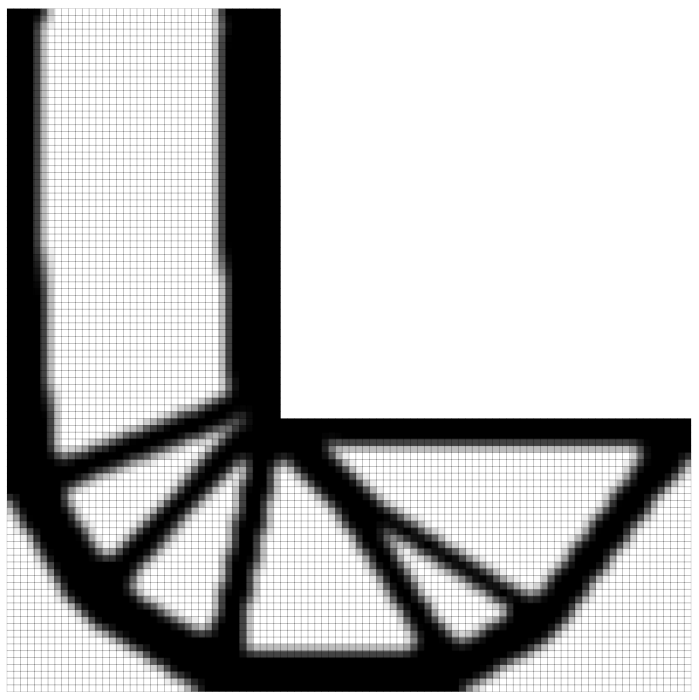

(a) No stability constraint

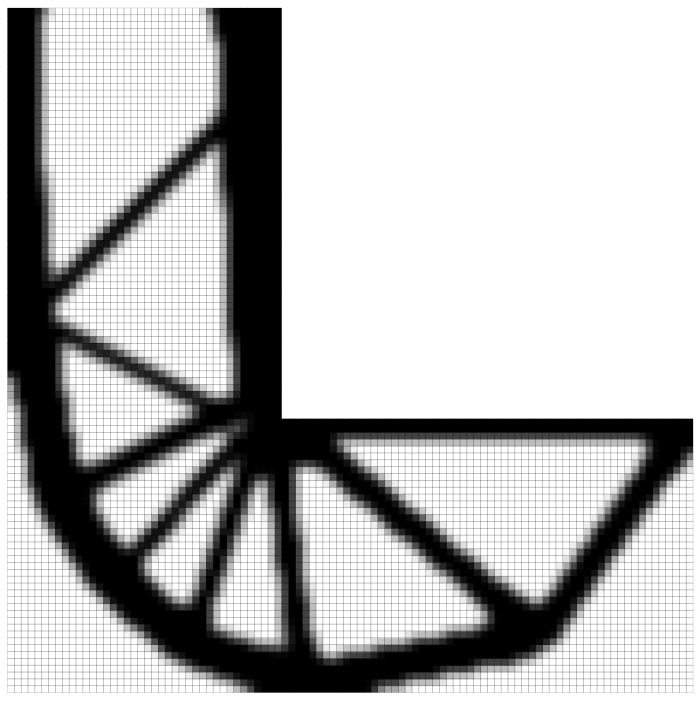

(b) Stability constraint $\hat{\lambda}$ = 0.001

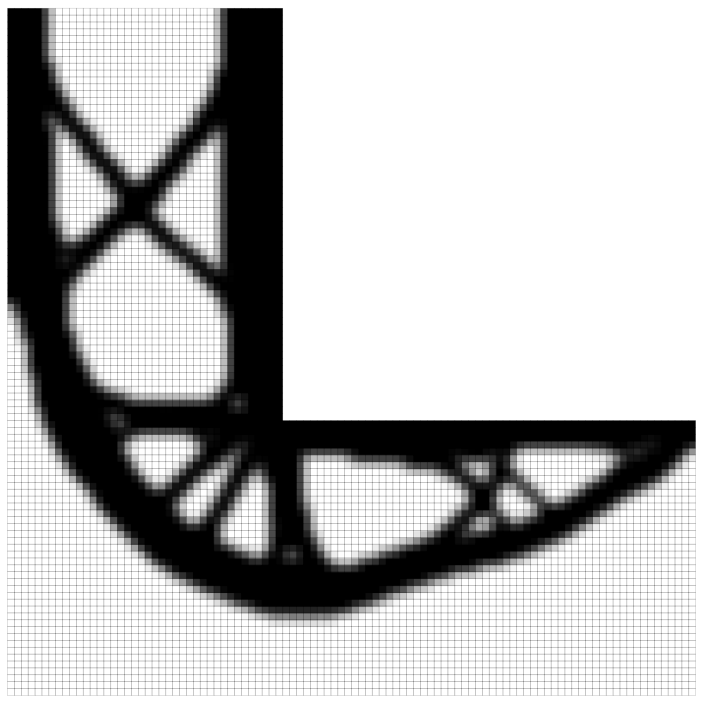

(c) Stability constraint $\hat{\lambda}$ = 0.002

Figure 27. Optimized topologies of the L-shape domain for different stability constraints.

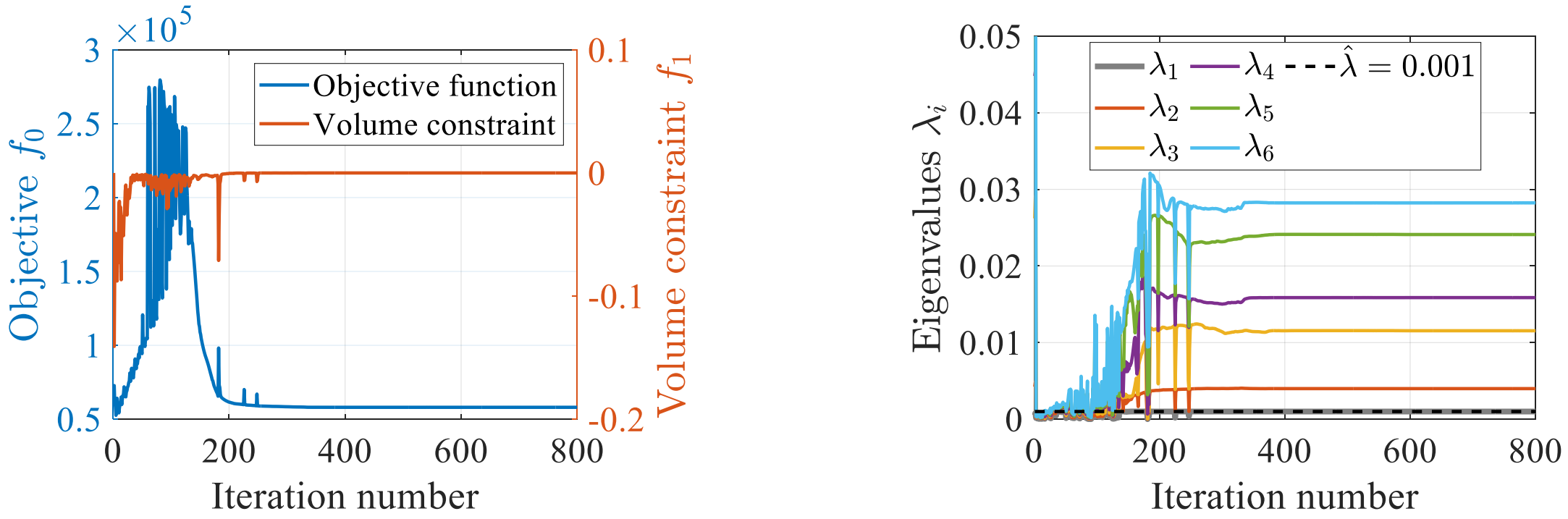

(a) Stability constraint $\hat{\lambda}$ = 0.001

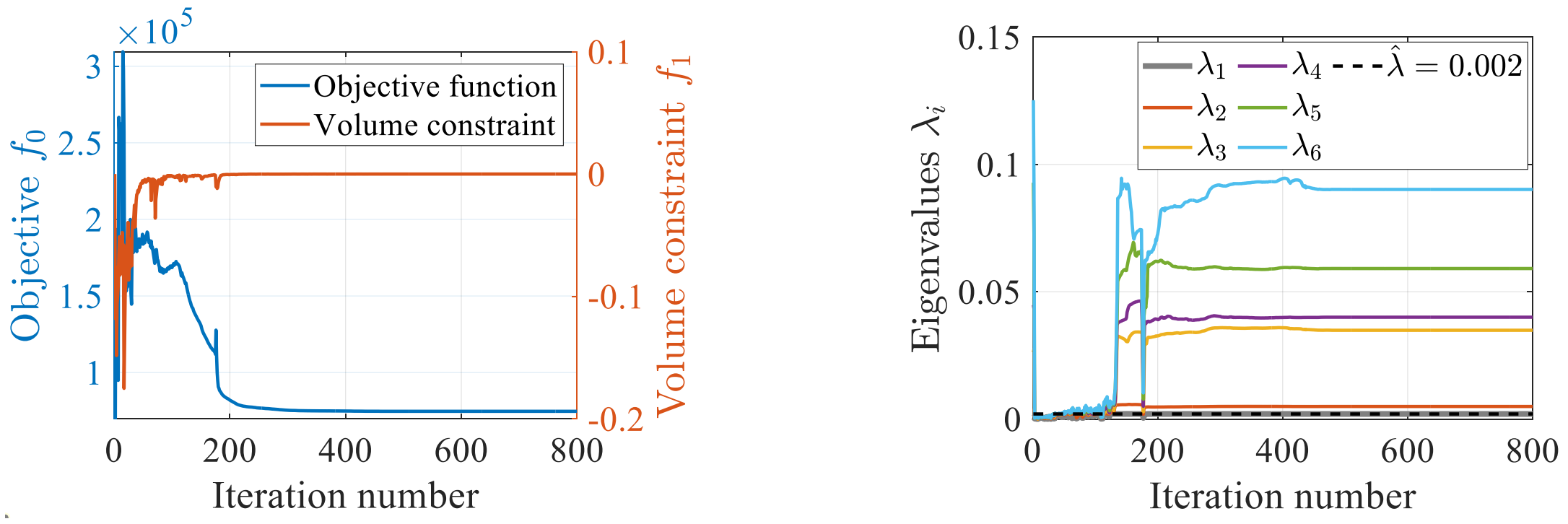

(b) Stability constraint $\hat{\lambda} = 0.002$

Figure 28. Optimization histories of objective, material volume constraint function values, as well as the first six eigenvalues of the tangent stiffness matrix of the designs at the target load for the different stability threshold values in the L-shape problem.

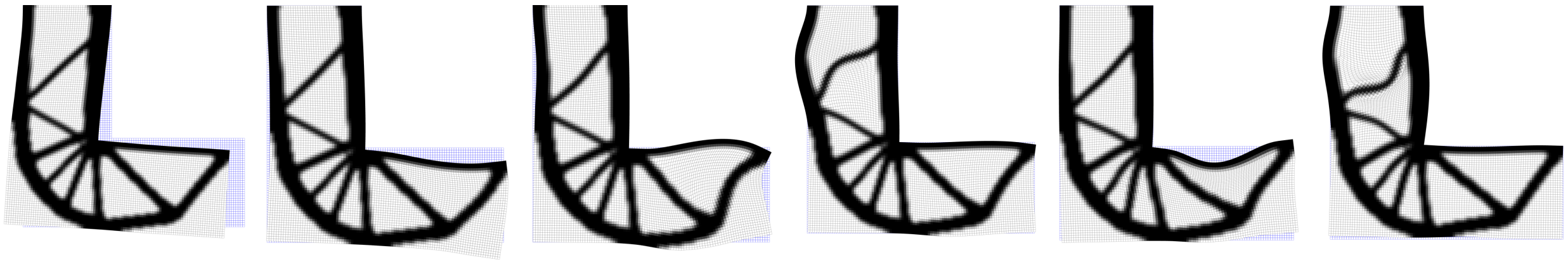

(a) Stability constraint $\hat{\lambda} = 0.001$

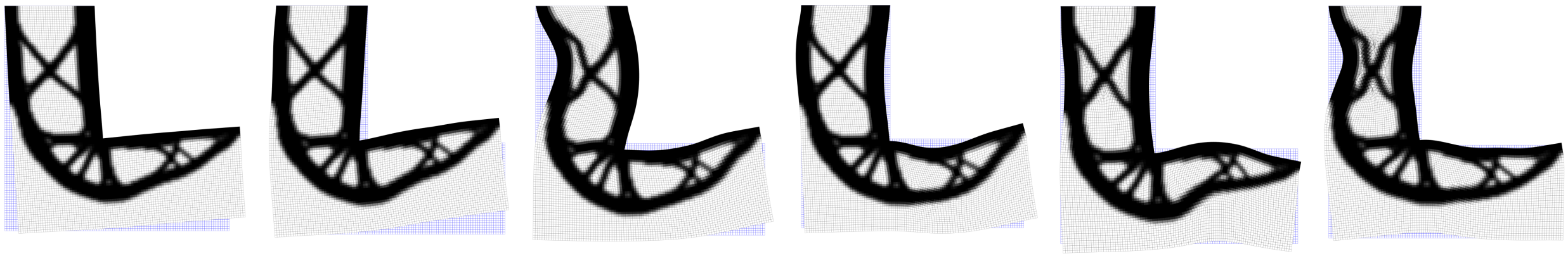

(b) Stability constraint $\hat{\lambda} = 0.002$

Figure 29. Eigenmodes of the optimized stable designs at the target load: from the left to the right are $\boldsymbol{\phi}_1$,…, $\boldsymbol{\phi}_6$. (Note: appropriate scales are applied to the eigenvectors for visualization purposes)

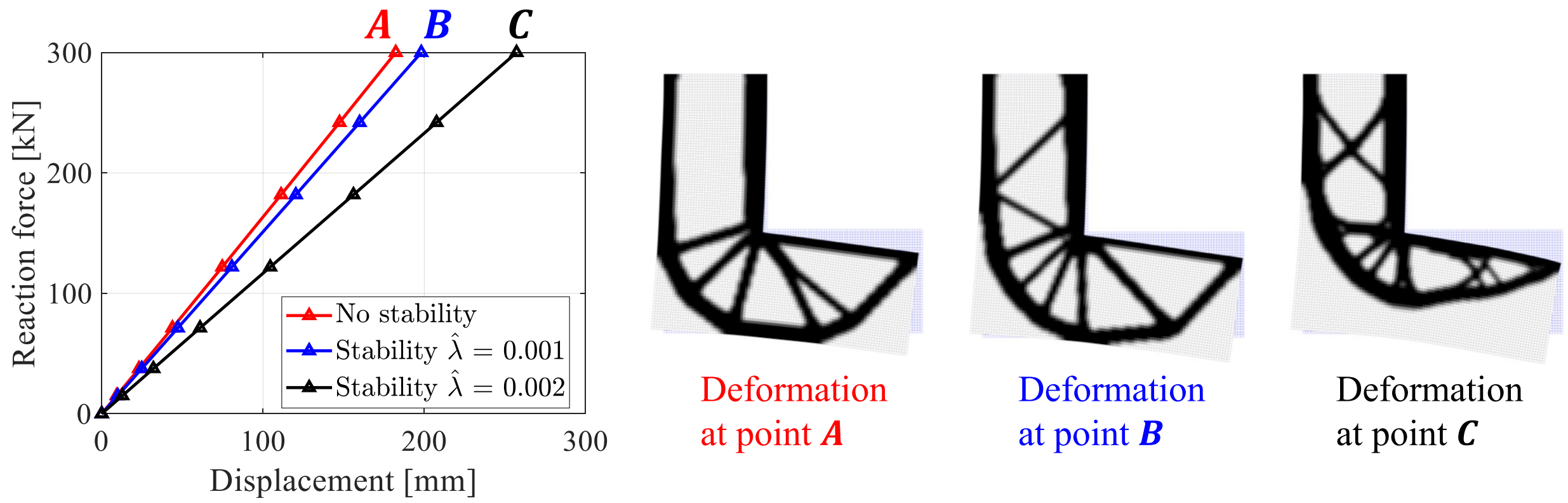

Figure 30. Load-displacement curves of the three optimized L-shape topologies and their deformed shapes at the target load.

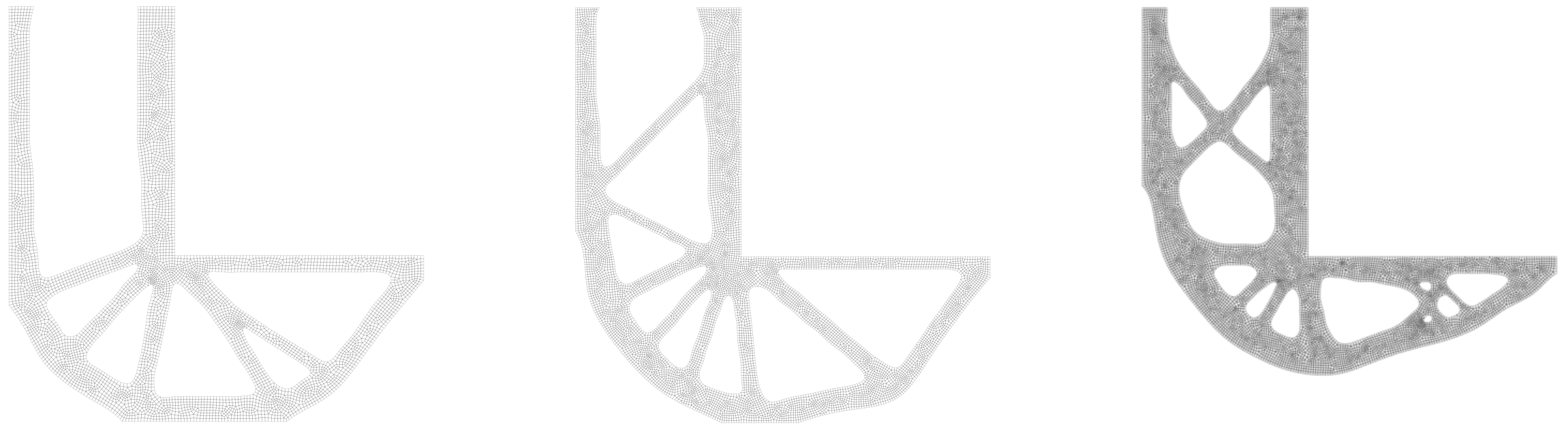

(a) No stability constraint

(b) Stability constraint $\hat{\lambda}$ = 0.001

(c) Stability constraint $\hat{\lambda}$ = 0.002

Figure 31. FE meshes of the B-spline fitted designs for the three cases in Figure 27.

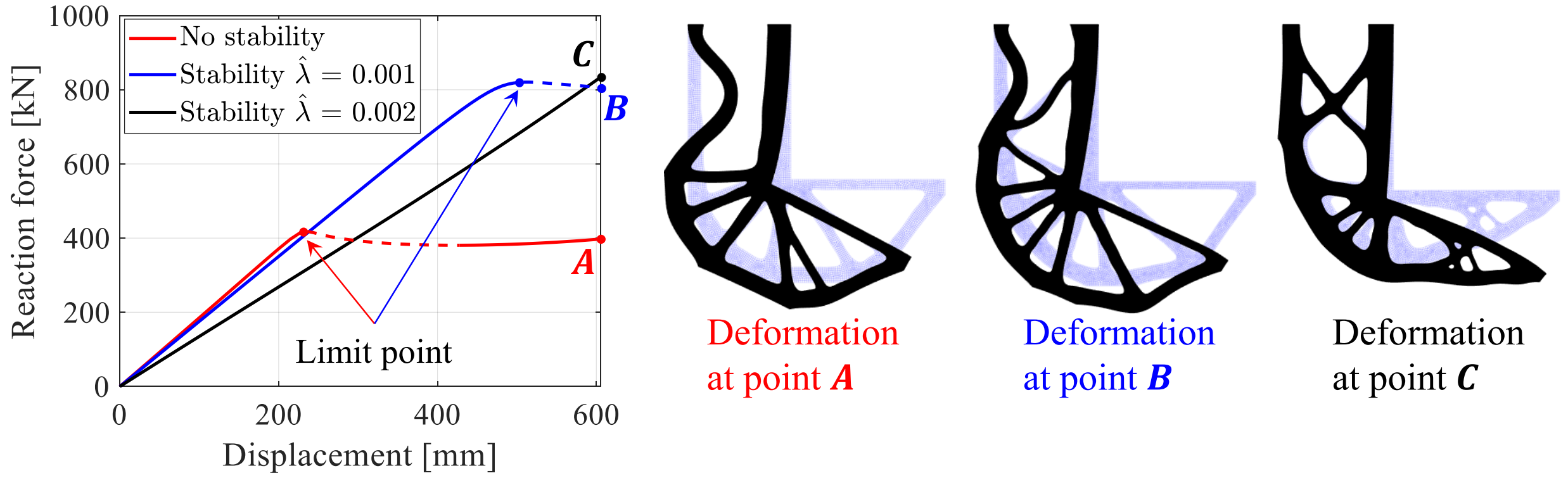

Figure 32. Load-displacement curves of the three B-spline fitted optimized L-shape topologies and their deformed shapes (solid line corresponds to stable region, while dashed line corresponds to unstable region).

### 7.4 Symmetric compressive block

The last example considers a square block subject to compressive loads from its four sides, as shown in Figure 33. The design is discretized by a 100×100 FE mesh and the point load on each side is equally distributed to the three middle nodes. Due to the symmetry of the load and boundary conditions as well as the geometry, the optimal design is expected to have high symmetry. Therefore, only the eighth part of the design domain is optimized but the entire domain is included in the FEA to capture asymmetric buckling modes. The allowable volume fraction is $V_f$ = 0.2, and the density filter radius $r_{min}$ = 20 mm. Three cases are considered – one without stability constraint and two with different stability thresholds $\hat{\lambda}$ = 0.001 and $\hat{\lambda}$ = 0.002.

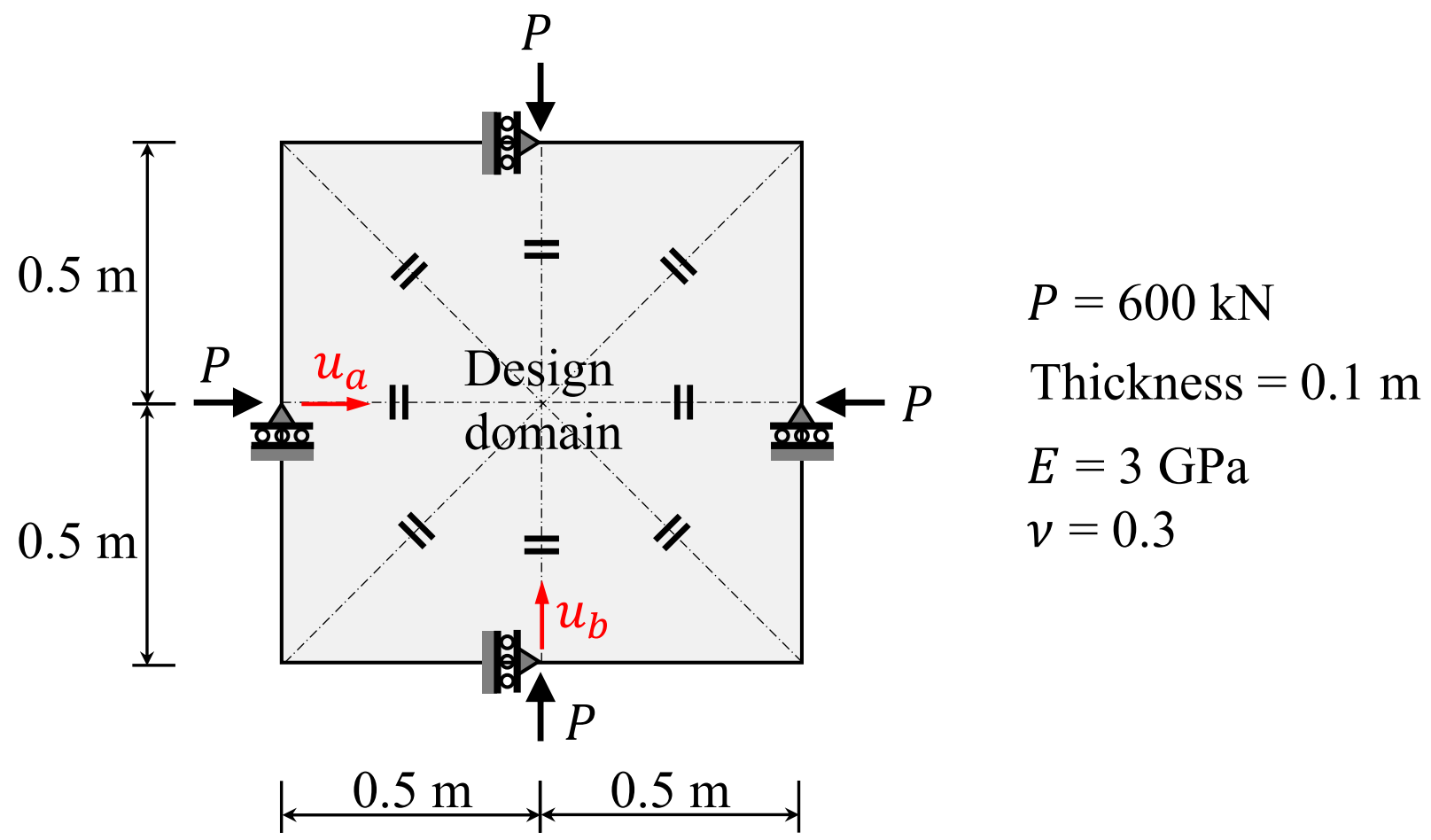

Figure 33. Sketch of the symmetric compressive block problem.

Table 3. Three cases of topology optimization of the symmetric compressive block with different stability constraints

| Case number | Case-1 | Case-2 | Case-3 |
|---|---|---|---|
| Stability constraint | No stability constraint | $\hat{\lambda}$ = 0.001 | $\hat{\lambda}$ = 0.002 |
| Minimized objective value $f_0$ [kN·m] | 24.71 | 25.09 | 26.43 |
| First eight eigenvalues of the optimized topology | $\begin{bmatrix}\lambda_1\\\lambda_2\\\lambda_3\\\lambda_4\\\lambda_5\\\lambda_6\\\lambda_7\\\lambda_8\end{bmatrix}=\begin{bmatrix}-0.00047528\\-0.00047528\\0.00333745\\0.00387963\\0.02189753\\0.08101026\\0.08101026\\0.12860280\end{bmatrix}$ | $\begin{bmatrix}\lambda_1\\\lambda_2\\\lambda_3\\\lambda_4\\\lambda_5\\\lambda_6\\\lambda_7\\\lambda_8\end{bmatrix}=\begin{bmatrix}0.00100007\\0.00100007\\0.00100012\\0.00658858\\0.02030942\\0.06050304\\0.06050304\\0.06557759\end{bmatrix}$ | $\begin{bmatrix}\lambda_1\\\lambda_2\\\lambda_3\\\lambda_4\\\lambda_5\\\lambda_6\\\lambda_7\\\lambda_8\end{bmatrix}=\begin{bmatrix}0.00200017\\0.00200017\\0.00302139\\0.00664495\\0.01716183\\0.05217245\\0.05217245\\0.07241556\end{bmatrix}$ |

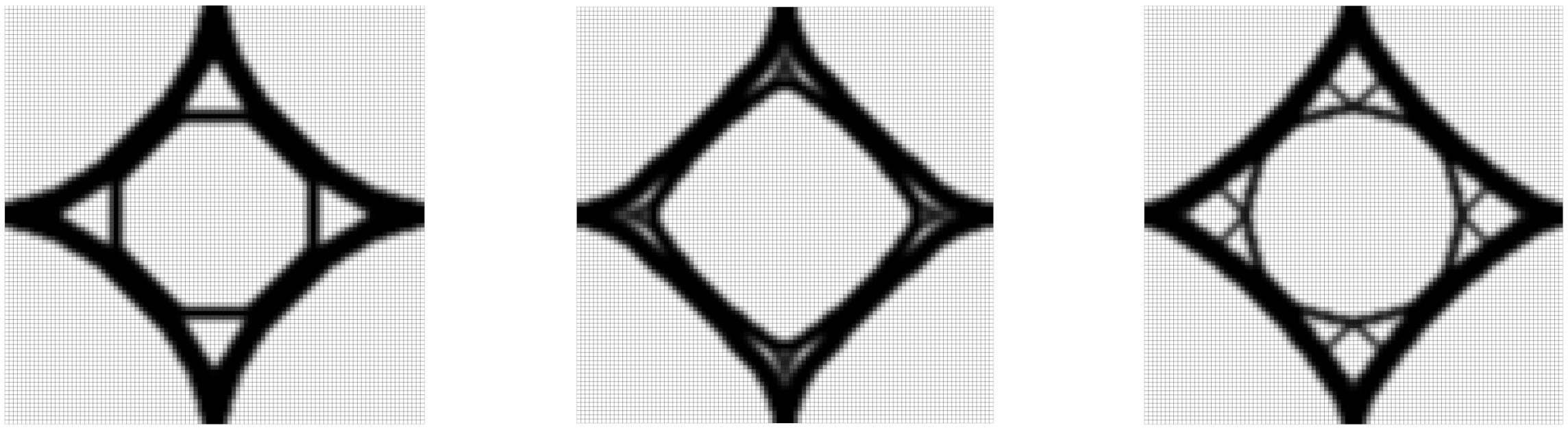

(a) No stability constraint

(b) Stability constraint $\hat{\lambda}$ = 0.001

(c) Stability constraint $\hat{\lambda}$ = 0.002

Figure 34. Optimized topologies of the symmetric compressive block domain for different stability constraint values.

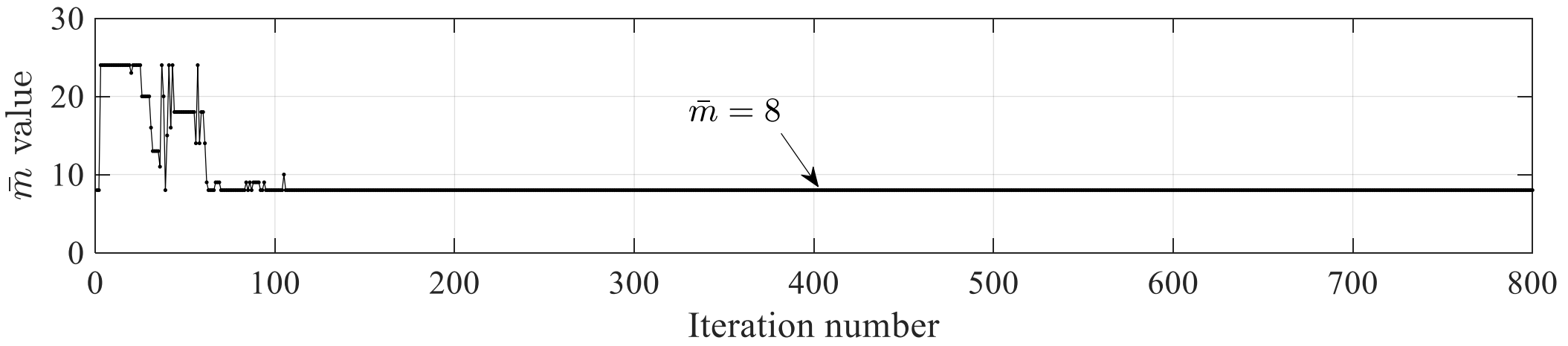

(a) $\hat{\lambda}$ = 0.001

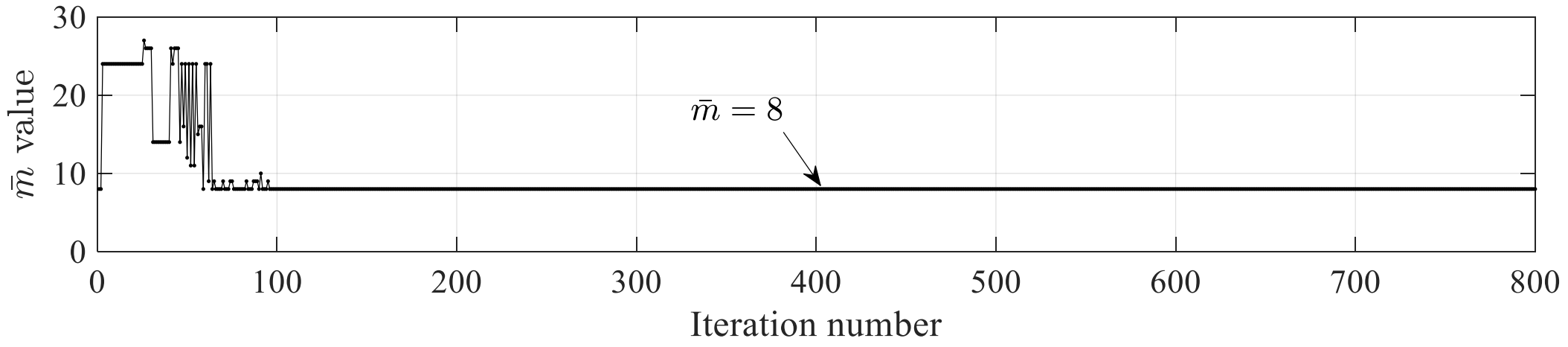

(b) $\hat{\lambda}$ = 0.002

Figure 35. Number of eigenvalues included in the optimization process, i.e., the value of $\bar{m}$ of the intermediate designs in the symmetric compressive block problem for different stability constraint values.

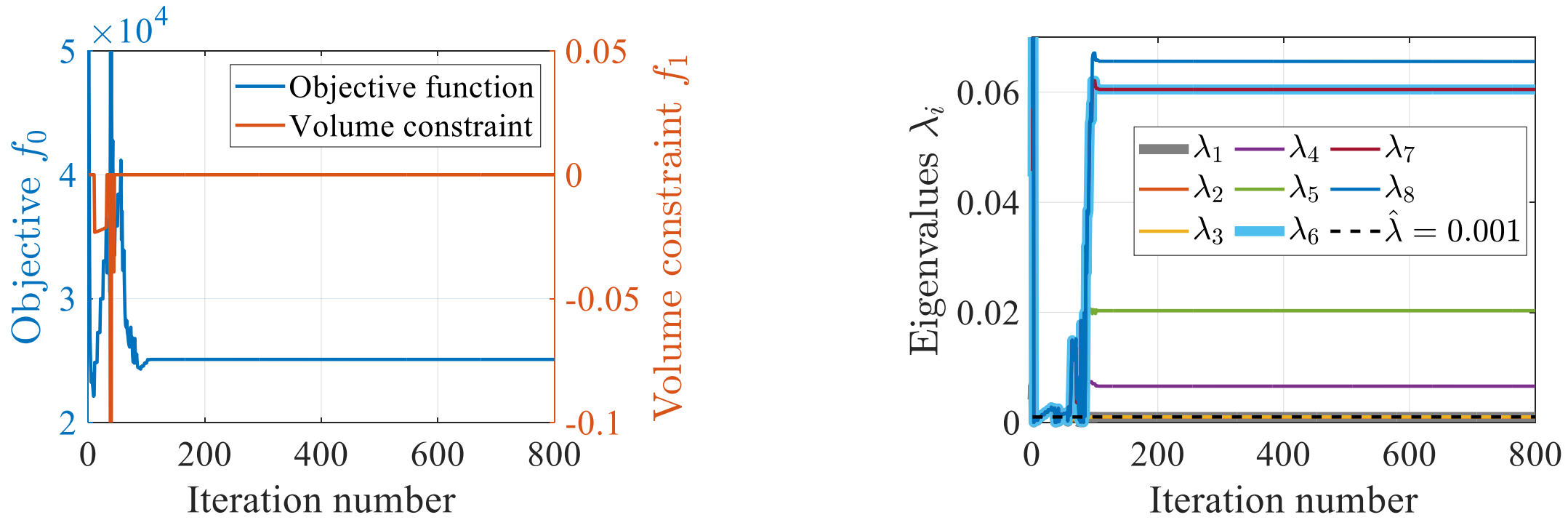

(a) Stability constraint $\hat{\lambda} = 0.001$

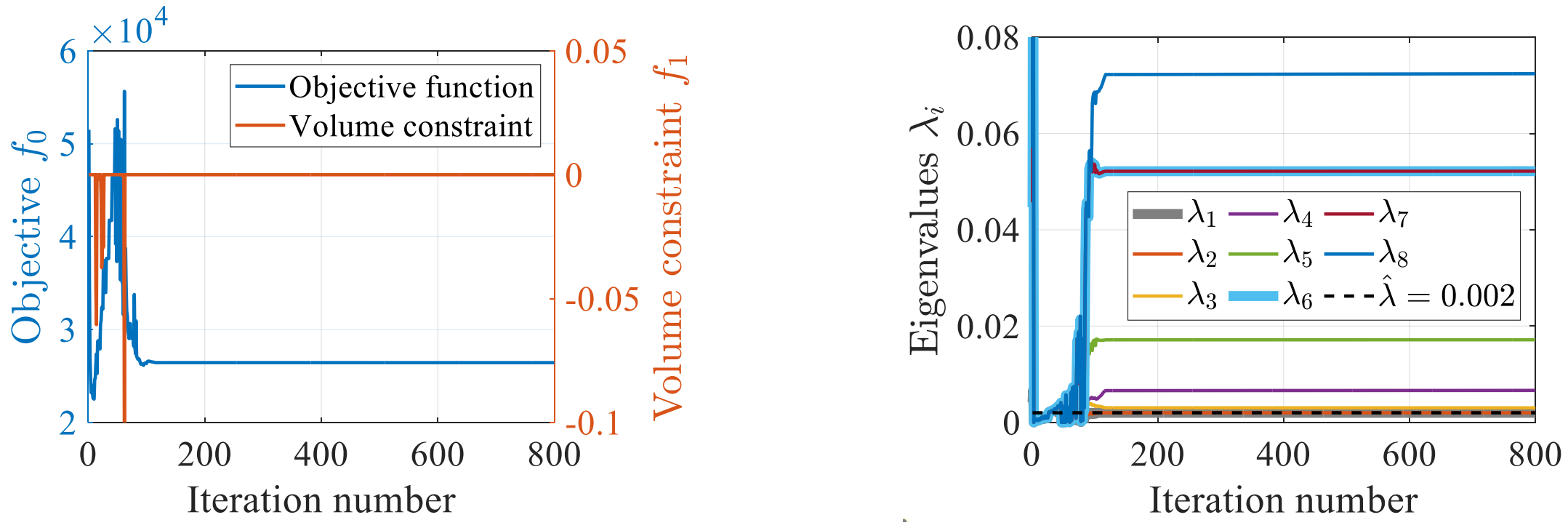

(b) Stability constraint $\hat{\lambda} = 0.002$

Figure 36. Optimization histories of objective, material volume constraint function values, as well as the first eight eigenvalues of the tangent stiffness matrix of the designs at the target load for the different stability threshold values in the symmetric compressive block problem.

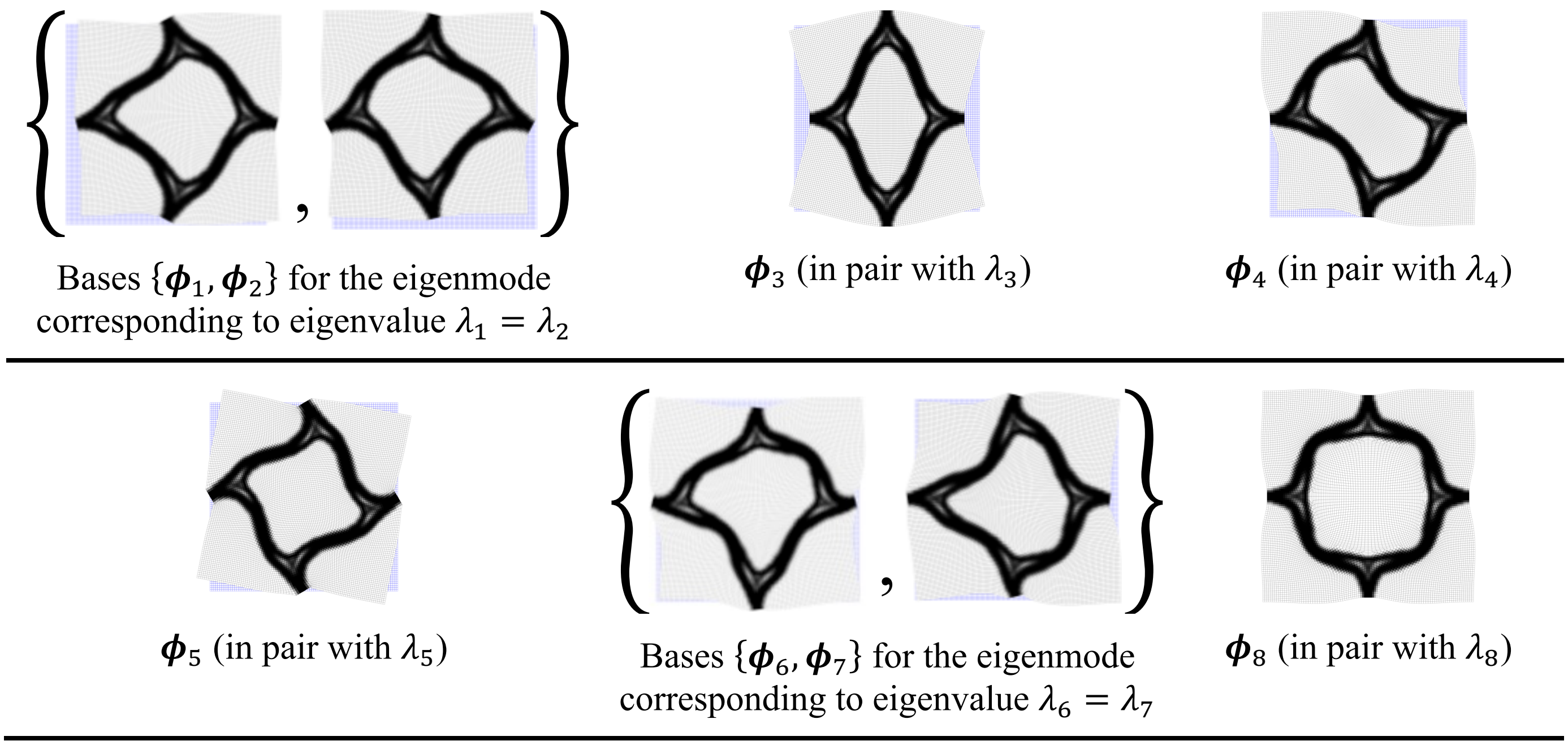

(a) Stability constraint $\hat{\lambda} = 0.001$

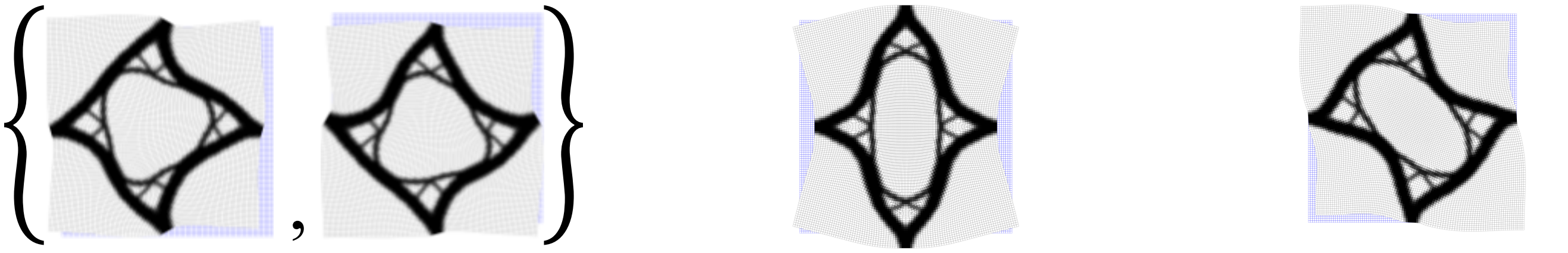

Bases $\{\boldsymbol{\phi}_1, \boldsymbol{\phi}_2\}$ for the eigenmode corresponding to eigenvalue $\lambda_1 = \lambda_2$

$\boldsymbol{\phi}_3$ (in pair with $\lambda_3$)

$\boldsymbol{\phi}_4$ (in pair with $\lambda_4$)

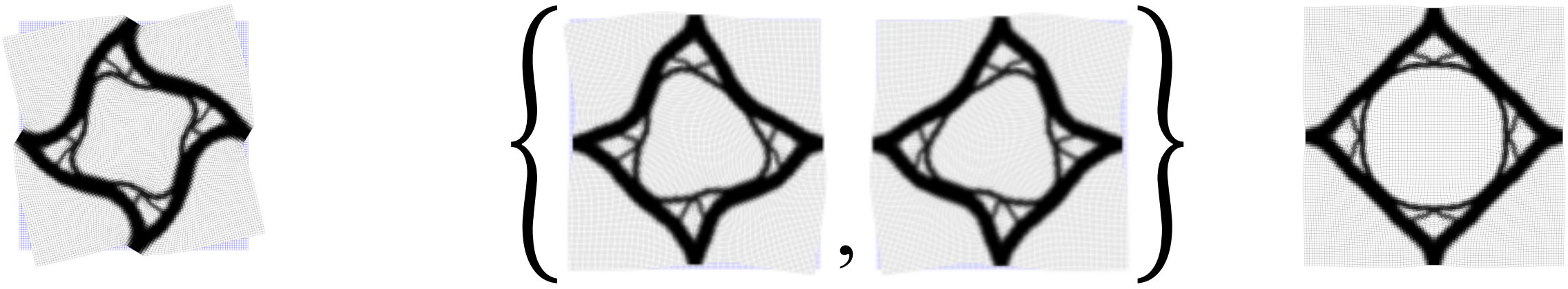

$\boldsymbol{\phi}_5$ (in pair with $\lambda_5$)

Bases $\{\boldsymbol{\phi}_6, \boldsymbol{\phi}_7\}$ for the eigenmode corresponding to eigenvalue $\lambda_6 = \lambda_7$

$\boldsymbol{\phi}_8$ (in pair with $\lambda_8$)

(b) Stability constraint $\hat{\lambda} = 0.002$

Figure 37. Eigenmodes of the optimized stable designs at the target load (Note: appropriate scales are applied to the eigenvectors for visualization purposes)

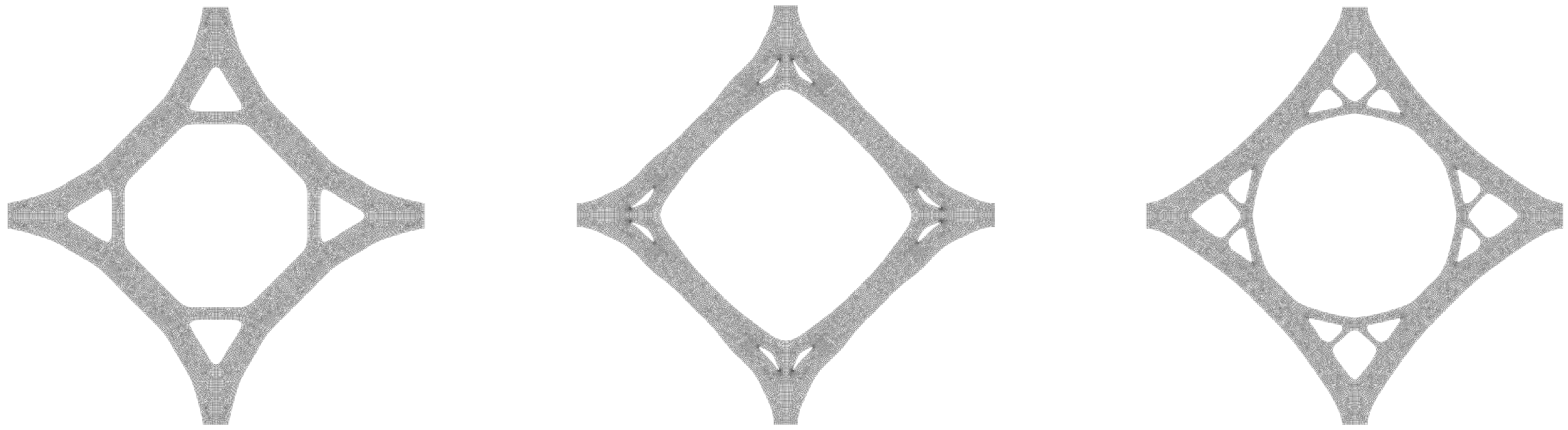

(a) No stability constraint

(b) Stability constraint $\hat{\lambda} = 0.001$

(c) Stability constraint $\hat{\lambda} = 0.002$

Figure 38. FE meshes of the B-spline fitted designs for the three cases in Figure 34.

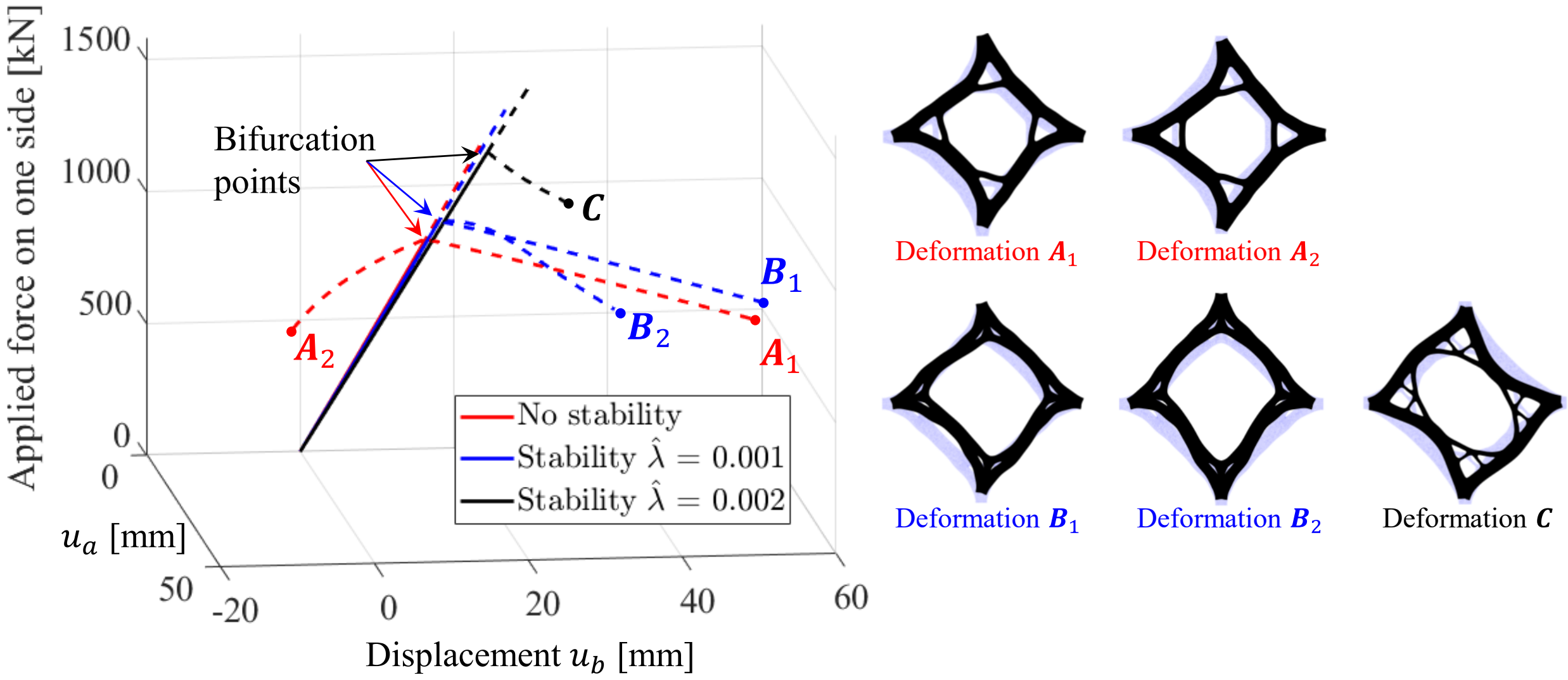

Figure 39. Load-displacement curves of the three B-spline fitted optimized symmetric compressive block topologies and their deformed shapes (solid line corresponds to stable region, while dashed line corresponds to unstable region).

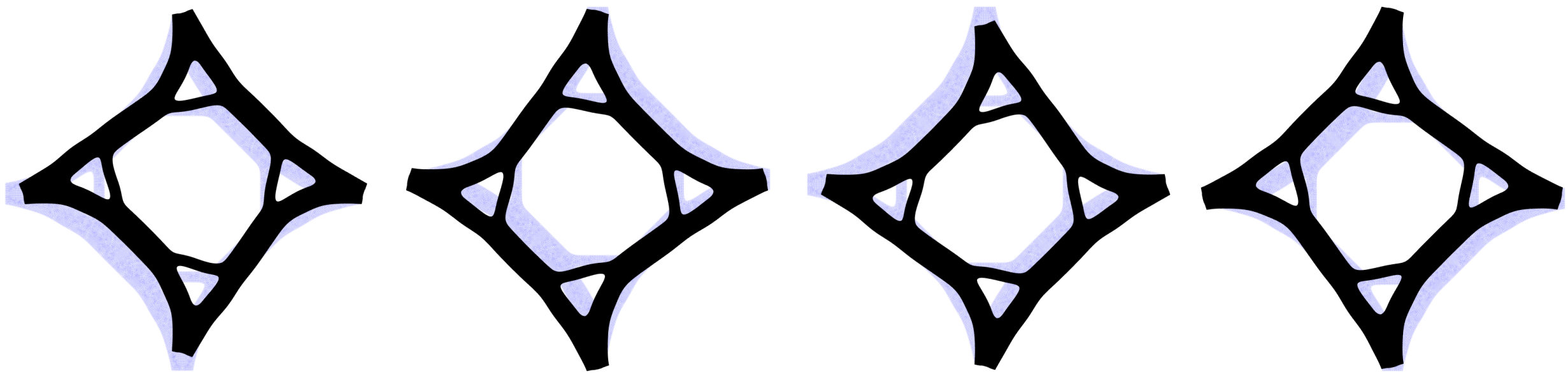

(a) Two directions of buckling mode 1 (b) Two directions of buckling mode 2

Figure 40. Illustration of the symmetric solutions related to the deformation $\boldsymbol{A}_1$ in Figure 39.

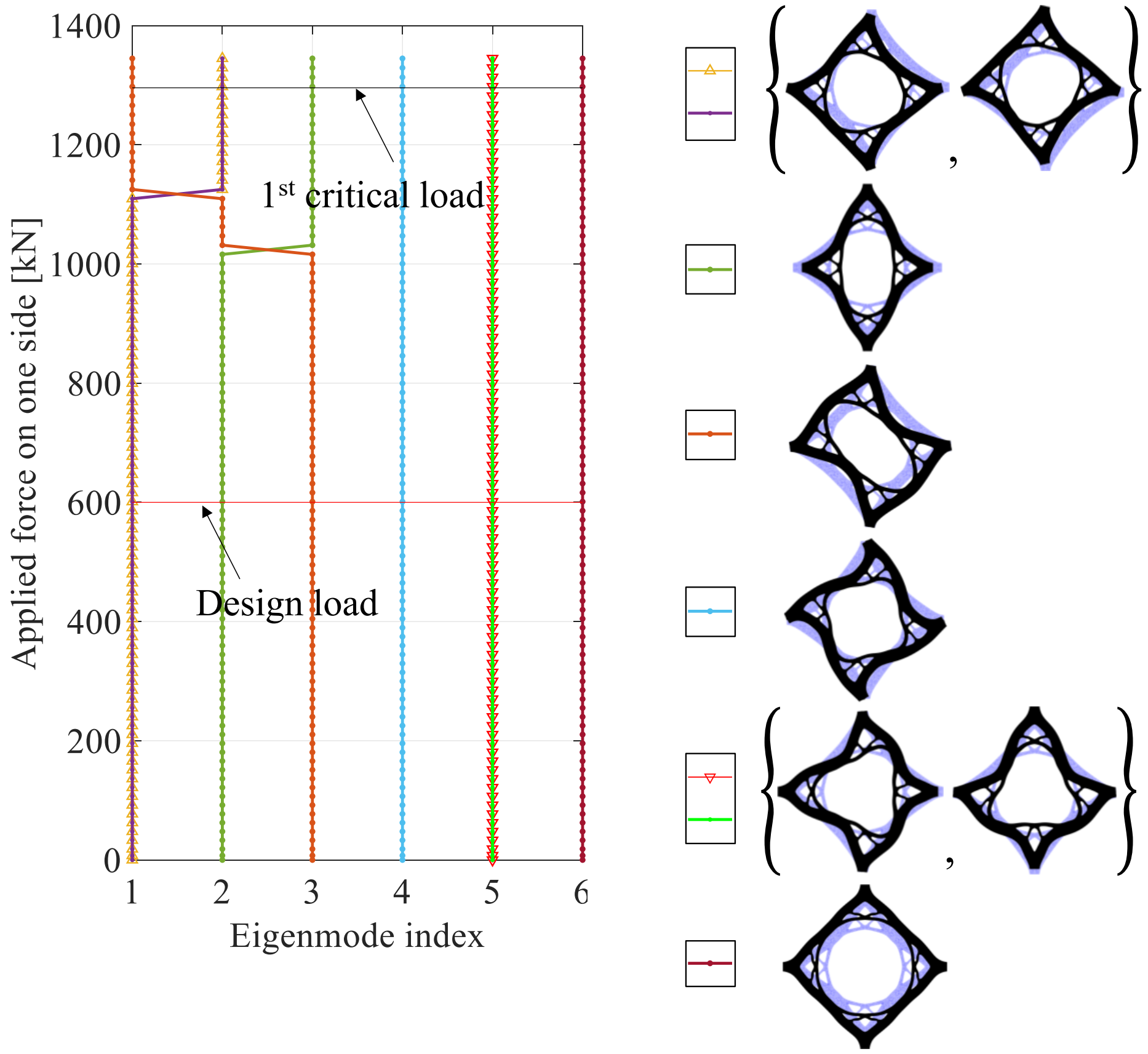

Figure 41. Mode switching during the loading process of the fitted design with stability threshold $\hat{\lambda}$ = 0.002 (Note: the curves above the 1st critical load are following the primary branch).

The optimized topologies of the three cases are shown in Figure 34, with their objective function values and the first eight eigenvalues given in Table 3. From Table 3 it can be observed that the design in case 1 without stability constraint already lost stability before reaching the target load. Due to symmetry, there always are multiple eigenvalues in the considered first six clusters of eigenvalues during the optimization, and therefore, the sub-optimization in Eq. (41) is activated throughout the optimization process for case 2 and case 3. Figure 35 plots the number of eigenvalues included ($\overline{m}$) during the optimization process, where it can be seen that in roughly the first hundred iterations there can be as many as 27 eigenvalues (Figure 35b) in the first six clusters of different eigenvalues and $\overline{m}$ = 8 for the rest of the iterations. Figure 36 presents the optimization history of the objective function, volumetric constraint function, and the first eight eigenvalues, where all the constraints are satisfied, and the objective is decreasing. The constrained eigenmodes of the final optimized designs are depicted in Figure 37 which are all "real" modes.

The optimized topologies in Figure 34 are further fitted using B-splines and discretized with FE meshes as shown in Figure 38. Under the same load pattern, with arc-length continuation and branch-switching methods in Appendix D2, the load-displacement curves of the three designs are plotted in Figure 39, where the 1st bifurcation loads for cases 1, 2, and 3 are 912.49 kN, 992.61 kN, and 1295.55 kN, respectively. In Figure 39, it should be noted that for simplicity, out of every two/four symmetric solutions only one is plotted. As an illustration, Figure 40 shows four different solutions that are essentially symmetric. Out of these four solutions only the one corresponding to the deformation $\boldsymbol{A}_1$ is plotted in Figure 39. In addition, the load-displacement curve of case 2 ($\hat{\lambda} = 0.002$) reveals that the bifurcated branch at the 1st bifurcation point does not follow the modes associated with the smallest eigenvalue at the design load (Figure 37b), as case 1 does. To investigate this further, Figure 41 traces each eigenmode and its associated eigenvalue in the eigenvalue clusters during the loading process. As shown in Figure 41, before the external load reaches 1000 kN, the order of the eigenmodes is the same as that in Figure 37b. However, mode switching happens between the load 1000 kN and 1200 kN, after which the eigenvalue associated with the initial eigenmode $\boldsymbol{\phi}_4$ becomes the smallest while the multiple eigenvalues in pair with the eigenmode lying on the hyperplane spanned by the initial eigenbases $\boldsymbol{\phi}_1$ and $\boldsymbol{\phi}_2$ switch to the second cluster.

## 8 Conclusions

This study presents a computational framework for geometrically nonlinear topology optimization with nonlinear stability constraints. Unlike the linear buckling constraints that may only provide accurate results when the deformations are small, the nonlinear stability constraints used in the current study provide accurate control over the potential buckling at large deformations. The nonlinear stability analysis at finite deformations is realized with the help of the proposed pseudo-mass matrix which can effectively remove spurious buckling modes related to void and low-density elements. This pseudo-mass matrix eventually degenerates into an identity matrix when all the elements are solid. The lack of differentiability of multiple (repeated) eigenvalues is addressed by computing the directional derivatives using the perturbation strategy proposed by Seyranian et al. [21] and the framework is extended to incorporate the nonlinear tangent stiffness and pseudo-mass matrices. A novel optimization formulation is proposed to accommodate both the simple and multiple eigenvalue scenarios that consist of outer loops optimizing design variables and inner loops optimizing the incremental design variables with linearized objective and constraint functions. The mesh distortion issue due to geometric nonlinearity is addressed by the adaptive linear energy interpolation scheme proposed by the authors in [16].

The numerical examples demonstrate the effectiveness of the developed framework for handling design optimization problems with both simple and multiple eigenvalues. From the presented results it can be concluded that by incorporating the nonlinear buckling constraints, the structural stability of the optimized designs up to the target load can be guaranteed. Moreover, increasing the stability threshold ($\hat{\lambda}$) value in the stability constraints can lead to optimized structures with a higher critical load. As the results suggest, higher stability performance can be achieved using the proposed framework but at the expense of lower stiffness. It is also worth mentioning that due to mode switching the eigenmode(s) corresponding to the smallest eigenvalue in the stability constraint might not be the buckling mode of the optimized structure at its 1st critical load which is often greater than the target design load.

## Acknowledgments

The first author would like to acknowledge the financial support from the Fundamental Research Funds for the Central Universities (No. 3205002207A1). The second author would like to acknowledge the support from the US National Science Foundation through grant CMMI-1762277. Any opinions, findings, conclusions, and recommendations expressed in this paper are those of the authors and do not necessarily reflect the views of the sponsors.

## Appendix A: Derivatives for the sensitivity analysis of simple/multiple eigenvalues

This appendix gives details on the calculation of the terms needed for the sensitivity analysis of simple/multiple eigenvalues, i.e., the terms given in Eqns. (25) and (39).

### A 1. Derivatives $\left.\dfrac{\partial(\boldsymbol{\phi}_s^T \boldsymbol{K}_T \boldsymbol{\phi}_k)}{\partial \boldsymbol{\rho}}\right|_{\boldsymbol{\phi}_s,\boldsymbol{\phi}_k \text{ fixed}}$

As the density field is element-wise constant, the derivatives can be computed in an element-by-element manner. That is, the eigenvectors $\boldsymbol{\phi}_s$ are first decomposed in an element-by-element format $\boldsymbol{\phi}_s^e$ ($e = 1, \dots, n_{ele}$), and then the derivatives can be computed as

$$\left.\frac{\partial(\boldsymbol{\phi}_s^T \boldsymbol{K}_T \boldsymbol{\phi}_k)}{\partial \rho_e}\right|_{\boldsymbol{\phi}_s,\boldsymbol{\phi}_k \text{ fixed}} = \boldsymbol{\phi}_s^{e\,T} \frac{\partial \boldsymbol{k}_T^e}{\partial \rho_e} \boldsymbol{\phi}_k^e \quad , \quad e = 1, \dots, n_{ele} \tag{A 1}$$

where using Eqns. (14), (12), and (3) the derivative $\partial \boldsymbol{k}_T^e / \partial \rho_e$ for the $e^{\text{th}}$ element is calculated by

$$\begin{aligned}\frac{\partial \boldsymbol{k}_T^e}{\partial \rho_e} = {} & 2\eta \frac{\partial \eta}{\partial \rho_e} \int_{\Omega_0^e} \boldsymbol{B}^T [\mathbb{A}] \boldsymbol{B} \, dV + \eta^2 \int_{\Omega_0^e} \boldsymbol{B}^T \left[ \left.\frac{\partial \mathbb{A}}{\partial \rho_e}\right|_{\boldsymbol{F} \text{ fixed}} + \frac{\partial \mathbb{A}}{\partial \boldsymbol{F}} : \frac{\partial \boldsymbol{F}}{\partial \rho_e} \right] \boldsymbol{B} \, dV \\ & - 2\eta \frac{\partial \eta}{\partial \rho_e} \int_{\Omega_0^e} \boldsymbol{B}_L^T [\mathbb{C}] \boldsymbol{B}_L \, dV + (1 - \eta^2) \int_{\Omega_0^e} \boldsymbol{B}_L^T \left[ \frac{\partial \mathbb{C}}{\partial \rho_e} \right] \boldsymbol{B}_L \, dV \end{aligned} \tag{A 2}$$

in which $\partial \eta / \partial \rho_e$ is computed using Eq. $(12)_2$, $\partial \mathbb{A} / \partial \rho_e |_{\boldsymbol{F} \text{ fixed}}$ and $\partial \mathbb{C} / \partial \rho_e$ are straightforward as they are proportional to Young's modulus, and the derivative $\partial \mathbb{A} / \partial \boldsymbol{F}$ for the considered neo-Hookean hyperelastic model in Eq. (8) is given by

$$\frac{\partial \mathbb{A}}{\partial \boldsymbol{F}} = \kappa(2J-1)J\frac{\partial \boldsymbol{F}^{-T}}{\partial \boldsymbol{F}} \otimes \boldsymbol{F}^{-T} + \kappa(J-1)J\frac{\partial^2 \boldsymbol{F}^{-T}}{\partial \boldsymbol{F}\partial \boldsymbol{F}} + \kappa(4J-1)J\boldsymbol{F}^{-T} \otimes \boldsymbol{F}^{-T} \otimes \boldsymbol{F}^{-T} + \kappa(2J-1)J\frac{\partial(\boldsymbol{F}^{-T} \otimes \boldsymbol{F}^{-T})}{\partial \boldsymbol{F}} - \frac{\mu}{3}\frac{\partial \boldsymbol{F}^{-T}}{\partial \boldsymbol{F}} \otimes \frac{\partial \bar{I}_1}{\partial \boldsymbol{F}} - \frac{\mu}{3}\bar{I}_1\frac{\partial^2 \boldsymbol{F}^{-T}}{\partial \boldsymbol{F}\partial \boldsymbol{F}} - \frac{\mu}{3}\boldsymbol{F}^{-T} \otimes \frac{\partial^2 \bar{I}_1}{\partial \boldsymbol{F}\partial \boldsymbol{F}} - \frac{\mu}{3}\frac{\partial \boldsymbol{F}^{-T}}{\partial \boldsymbol{F}} \boxdot \frac{\partial \bar{I}_1}{\partial \boldsymbol{F}} - \frac{2}{3}\mu J^{-\frac{2}{3}}\mathbb{I}_4 \otimes \boldsymbol{F}^{-T} + \frac{4}{9}\mu J^{-\frac{2}{3}}\boldsymbol{F} \otimes \boldsymbol{F}^{-T} \otimes \boldsymbol{F}^{-T} - \frac{2}{3}\mu J^{-\frac{2}{3}}\frac{\partial(\boldsymbol{F} \otimes \boldsymbol{F}^{-T})}{\partial \boldsymbol{F}} \tag{A 3}$$

where

$$\left[\frac{\partial^2 \boldsymbol{F}^{-T}}{\partial \boldsymbol{F}\partial \boldsymbol{F}}\right]_{ijklpq} = F_{qi}^{-1}F_{lp}^{-1}F_{jk}^{-1} + F_{qk}^{-1}F_{jp}^{-1}F_{li}^{-1}$$

$$\left[\frac{\partial(\boldsymbol{F}^{-T} \otimes \boldsymbol{F}^{-T})}{\partial \boldsymbol{F}}\right]_{ijklpq} = -F_{qi}^{-1}F_{jp}^{-1}F_{lk}^{-1} - F_{qk}^{-1}F_{lp}^{-1}F_{ji}^{-1} \tag{A 4}$$

$$\left[\frac{\partial(\boldsymbol{F} \otimes \boldsymbol{F}^{-T})}{\partial \boldsymbol{F}}\right]_{ijklpq} = \delta_{ip}\delta_{jq}F_{lk}^{-1} - F_{ij}F_{qk}^{-1}F_{lp}^{-1}$$

$$\frac{\partial^2 \bar{I}_1}{\partial \boldsymbol{F}\partial \boldsymbol{F}} = -\frac{2}{3}\boldsymbol{F}^{-T} \otimes \frac{\partial \bar{I}_1}{\partial \boldsymbol{F}} - \frac{2}{3}\bar{I}_1\frac{\partial \boldsymbol{F}^{-T}}{\partial \boldsymbol{F}} - \frac{4}{3}J^{-2/3}\boldsymbol{F} \otimes \boldsymbol{F}^{-T} + 2J^{-2/3}\mathbb{I}_4$$

and the calculation of the terms $\partial \bar{I}_1/\partial \boldsymbol{F}$ and $\partial \boldsymbol{F}^{-T}/\partial \boldsymbol{F}$ can be found in Eqns. (9) and (11), respectively, and the operator $\boxdot$ is defined by $(\mathbb{H} \boxdot \boldsymbol{H})_{ijklpq} \stackrel{\text{def}}{=} \mathbb{H}_{ijpq}H_{kl}$ where $\mathbb{H}$ is a 4th-order tensor and $\boldsymbol{H}$ is a 2nd-order tensor.

In Eq. (A 2), the dependence of the deformation gradient $\boldsymbol{F}$ on the element density variable $\rho_e$ is determined by the linear energy interpolation, see Eq. (12), and the derivative is

$$\left[\frac{\partial \boldsymbol{F}}{\partial \rho_e}\right] = \frac{\partial \eta}{\partial \rho_e}\boldsymbol{B}\boldsymbol{u}_e \tag{A 5}$$

## A 2. Derivatives $\left.\frac{\partial(\boldsymbol{\phi}_s^T \boldsymbol{K}_T \boldsymbol{\phi}_k)}{\partial \boldsymbol{u}}\right|_{\boldsymbol{\phi}_s,\boldsymbol{\phi}_k \text{ fixed}}$

The calculation of this term can be carried out in an element-by-element format by FE assembly, i.e.,

$$\left.\frac{\partial(\boldsymbol{\phi}_s^T \boldsymbol{K}_T \boldsymbol{\phi}_k)}{\partial \boldsymbol{u}}\right|_{\boldsymbol{\phi}_s,\boldsymbol{\phi}_k \text{ fixed}} = \mathop{\mathcal{A}}_{e=1}^{n_{ele}} \frac{\partial\left(\boldsymbol{\phi}_s^{e^T} \boldsymbol{k}_T^e \boldsymbol{\phi}_k^e\right)}{\partial \boldsymbol{u}_e} \tag{A 6}$$

in which the term $\boldsymbol{\phi}_s^{e^T} \boldsymbol{k}_T^e \boldsymbol{\phi}_k^e$ is first expanded as

$$\boldsymbol{\phi}_s^{e^T} \boldsymbol{k}_T^e \boldsymbol{\phi}_k^e = \int_{\Omega_0^e} \eta^2 (\boldsymbol{B}\boldsymbol{\phi}_s^e)^T [\mathbb{A}] (\boldsymbol{B}\boldsymbol{\phi}_k^e)\, dV + \int_{\Omega_0^e} (1-\eta^2)(\boldsymbol{B}_L\boldsymbol{\phi}_s^e)^T [\mathbb{C}] (\boldsymbol{B}_L\boldsymbol{\phi}_k^e)\, dV \tag{A 7}$$

and can be further rewritten to

$$\begin{aligned}\boldsymbol{\phi}_s^{e^T} \boldsymbol{k}_T^e \boldsymbol{\phi}_k^e &= \int_{\Omega_0^e} \eta^2 [\mathbb{A}] : [(\boldsymbol{B}\boldsymbol{\phi}_k^e)(\boldsymbol{B}\boldsymbol{\phi}_s^e)^T]\, dV \\ &\quad + \int_{\Omega_0^e} (1-\eta^2)[\mathbb{C}] : [(\boldsymbol{B}_L\boldsymbol{\phi}_k^e)(\boldsymbol{B}_L\boldsymbol{\phi}_s^e)^T]\, dV\end{aligned} \tag{A 8}$$

where the double-dot product ':' works similarly as tensor contraction but applies between two matrices, i.e., $[\boldsymbol{A}]:[\boldsymbol{B}] = \sum_{i,j} A_{ij}B_{ij}$. Hence, the derivative $\partial\left(\boldsymbol{\phi}_s^{e^T} \boldsymbol{k}_T^e \boldsymbol{\phi}_k^e\right)/\partial \boldsymbol{u}_e$ can be computed by

$$\frac{\partial\left(\boldsymbol{\phi}_s^{e^T} \boldsymbol{k}_T^e \boldsymbol{\phi}_k^e\right)}{\partial \boldsymbol{u}_e} = \int_{\Omega_0^e} \eta^2 [(\boldsymbol{B}\boldsymbol{\phi}_k^e)(\boldsymbol{B}\boldsymbol{\phi}_s^e)^T] : \left[\frac{\partial \mathbb{A}}{\partial \boldsymbol{u}_e}\right] dV \tag{A 9}$$

since the linear part is not a function of the displacement. In Eq. (A *8*), the matrix $[\mathbb{A}]$ is 4×4 matrix for 2D (and 9×9 for 3D) case. Accordingly, $[(\boldsymbol{B}\boldsymbol{\phi}_k^e)(\boldsymbol{B}\boldsymbol{\phi}_s^e)^T]$ is of the same size as the matrix $[\mathbb{A}]$. To simplify the numerical implementations, the matrices $[(\boldsymbol{B}\boldsymbol{\phi}_k^e)(\boldsymbol{B}\boldsymbol{\phi}_s^e)^T]$ and $[\mathbb{A}]$ for 2D case can be reshaped to a vector of size 16×1, and the double-dot product becomes a dot product between two vectors. Therefore, the matrix form of the tensor $[\partial\mathbb{A}/\partial\boldsymbol{u}_e]$ is of size 16×8. The derivative $\partial\mathbb{A}/\partial\boldsymbol{u}_e$ is derived as

$$\frac{\partial\mathbb{A}}{\partial\boldsymbol{u}_e} = \frac{\partial\mathbb{A}}{\partial\boldsymbol{F}} : \frac{\partial\boldsymbol{F}}{\partial\boldsymbol{u}_e} \tag{A 10}$$

where the derivative $\partial\boldsymbol{F}/\partial\boldsymbol{u}_e$ in a matrix form is simply $[\partial\boldsymbol{F}/\partial\boldsymbol{u}_e] = \eta\boldsymbol{B}$.

### A 3. Derivatives $\left.\frac{\partial(\boldsymbol{\phi}_s^T\boldsymbol{S}_M\boldsymbol{\phi}_k)}{\partial\boldsymbol{\rho}}\right|_{\boldsymbol{\phi}_s,\boldsymbol{\phi}_k \text{ fixed}}$

First, the term $\boldsymbol{\phi}_s^T\boldsymbol{S}_M\boldsymbol{\phi}_k$ is expanded by arranging the eigenvector $\boldsymbol{\phi}_s$ in a nodal format. That is, the eigenvector $\boldsymbol{\phi}_s$ is reshaped to a matrix of size $2 \times \#node$ for 2D case and $3 \times \#node$ for 3D case. Each column represents the displacement of a node in the corresponding eigenmode. We denote the new format of the eigenvector as

$$\boldsymbol{\phi}_s = [\widehat{\boldsymbol{\phi}}_s^1 \quad \widehat{\boldsymbol{\phi}}_s^2 \quad \cdots \quad \widehat{\boldsymbol{\phi}}_s^{\#node}]_{d\times\#node} \tag{A 11}$$

As a result, the term $\boldsymbol{\phi}_s^T\boldsymbol{S}_M\boldsymbol{\phi}_k$ can be expressed as

$$\boldsymbol{\phi}_s^T\boldsymbol{S}_M\boldsymbol{\phi}_k = \sum_{i=1}^{\#node} \widetilde{m}_i {\widehat{\boldsymbol{\phi}}_s^{i}}^T \widehat{\boldsymbol{\phi}}_k^i \tag{A 12}$$

The density vector $\boldsymbol{\rho}$ is decomposed into nodal density vectors $\widehat{\boldsymbol{\rho}}_i$ ($i = 1, \dots, \#node$), where

$$\widehat{\boldsymbol{\rho}}_i = \begin{bmatrix} \vdots \\ \rho_r \\ \vdots \end{bmatrix}, \quad r \in \mathcal{B}_i \tag{A 13}$$

maybe of different sizes depending on the number of element IDs inside $\mathcal{B}_i$, see explanations following Eq. (18). From Eqns. (18) and (19), it is clear that

$$\frac{\partial\widetilde{m}_i}{\partial\widehat{\boldsymbol{\rho}}_i} = \frac{\partial\widetilde{m}_i}{\partial\varpi_i}\frac{\partial\varpi_i}{\partial\widehat{\boldsymbol{\rho}}_i} \tag{A 14}$$

with

$$\frac{\partial \widetilde{m}_i}{\partial \varpi_i} = \begin{cases} p_m(1-\hat{\epsilon})\varpi_i^{p_m-1} & \text{if } \varpi_i \le \varpi_L \\ a_1 + 2a_2\varpi_i + 3a_3\varpi_i^2 & \text{if } \varpi_L < \varpi_i < \varpi_H \\ 0 & \text{if } \varpi_i \ge \varpi_H \end{cases}$$

$$\frac{\partial \varpi_i}{\partial \widehat{\boldsymbol{\rho}}_i} = \frac{1}{q}\left(\sum_{r\in\mathcal{B}_i} \rho_r^q\right)^{\frac{1}{q}-1} \begin{bmatrix} \vdots \\ q\rho_r^{q-1} \\ \vdots \end{bmatrix}, \quad r \in \mathcal{B}_i$$

and that $\partial\widetilde{m}_i/\partial\widehat{\boldsymbol{\rho}}_j = \mathbf{0}$ if $i \neq j$.

Therefore, it is straightforward that

$$\left.\frac{\partial(\boldsymbol{\phi}_s^T \boldsymbol{S}_M \boldsymbol{\phi}_k)}{\partial \boldsymbol{\rho}}\right|_{\boldsymbol{\phi}_s,\boldsymbol{\phi}_k \text{ fixed}} = \mathop{\widehat{\mathcal{A}}}_{\imath=1}^{\#node} \frac{\partial(\boldsymbol{\phi}_s^T \boldsymbol{S}_M \boldsymbol{\phi}_k)}{\partial \widehat{\boldsymbol{\rho}}_i} \tag{A 15}$$

where $\hat{\mathcal{A}}$ is an assembly operator that adds the terms of the same element number together. According to Eq. (A *12*), the derivative $\partial(\boldsymbol{\phi}_s^T \boldsymbol{S}_M \boldsymbol{\phi}_k)/\partial\widehat{\boldsymbol{\rho}}_i$ can be computed as

$$\frac{\partial(\boldsymbol{\phi}_s^T \boldsymbol{S}_M \boldsymbol{\phi}_k)}{\partial \widehat{\boldsymbol{\rho}}_i} = \left(\widehat{\boldsymbol{\phi}}_s^{i\,T} \widehat{\boldsymbol{\phi}}_k^i\right) \frac{\partial \widetilde{m}_i}{\partial \widehat{\boldsymbol{\rho}}_i} \tag{A 16}$$

## A 4. Derivative $\partial\boldsymbol{R}/\partial\boldsymbol{\rho}$

The derivative $\partial\boldsymbol{R}/\partial\boldsymbol{\rho}$ is calculated by FE assembly as

$$\frac{\partial \boldsymbol{R}}{\partial \boldsymbol{\rho}} = \mathop{\mathcal{A}}_{e=1}^{n_{ele}} \frac{\partial \boldsymbol{R}_e}{\partial \boldsymbol{\rho}} \quad \text{with} \quad \frac{\partial \boldsymbol{R}_e}{\partial \boldsymbol{\rho}} = \begin{bmatrix} \frac{\partial \boldsymbol{R}_e}{\partial \rho_1} & \frac{\partial \boldsymbol{R}_e}{\partial \rho_2} & \cdots & \frac{\partial \boldsymbol{R}_e}{\partial \rho_{n_{ele}}} \end{bmatrix} \tag{A 17}$$

in which

$$\begin{aligned} \frac{\partial \boldsymbol{R}_e}{\partial \rho_e} = \frac{\partial \eta}{\partial \rho_e} \int_{\Omega_0^e} \boldsymbol{B}^T \boldsymbol{P}\, dV + \int_{\Omega_0^e} \eta \boldsymbol{B}^T \left( \left.\frac{\partial \boldsymbol{P}}{\partial \rho_e}\right|_{\boldsymbol{F} \text{ fixed}} + \mathbb{A}{:}\frac{\partial \boldsymbol{F}}{\partial \rho_e} \right) dV \\ - 2\eta \frac{\partial \eta}{\partial \rho_e} \int_{\Omega_0^e} \boldsymbol{B}_L^T [\mathbb{C}{:}\,\boldsymbol{\varepsilon}]\, dV + \int_{\Omega_0^e} (1-\eta^2) \boldsymbol{B}_L^T \left[\frac{\partial \mathbb{C}}{\partial \rho_e}{:}\,\boldsymbol{\varepsilon}\right] dV \end{aligned} \tag{A 18}$$

and $\partial \boldsymbol{R}_e / \partial \rho_j = \boldsymbol{0}$ if $e \neq j$. The derivatives $(\partial \boldsymbol{P} / \partial \rho_e)|_{\boldsymbol{F}\ \text{fixed}}$ is straightforward as the 1st PK stress $\boldsymbol{P}$ is proportional to Young's modulus.

### A 5. Derivative $\partial \boldsymbol{R} / \partial \boldsymbol{u}$

By definition, the derivative $\partial \boldsymbol{R} / \partial \boldsymbol{u}$ gives the tangent stiffness matrix, i.e.,

$$\frac{\partial \boldsymbol{R}}{\partial \boldsymbol{u}} = \boldsymbol{K}_T \tag{A 19}$$

where $\boldsymbol{K}_T$ is given in Eq. (*14*).

## Appendix B: Sensitivity analysis of end compliance and volume fraction

### B 1. Sensitivity of the objective $f_0$

The sensitivity analysis of the objective function is carried out using the adjoint method

$$\frac{df_0}{d\boldsymbol{\rho}} = \frac{\partial f_0}{\partial \boldsymbol{\rho}} + \boldsymbol{\chi}^T \frac{\partial \boldsymbol{R}}{\partial \boldsymbol{\rho}} \tag{B 1}$$

where $\boldsymbol{\chi}$ is the adjoint variable vector of the same size as $\boldsymbol{R}$ and is calculated as

$$\boldsymbol{\chi}^T = -\frac{\partial f_0}{\partial \boldsymbol{u}} \left[\frac{\partial \boldsymbol{R}}{\partial \boldsymbol{u}}\right]^{-1} \tag{B 2}$$

in which the derivatives $\partial \boldsymbol{R}/\partial \boldsymbol{\rho}$ and $\partial \boldsymbol{R}/\partial \boldsymbol{u}$ are given in Appendix A4 and A5, and the remaining terms to be computed are $\partial f_0/\partial \boldsymbol{\rho}$ and $\partial f_0/\partial \boldsymbol{u}$ and are given by

$$\begin{aligned} \frac{\partial f_0}{\partial \boldsymbol{\rho}} &= \boldsymbol{0} \\ \frac{\partial f_0}{\partial \boldsymbol{u}} &= \boldsymbol{F}_{ext}^T \end{aligned} \tag{B 3}$$

### B 2. Sensitivity of the volume fraction $f_1$

The volume fraction function is simply only a function of the density variables and are not dependent on the structural response. According to Eqns. (40) and (43),

$$\frac{df_1}{d\boldsymbol{\rho}} = \frac{1}{V_f}\frac{1}{V_s}\tilde{\boldsymbol{v}}^T \tag{B 4}$$

## Appendix C: Sensitivity verification

In this appendix, the accuracy of the sensitivity derivations given in Section 5, Appendices A and B is verified using a double-clamped beam example with the central difference method (CDM) wherein a perturbation value of $h = 10^{-5}$ is used. The dimension, load and boundary conditions, and material properties of the double-clamped beam are given in Figure 9. The design domain is discretized by a 30×10 FE mesh with (random) density distribution and element numbering shown in Figure 42. The parameters in SIMP are chosen as $p = 3$, $p_L = 6$, $\epsilon = 10^{-8}$, and the parameters used in the pseudo-mass construction are the same as those used in Section 4.2.

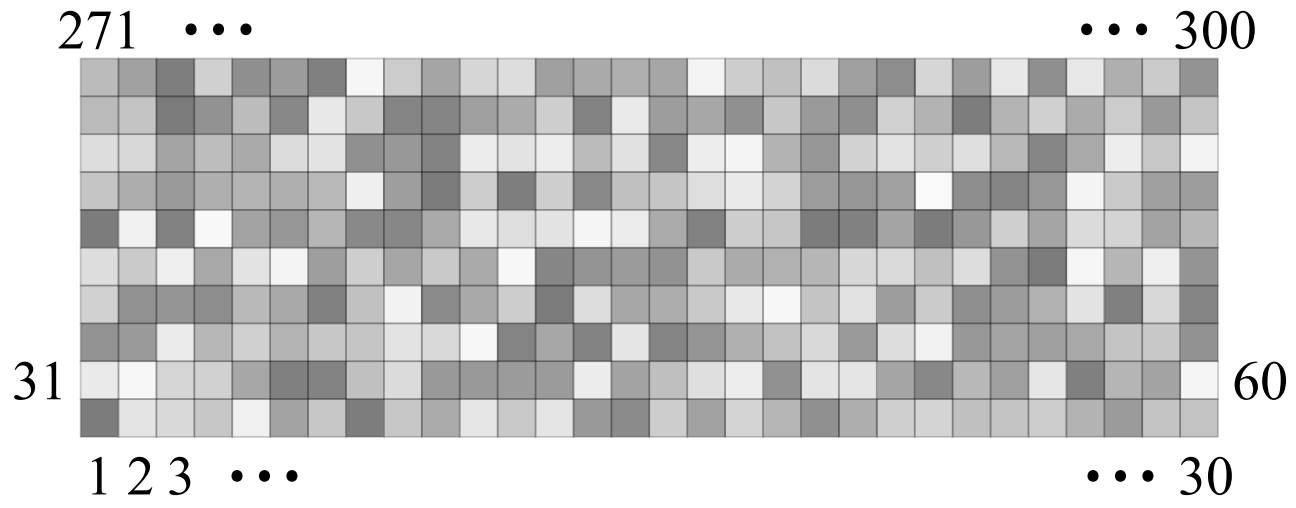

Figure 42. FE mesh, element numbering, and density distribution of the double-clamped beam for sensitivity verification.

With the density distribution shown in Figure 42, the first six eigenvalues are all simple. Figure 43 and Figure 44 show the verification results of the sensitivities of the eigenvalues ($df_i/d\boldsymbol{\rho}$ with $f_i$ in Eq. $(40)_3$) as well as the objective function ($df_0/d\boldsymbol{\rho}$), where the design sensitivities obtained from the proposed adjoint method are compared with those obtained via CDM. As shown in Figure 43 and Figure 44, the sensitivities computed using the adjoint method match closely with those from the CDM with relative errors between $10^{-4}$ to $10^{-10}$. It should be remarked that, although only the simple eigenvalue case is examined in the sensitivity verification, the ingredients used therein are the same as those in the multiple eigenvalue case, see Eqns. (25) and (39), and Appendix A. Hence, the adjoint sensitivity analysis is correct and can be safely used in topology optimization studies.

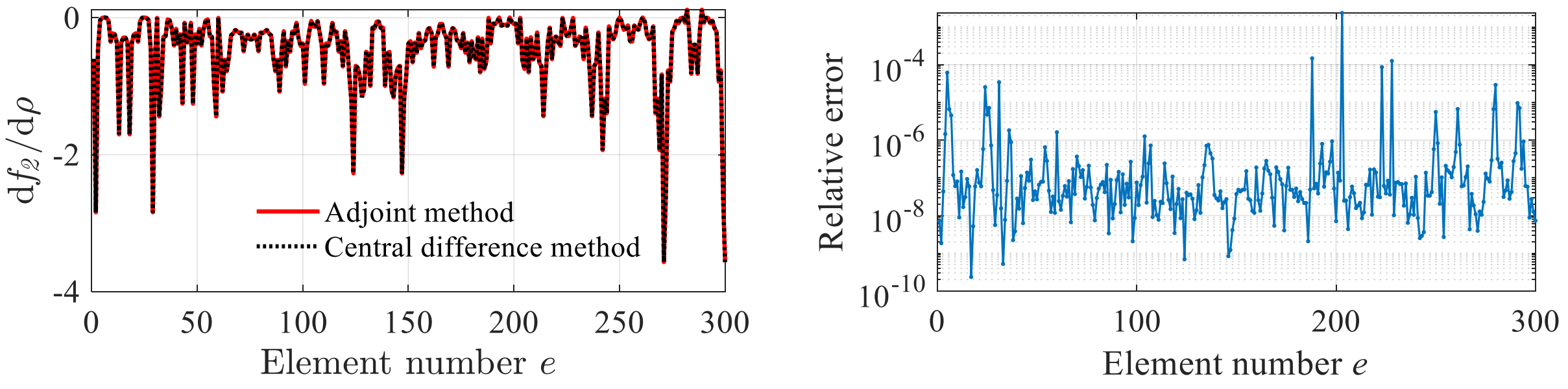

(a) Derivative of $df_2/d\boldsymbol{\rho}$

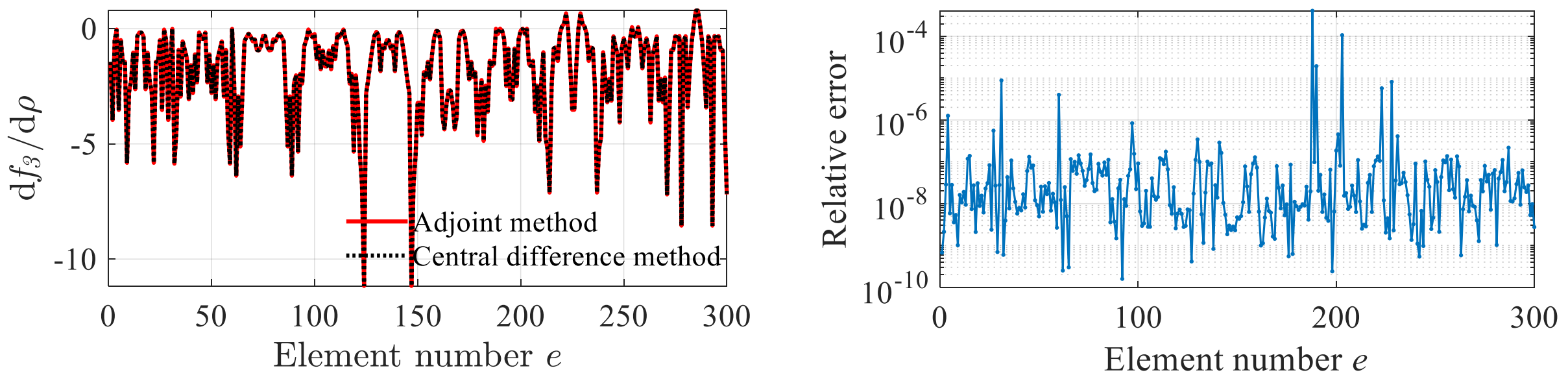

(b) Derivative of $df_3/d\boldsymbol{\rho}$

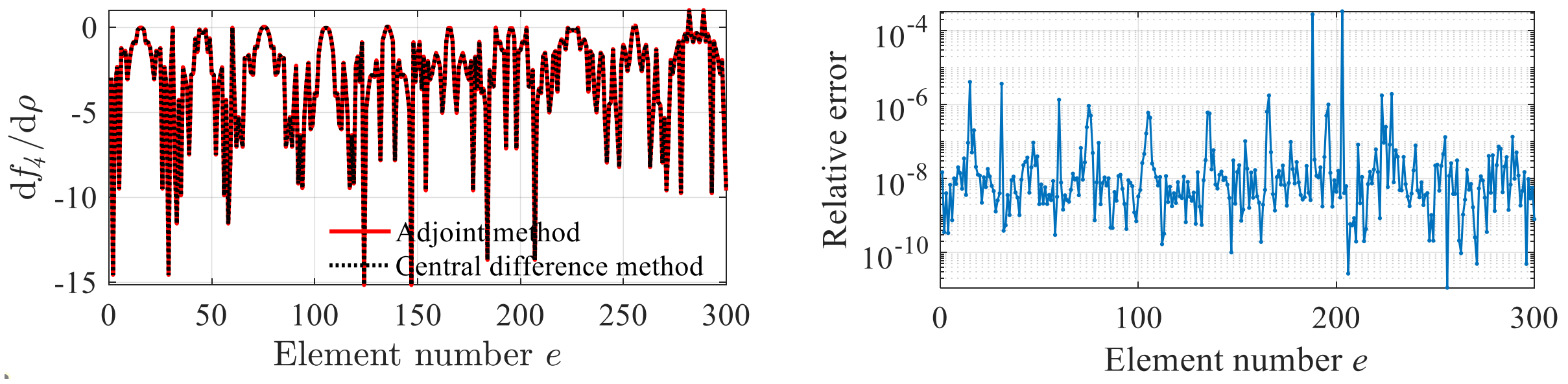

(c) Derivative of $df_4/d\boldsymbol{\rho}$

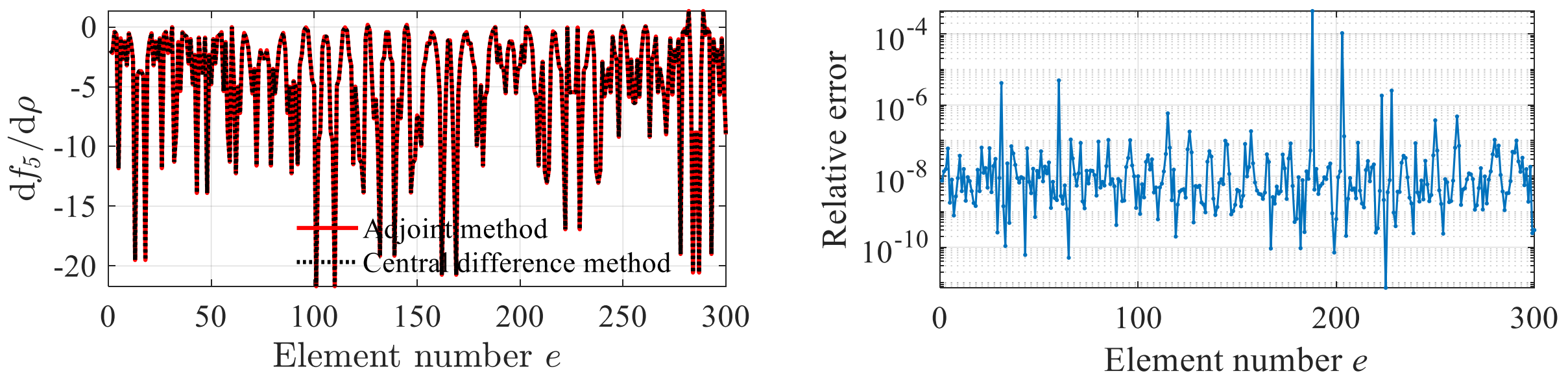

(d) Derivative of $df_5/d\boldsymbol{\rho}$

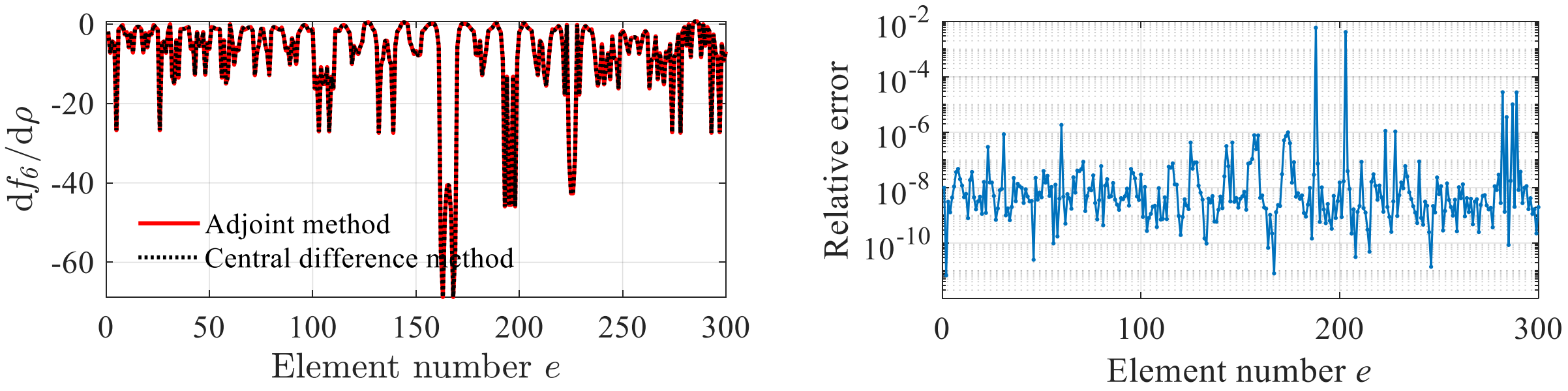

(e) Derivative of $df_6/d\boldsymbol{\rho}$

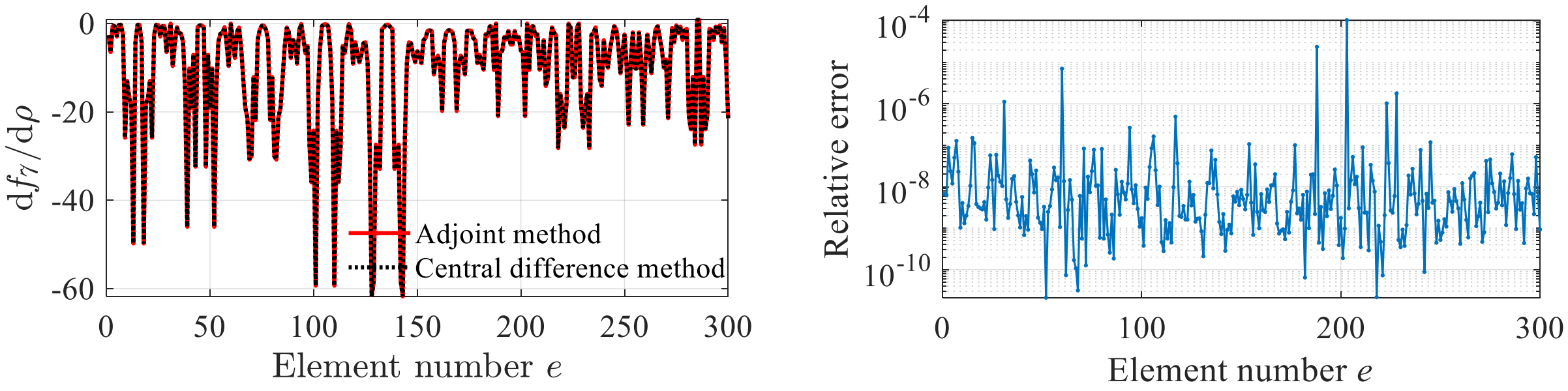

(f) Derivative of $df_7/d\boldsymbol{\rho}$

Figure 43. Comparison of the sensitivities of eigenvalue functions $f_i$ ($i$ = 2, …, 7) computed from the adjoint method and central difference method. Left: comparison between the two methods; Right: relative error of adjoint method w.r.t. central difference method.

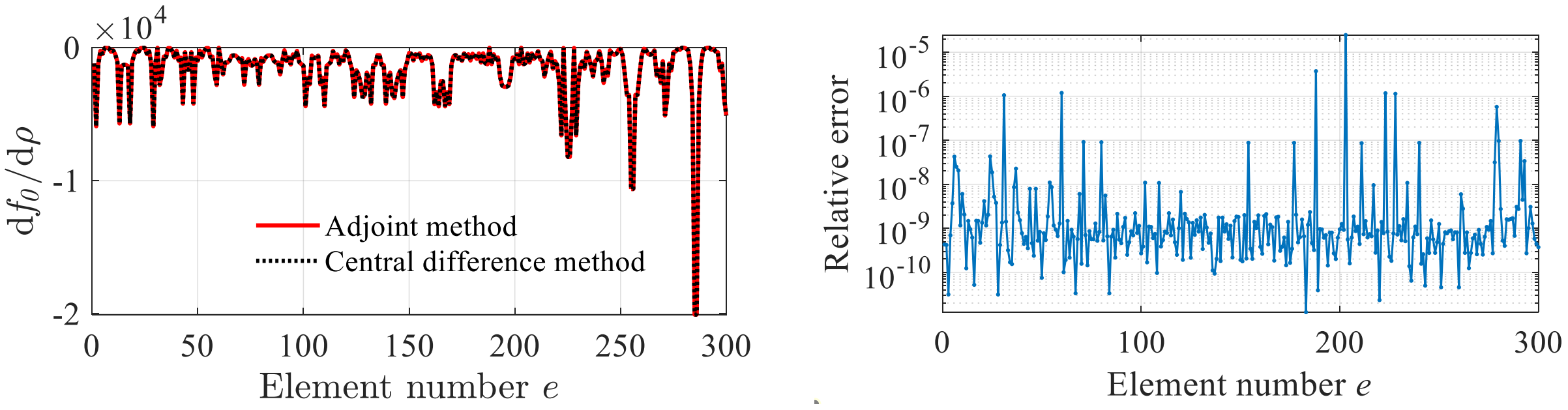

Figure 44. Comparison of the sensitivities of the objective function $f_0$ computed from the adjoint method and central difference method. Left: comparison between the two methods; Right: relative error of adjoint method w.r.t. central difference method.

## Appendix D: Branch-switching bifurcations

When the critical point is a limit point, the solution is unique (i.e., principal solution) and the arc-length continuation can be applied to trace the snap-through or possibly snap-back path. However, when the critical point is a bifurcation point, post-buckling analysis method has to be capable of switching from the primary branch to the secondary branch(es). Available treatments for branch-switching following simple/multiple bifurcation points can be found in Refs. [60-64]. In this appendix, a brief description of the approaches that are adopted in this study for handling branch-switching is presented.

### D 1. Simple bifurcation point

At the simple bifurcation point, there is only one bifurcated branch beside the principal/primary branch, and switching from the primary to the secondary branch can be straightforwardly accomplished by using a perturbed solution as the initial guess of the NR iteration for the 1st point on the secondary path emanating from the bifurcation point [60]. The perturbation considers the addition of a scaled eigenvector on the solution at the bifurcation point, i.e.,

$$\boldsymbol{u}^p = \boldsymbol{u}_{cr} + \zeta \frac{\boldsymbol{\phi}_1}{\|\boldsymbol{\phi}_1\|} \quad \text{(D 1)}$$

where $\boldsymbol{u}^p$ serves as the predictor in NR iteration for the first point on the secondary path following the bifurcation point. Here $\boldsymbol{u}_{cr}$ represents the solution at the critical (bifurcation) point and $\boldsymbol{\phi}_1$ is the eigenvector corresponding to the simple zero eigenvalue at the bifurcation point. The scaling factor $\zeta$, as suggested in the literature [61, 64], can be estimated by

$$\zeta = \pm \frac{\|\boldsymbol{u}_{cr}\|}{\tau} \quad \text{(D 2)}$$

where $\tau$ is a factor controlling the magnitude of the perturbation and is suggested to be chosen roughly of the order of 100 [61]. The sign of $\zeta$ indicates the direction to follow of the secondary branch. At simple bifurcation point, there exists three directions – one following the primary branch and the other two associated with the secondary branches. In the post-buckling analysis, with symmetry of the two directions, only one is followed and presented.

### D 2. Multiple bifurcation point

Compared to the simple bifurcation case, seeking the post-bifurcation branches at a multiple bifurcation point is much more computationally involved as the number of branches is not *a priori* known. Although the perturbation in Eq. (D 1) can be adopted by replacing the scaled eigenvector with the linear combination of the multiple eigenvectors [61], it is hard to obtain a full set of branches. A review of the existing branch-switching algorithms is given by Kouhia and Mikkola [63]. Among these algorithms, the method developed by Huitfeldt [65] by traversing a branch connecting curve based on the local perturbation algorithm is considered robust for computing all equilibrium branches. This method belongs to the class of *generalized path-following methods* [64] that can be seen as an extension of the traditional equilibrium path following to a multi-dimensional solution manifold exploring [14, 62, 64] via a multi-parametric setting (e.g., load factor, material properties, geometric dimensions, etc.) rather than only one-parameter (i.e., load factor) setting.

Following Refs. [62, 64], the global nonlinear equilibrium to be solved, when combined with an auxiliary equation and arc-length approach, can be expressed as

$$\boldsymbol{R}(\boldsymbol{u},\gamma,\gamma_s) = \begin{bmatrix} \boldsymbol{R}_u(\boldsymbol{u},\gamma,\gamma_s) \\ R_{aux}(\boldsymbol{u}) \\ R_{arc}(\boldsymbol{u}) \end{bmatrix} = \boldsymbol{0} \quad \text{with}$$

$$\boldsymbol{R}_u(\boldsymbol{u},\gamma,\gamma_s) = \boldsymbol{F}_{int}(\boldsymbol{u}) - \boldsymbol{F}_{ext}(\gamma,\gamma_s) \quad \text{with} \quad \boldsymbol{F}_{ext}(\gamma,\gamma_s) = \gamma\widehat{\boldsymbol{P}} + \gamma_s\widehat{\boldsymbol{P}}_s \tag{D 3}$$

$$R_{aux}(\boldsymbol{u}) = (\boldsymbol{u}-\boldsymbol{u}_{cr})^T(\boldsymbol{u}-\boldsymbol{u}_{cr}) - r^2$$

$$R_{arc}(\boldsymbol{u}) = \Delta\boldsymbol{u}^T\Delta\boldsymbol{u} - \ell^2$$

where $\widehat{\boldsymbol{P}}$ is the original external load vector (i.e., the term $\boldsymbol{F}_{ext}$ in Eq. (7)) scaled by a load factor $\gamma$, $\widehat{\boldsymbol{P}}_s$ stands for a disturbing force vector with $\gamma_s$ used to control the magnitude; $\boldsymbol{u}_{cr}$ denotes the known displacement solution at the critical point; $r$ serves as the radius of a cylinder enclosing the critical point $\boldsymbol{u}_{cr}$; $\ell$ specifies an arc-length parameter. Here $\boldsymbol{R}_u(\boldsymbol{u}, \gamma, \gamma_s) = \boldsymbol{0}$ defines a perturbed equilibrium state that is parameterized by two parameters $\gamma$ and $\gamma_s$, representing a 2-dimensional solution manifold in $\mathbb{R}^{n+2}$ with $\boldsymbol{u} \in \mathbb{R}^n$. The cylindrical auxiliary equation $R_{aux} = 0$ has been shown to outperform the spherical equation (e.g., $R_{aux}(\boldsymbol{u}, \gamma, \gamma_s) = (\boldsymbol{u} - \boldsymbol{u}_{cr})^T(\boldsymbol{u} - \boldsymbol{u}_{cr}) + (\gamma - \gamma_{cr})^2 + \gamma_s^2 - r^2$ with $\gamma_{cr}$ the load factor at the critical point) in Eriksson [62]. The intersection of $\boldsymbol{R}_u(\boldsymbol{u}, \gamma, \gamma_s) = \boldsymbol{0}$ and $R_{aux}(\boldsymbol{u}) = 0$ defines a closed one-dimensional curve in $\mathbb{R}^{n+2}$, the so-called *branch connecting curve* (BCC). It is clear that the subset of the solutions with $\gamma_s = 0$ are the solutions of the original equilibrium problem in Eq. (7). The curve passes exactly one point on each branch and the perturbation parameter $\gamma_s$ changes sign when passing such a point. Therefore, the goal is to locate the zero points of $\gamma_s$ while tracing the BCC. The BCC curve tracing is fulfilled by a cylindrical arc-length method [57] by adding $R_{arc}(\boldsymbol{u}) = 0$ where $\Delta\boldsymbol{u}$ is the displacement increment from the last step to the current solution step.

*Remarks:*

1. As suggested in Eriksson [62], the radius of the cylinder defined in $R_{aux}(\boldsymbol{u}) = 0$ is chosen as $r \approx 0.1\|\boldsymbol{u}_{cr}\|_2$.

2. For the selection of the disturbing force $\widehat{\boldsymbol{P}}_s$, in contrast to Eriksson [62] where a unit vector for the dominant component of the eigenmode is suggested, it is found that a random vector orthogonalized by the original force $\hat{\boldsymbol{P}}$ is more robust in achieving all the branches by traversing the BCC, i.e.,

$$\hat{\boldsymbol{P}}_s = \left(\boldsymbol{I} - \frac{1}{\|\hat{\boldsymbol{P}}\|}\hat{\boldsymbol{P}} \otimes \hat{\boldsymbol{P}}\right)\boldsymbol{\xi} \quad \text{where } \boldsymbol{\xi} = \begin{bmatrix} \xi_1 \\ \vdots \\ \xi_n \end{bmatrix} \text{ with } \xi_i \sim U(-1,1)\,,\ i = 1, \dots, n \qquad \text{(D 4)}$$

in which $U(-1,1)$ denotes a uniform distribution ranging from -1 to 1. $\hat{\boldsymbol{P}}_s$ can be further normalized by dividing by its norm $\|\hat{\boldsymbol{P}}_s\|_2$.

3. For determining the solution points with $\gamma_s = 0$, as mentioned in Refs [62, 64], a linear interpolation between the positive and negative $\gamma_s$'s can be a good approximation. However, with a relatively big arc-length parameter $\ell$, the accuracy of the solution $(\boldsymbol{u}, \gamma)$ might be sabotaged. In this scenario, NR iteration can be applied to improve the solution accuracy.

4. The arc-length continuation terminates when the BCC curved is closed, i.e., the last found solution point [$(\boldsymbol{u}, \gamma)$ at $\gamma_s = 0$] is the same as the first found solution.

5. Each of the bifurcated branches has two directions (see Figure 40a or Figure 40b). Using the presented BCC traversing approach with an appropriately selected disturbing load vector, it is expected that all the secondary branches in both directions can be computed.

6. The system of nonlinear equations in Eq. (D 3) is solved for the first point on each secondary branch. After that, the cylindrical arc-length method is used again starting from the found point to trace along each of the secondary branches to finish the post-buckling analysis.